\documentclass[twocolumn]{aastex63}

\newcommand{\oi}{O\,{\scriptsize I}}
\newcommand{\oii}{O\,{\scriptsize II}}
\newcommand{\oiii}{O\,{\scriptsize III}}
\newcommand{\caii}{Ca\,{\scriptsize II}}

\newcommand{\ciii}{C\,{\scriptsize III}}
\newcommand{\sii}{S\,{\scriptsize II}}

\newcommand{\nii}{N\,{\scriptsize II}}

\newcommand{\niv}{N\,{\scriptsize IV}}
\newcommand{\neiii}{Ne\,{\scriptsize III}}
\newcommand{\nv}{N\,{\scriptsize V}}
\newcommand{\civ}{C\,{\scriptsize IV}}
\newcommand{\hi}{H\,{\scriptsize I}}

\newcommand{\hei}{He\,{\scriptsize I}}
\newcommand{\heii}{He\,{\scriptsize II}}
\newcommand{\feii}{Fe\,{\scriptsize II}}
\newcommand{\hdelta}{H$\delta$}
\newcommand{\hbeta}{H$\beta$}
\newcommand{\halpha}{H$\alpha$}
\newcommand{\lya}{Ly$\alpha$}

\newcommand{\jwst}{JWST}

\usepackage{rotating}
\usepackage{amsmath}

\graphicspath{{./}{figures/}}

\shorttitle{The Spectral Features and Properties of High-Redshift Galaxies}
\shortauthors{Roberts-Borsani et al.}

\begin{document}

\title{Between the Extremes: A JWST Spectroscopic Benchmark for High-redshift Galaxies Using $\sim$500 Confirmed Sources at $z\geqslant5$}

\correspondingauthor{Guido Roberts-Borsani}
\email{guido.roberts-borsani@unige.ch}
\author[0000-0002-4140-1367]{Guido Roberts-Borsani}
\affiliation{Department of Astronomy, University of Geneva, Chemin Pegasi 51, 1290 Versoix, Switzerland}

\author[0000-0002-8460-0390]{Tommaso Treu}
\affiliation{Department of Physics and Astronomy, University of California, Los Angeles, 430 Portola Plaza, Los Angeles, CA 90095, USA}

\author[0000-0003-3509-4855]{Alice Shapley}
\affiliation{Department of Physics and Astronomy, University of California, Los Angeles, 430 Portola Plaza, Los Angeles, CA 90095, USA}

\author[0000-0003-3820-2823]{Adriano Fontana}
\affiliation{INAF - Osservatorio Astronomico di Roma, Via Frascati 33, 00078 Monteporzio Catone, Rome, Italy}

\author[0000-0001-8940-6768 ]{Laura Pentericci}
\affiliation{INAF - Osservatorio Astronomico di Roma, Via Frascati 33, 00078 Monteporzio Catone, Rome, Italy}

\author[0000-0001-9875-8263]{Marco Castellano}
\affiliation{INAF - Osservatorio Astronomico di Roma, Via Frascati 33, 00078 Monteporzio Catone, Rome, Italy}

\author[0000-0002-8512-1404]{Takahiro Morishita}
\affiliation{IPAC, California Institute of Technology, MC 314-6, 1200 E. California Boulevard, Pasadena, CA 91125, USA}

\author[0000-0003-1383-9414]{Pietro Bergamini}
\affiliation{Dipartimento di Fisica, Universit\`a degli Studi di Milano, Via Celoria 16, I-20133 Milano, Italy}
\affiliation{INAF - Osservatorio di Astrofisica e Scienza dello Spazio di Bologna, via Gobetti 93/3, I-40129 Bologna, Italy}

\author[0000-0002-6813-0632]{Piero Rosati}
\affiliation{INAF - Osservatorio di Astrofisica e Scienza dello Spazio di Bologna, via Gobetti 93/3, I-40129 Bologna, Italy}
\affiliation{Dipartimento di Fisica e Scienze della Terra, Universit\`a degli Studi di Ferrara, Via Saragat 1, I-44122 Ferrara, Italy}

\begin{abstract}
The exceptional spectra of the most luminous $z>10$ sources observed so far have challenged our understanding of early galaxy evolution, requiring a new observational benchmark for meaningful interpretation. As such, we construct spectroscopic templates representative of high-redshift, star-forming populations, using 482 confirmed sources at $z=5.0-12.9$ with JWST/NIRSpec prism observations, and report on their average properties. We find $z=5-11$ galaxies are dominated by blue UV continuum slopes ($\beta=-2.3$ to $-2.7$) and reduced Balmer indices, characteristic of dust-poor and young systems, with a shift towards bluer slopes and younger ages with redshift. The evolution is mirrored by ubiquitous \ciii] detections across all redshifts (rest-frame equivalent widths of $5-14$ \AA), which increase in strength towards early times. Rest-frame optical lines reveal elevated ratios ($O32=7-31$, $R23=5-8$, and $Ne3O2=1-2$) and subsolar metallicities (log\,(O/H)$=7.3-7.9$), typical of ionization conditions and metallicities rarely observed in $z\sim0$ populations. Within our sample, we identify 57 \lya\ emitters, which we stack and compare to a matched sample of nonemitters. The former are characterized by more extreme ionizing conditions with enhanced \ciii], \civ, and \heii+[\oiii] line emission, younger stellar populations from Balmer jumps, and a more pristine interstellar medium seen through bluer UV slopes and elevated rest-frame optical line ratios. The novel comparison illustrates important intrinsic differences between the two populations, with implications for \lya\ visibility. The spectral templates derived here represent a new observational benchmark with which to interpret high-redshift sources, lifting our constraints on their global properties to unprecedented heights and extending out to the earliest of cosmic times.
\end{abstract}

\keywords{galaxies: high-redshift, galaxies: ISM, galaxies: star formation, cosmology: dark ages, reionization, first stars}

\section{Introduction}
Determining the properties of sources that emerged in the early Universe and transformed it into a fully ionized one represents one of the major milestones of modern extragalactic astronomy, and a prerequisite toward charting the evolution of galaxies from Cosmic Dawn to the present day. The most fundamental of these are the ages, masses, and metallicities of their stellar populations, the ionizing conditions and metal and dust contents of their interstellar medium (ISM), and the presence or otherwise of supermassive black holes. Stellar and emission-line features at rest-frame optical wavelengths ($\lambda\simeq$ 3800-6700 \AA) underpin most of these properties and are critical in determining if and how they evolve between sources at the end of the Reionization Era ($z\lesssim5.3-5.8$; \citealt{eilers18,bosman22,Zhu2023}) and those firmly within it ($z\gtrsim7.5$; \citealt{treu13,mason19}). Direct probes of the rest-frame optical spectrum in representative $z\geqslant5$ sources, therefore, represents a necessary step forward in deducing whether Reionization Era sources reflect distinct populations compared to their lower-redshift counterparts.

Until recently, inferences of such properties in $z\sim5-13$ sources -- probing the aftermath, the conclusion, and the unfolding of the reionization process -- have relied primarily on photometric studies of the rest-frame ultraviolet (UV) with imaging from the Hubble Space Telescope (HST) and (when available) the rest-frame optical from the Spitzer Space Telescope (e.g., \citealt{labbe13,smit14,bouwens15,strait20,rb22,yang22,whitler23}), revealing properties rarely seen in the local Universe. Analyses of individual sources at $z\sim6-9$ have painted a picture of generally compact (a few kiloparsecs or even subkiloparsec), young ($\sim1-50$ Myr) and star-forming ($\sim$1-100 $M_{\odot}$/yr) sources with subsolar ($\sim0.2-0.8 Z_{\odot}$) metallicities, low dust contents ($A_{\rm v}<1$ mag), and extreme rest-frame optical line emission ([\oiii]$\lambda\lambda$4960,5008 \AA\ + H$\beta$ equivalent widths (EWs) in excess of $>1000$ \AA). Stacks of Spitzer photometry (e.g., \citealt{labbe13,stefanon23}), yielding higher signal-to-noise ratio (SNR) measurements out to $z\sim10$, reaffirmed such a picture and provided an early benchmark for the interpretation of high-redshift sources. 

The advent of $>2 \mu$m filters from the James Webb Space Telescope's (\jwst) Near Infrared Camera (NIRCam) has extended the redshift frontier for galaxy formation studies, with the enhanced sensitivity and narrower bandpasses revealing bluer, compact, and star-forming sources with intense line emission out to $z\sim11-13$ (e.g., \citealt{topping22,topping23,cullen23,endsley23,laporte23,leethochawalit23a,ono23,santini23,withers23}). However, such analyses have generally relied on photometric redshifts, spectral energy distribution (SED) modeling using synthetic templates, and scaling relations calibrated to low-redshift samples.

Spectroscopy remains the ultimate tool for confirmation of galaxy redshifts and the interpretation of their global properties. While spectroscopic studies over statistical and magnitude-limited $z\geqslant5$ Lyman break galaxy (LBG) samples have already been attempted with ground-based facilities (e.g., \citealt{treu13,jung18,pentericci18,mason19,endsley21}), these have typically targeted the rest-frame UV Ly$\alpha$ emission line,  which is resonantly scattered from the line of sight in a neutral medium (with exceptions over smaller samples; e.g., \citealt{stark17,laporte17b,mainali18}, or gas tracers in the far-infrared; e.g., \citealt{bouwens21_rebels}). For the few luminous sources where spectroscopic redshifts could be determined, some characterization of their rest-frame optical properties (e.g., metallicities, specific star formation rates (SFRs), [\oiii]+H$\beta$ EW distributions) has been possible with Spitzer/IRAC photometry \citep{hashimoto18,laporte17,endsley21,endsley21b,rb23}. Direct spectroscopic characterization of the rest-frame optical has however, until recently, remained beyond the reach of $z\geqslant5$ sources and achievable only in ground-based surveys of lower-redshift ($z\sim1-4$) populations (e.g., the MOSDEF and VANDELS surveys; \citealt{kriek15,mclure18}). As such, constraints over the full rest-frame UV to optical of high redshifts sources remain sparse, and their average spectra and properties unknown.

\jwst's Near Infrared Spectrograph (NIRSpec) has proven remarkably efficient at confirming galaxy candidates out to $z\sim14$ and affirm their underlying properties across cosmic time (e.g., \citealt{arrabal23b,curtislake23,harikane23b,mascia23,morishita22,nakajima23,rb23_nature,sanders23,tang23,williams22,curti23,carniani24,rb24_borg}). The revelation of spectacular and perplexing spectra of GN-z11 \citep{bunker23} and GHZ2 \citep{castellano24}, however, has placed our already limited understanding of galaxy evolution at the earliest of times in doubt. Whether these sources comprise atypical or somewhat representative galaxies for their luminosities and epochs remains an open question, one where comparisons to photoionization models or low-redshift analogs have proved challenging. Indeed, comparisons to observations of high-redshift sources offer a more direct framework with which to interpret such observations, however the generally low SNR of individual spectra, relatively small samples sizes, and scatter in their inferred properties has made such comparisons challenging. Determining the ``typical'' spectrum of a high-redshift source (e.g., \citealt{shapley03}) with which to establish a representative benchmark and template to place individual sources into context, therefore, represents an ambitious challenge and one which would significantly enhance our capabilities in the interpretation of high-$z$ observations. 

The primary objectives of this paper are to (i) study the average spectra of a large number of high-redshift galaxies in order to understand their properties at a population level, while (ii) simultaneously providing high SNR composite templates of observed galaxy spectra to the community. The paper is organized as follows. In Section~\ref{sec:data} we present the spectroscopic and photometric data sets used in this study, and present our sample selection and stacking methods in Section~\ref{sec:sample}. In Section~\ref{sec:spectralfeatures} we present stacks of galaxy spectra as a function of global properties, and report on their main spectral features and line ratios. Section~\ref{sec:spectraprops} presents the derivation of global galaxy properties from those spectra and places them into context with comparisons to known objects and literature studies. Finally, we summarize our findings and present our conclusions in Section~\ref{sec:concl}. All stacked spectra presented in this study are available on the GLASS-JWST website\footnote{\url{https://glass.astro.ucla.edu/ers/}}, while redshifts are listed in Tables~\ref{tab:redshifts1}-\ref{tab:redshifts2} and line fluxes in Table~\ref{tab:line_ratios_redshift} in the Appendix.

\section{Data Sets}
\label{sec:data}
\subsection{NIRSpec Prism Spectroscopy}
\label{subsec:nirspec}
Given its large wavelength coverage, sensitivity to both line and continuum emission, and the number of observed sources, this analysis focuses on observations and objects where NIRSpec prism spectra are available. Our data sets derive primarily from publicly available programs or data releases (at the time of writing) using the multishutter array (MSA), and include observations over the Abell 2744, RXJ2129, and MACS 0647 lensing clusters through the GLASS-JWST Director's Discretionary Time (DDT) proposals (PID 2756; see \citealt{morishita22} and \citealt{rb23}), GO 3073 \citep{castellano24}, UNCOVER (PID 2561; see \citealt{bezanson22}), DDT 2767 (see \citealt{williams22}), and GO 1433 (see \citealt{hsiao23}), respectively, as well as the CANDELS-EGS, CANDELS-UDS, GOODS-North, and GOODS-South blank fields through CEERS (PID 1345; see \citealt{bagley23,finkelstein23}) and its associated DDT (PID 2750; see \citealt{arrabal23}), GO 2565, and the JADES survey (GTO 1181 and GTO 1211 for GOODS-North, GTO 1180, GTO 1210, and GO 3215 for GOODS-South; \citealt{bunker_jades}), respectively. In total, the data sets amount to 13 programs over seven unique fields, each affording low resolution ($R\simeq30-300$) spectroscopy with a continuous wavelength coverage from $\sim0.6-5.3 \mu$m. A list of the data sets along with their targeted fields is presented in Table~\ref{tab:data}.

With the exception of the JADES GTO 1210 program, we download, reduce, and extract all public NIRSpec/MSA spectra following the procedure described in \citet{morishita22}. In short, the count-rate (\texttt{\_rate.fits}) exposures are downloaded from the MAST archive and reduced with the \texttt{msaexp}\footnote{\url{https://github.com/gbrammer/msaexp}} code, which adopts the official STScI pipeline modules in addition to custom routines (including flat-fielding, snowball masking, 1/$f$ removal, and others). For a given object, the code then extracts the 2D spectrum from each exposure, rectifies and drizzles it to a common pixel grid, and background subtracts it using its associated dithers before combining these via an inverse-variance-weighted mean with outlier rejection. Individual exposures affected by catastrophic events (e.g., transient optical shorts or significant contamination) are discarded prior to combining. The final, rectified 2D spectrum is then used together with an optimal extraction procedure \citep{horne86} to extract a 1D spectrum with associated uncertainties. In the case of the JADES GTO 1210 spectra, we use the data products provided by the collaboration. All spectra are visually inspected for quality control and the 1D extraction repeated or optimized if required.

\begin{deluxetable*}{llllcl}
\tabletypesize{\footnotesize}
\tablewidth{0pt}
\tablecaption{The JWST/NIRSpec Prism Data Sets Used in This Study and the Number of Confirmed $z\geqslant5$ Sources Identified Here at the Time of Writing.}
\label{tab:data}
\tablehead{
\colhead{Prog. name} & 
\colhead{PI} & 
\colhead{Target} &
\colhead{Prog. ID} &
\colhead{$t_{\rm int}$} &
\colhead{$N_{\rm z\geqslant5}$}
}
\startdata
GLASS & Chen & Abell 2744 & DDT 2756 & 0.6/1.8 & 12 \\
GLASS & Castellano & Abell 2744 & GO 3073 & 1.8/9.7 & 44 \\
UNCOVER & Labb\'e & Abell 2744 & GO 2561 & 4.4/6.6 & 105 \\
CEERS & Finkelstein & CANDELS-EGS & ERS 1345 & 0.9/3.4 & 113 \\
CEERS & Arrabal & CANDELS-EGS & DDT 2750 & 5.1/0.2 & 22 \\
GO-Glazebrook & Glazebrook & CANDELS-UDS & GO 2565 & 0.5/0.5 & 4 \\
GO-Coe & Coe & MACS 0647 & GO 1433 & 1.8/3.6 & 19 \\
GO-Kelly & Kelly & RXJ2129 & GO 2767 & 1.2/2.4 & 4 \\
JADES & Eisenstein & GOODS-S & GTO 1180 & 1.0/3.1 & 57 \\
JADES & Luetzgendorf & GOODS-S & GTO 1210 & 9.3/7.8 & 47 \\
JADES & Eisenstein & GOODS-S & GO 3215 & 27.7/6.2 & 38 \\
JADES & Eisenstein & GOODS-N & GTO 1181 & 0.9/0.9 & 20 \\
JADES & Luetzgendorf & GOODS-N & GTO 1211 & 0.7/0.7 & 5 \\
\enddata
\tablecomments{The integration times refer to the median (left) and maximum (right) integration times (in hours) for the parent sample of $z\geqslant5$ sources used in this study.}
\vspace{-0.5cm}
\end{deluxetable*}

\subsection{Uniformly Processed HST and JWST Photometry}
\label{subsec:photo}
We also leverage deep JWST/NIRCam and/or HST/ACS+WFC3 imaging for spectrophotometric scaling and verification. The images are downloaded from the \texttt{Grizli} v6 Image Release\footnote{\url{https://github.com/gbrammer/grizli/blob/master/docs/grizli/image-release-v6.rst}}, a set of publicly available NIRCam and HST images deriving from public programs that were uniformly reduced over a consistent astrometric reference frame with the STScI pipelines and the \texttt{Grizli} code \citep{grizli}. The subset of NIRCam images used here derive from the programs listed in Table~\ref{tab:data} and the ERS 1324 program (GLASS-JWST; \citealt{treu22}), while the \text{HST} data derive from legacy surveys such as the Frontier Fields \citep{lotz17}, RELICS \citep{coe19}, and CANDELS \citep{grogin11} programs.

To derive photometric catalogs we use \texttt{Grizli}'s implementation of the Python wrapper for \texttt{SExtractor}, \texttt{SEP} \citep{bertin96,barbary2016}, in conjunction with an infrared detection image derived from an inverse-variance-weighted mean of the NIRCam F115W, F277W and F444W images, where available. In some cases, the reduced NIRSpec spectra lie outside of an NIRCam footprint, and in such cases we adopt the photometry from the HST images (using a stacked WFC3/F125W+F160W infrared image as reference). For the Extended Groth Strip (EGS) field, we extend the covered area with public CANDELS images\footnote{\url{https://archive.stsci.edu/hlsp/candels}} (ACS/F606W+F814W and WFC3/F125W+F160W), since these extend slightly beyond the \texttt{Grizli} images. Although Spitzer/IRAC data also exist for a large fraction of the total area considered, we opt not to use them given the comparatively shallower depth and poorer spatial resolution, which could lead to significant confusion in crowded fields. For galaxies residing in the JADES-Deep survey, we utilize the publicly released catalogs from \citet{rieke23} and \citet{eisenstein23}. In all cases we derive and/or utilize aperture photometry measured in fixed, $0''.5$ diameter circular apertures. The choice of aperture size is motivated to encapsulate a maximum of target flux while minimizing potential contamination -- we find the vast majority of our galaxy sample (see Section \ref{sec:sample}) are adequately encompassed without leaving room for contamination from nearby compact sources or light from extended objects in e.g., lensing clusters.

\section{Sample Selection and Methods}
\label{sec:sample}
\subsection{Redshift Measurements of $z\geqslant5$ Sources}
We verify the redshifts of all NIRSpec/prism galaxies identified and extracted in Section \ref{subsec:nirspec} using the 1D spectrum and the redshift-fitting module of \texttt{msaexp}, which evaluates a $\chi^{2}$ goodness of fit as a function of redshift using a series of \texttt{EAzY} \citep{brammer_eazy} galaxy and line templates at the nominal NIRSpec/prism resolution. We fit each galaxy using a redshift range between $z=0-15$ and visually inspect the resulting solutions. In total, our procedure yields 439 spectra at $z\geqslant5$ with secure redshifts, many of which have already been reported in the literature (e.g., \citealt{rb23,tang23,fujimoto23,williams22,arrabal23b,wang22,morishita22,mascia23,nakajima23,harikane23,napolitano24}) while a large number (approximately 69\%) represent novel confirmations. We further supplement our sample of galaxies with a further 48 sources confirmed at $z\geqslant5$ in the JADES-Deep survey and listed in \citet{bunker_jades}, increasing the total to 487 sources. For those sources where multiple observations or multiple lensed images exist (e.g., \citealt{rb23_nature,hsiao23}), we combine the (magnification-corrected, where applicable) observations via an inverse-variance-weighted mean, thus counting them as a single observations or source. In total, this results in a final sample of 482 confirmed sources.

A requirement of our selection is to be able to scale sources for an overall slit loss in the rest-frame UV (see Section \ref{subsec:datacheck} below). A number of objects in our sample suffer from a significant gap in the relevant portion of their NIRSpec spectrum, a lack of public photometry, or insufficient continuum flux to be able to do so. Therefore, those sources are discarded from our primary analysis and our fiducial sample is reduced to 454 unique sources. The full list of spectroscopic redshifts for our fiducial sample is tabulated in Table~\ref{tab:redshifts1}, along with several key spectrophotometric properties. Finally, in Table~\ref{tab:redshifts2} we also report the redshifts for those sources where a photometric scaling could not be performed, as well as for an additional 34 objects for which the redshift is not securely constrained -- while the \texttt{msaexp} code favors a high-$z$ solution, these sources either lack strong spectral features or SNR to unambiguously confirm their redshifts and are thus not included in any subsequent analysis.

\subsection{Bad Pixel Masking and Quality Control}
\label{subsec:datacheck}
We consider a number of additional checks for quality control. Firstly, all (1D and 2D) spectra are visually inspected by two authors (G.R.B. and T.T.) and pixels corresponding to clear contamination or artifacts are masked -- examples of these include detector gaps, bad background subtraction, and/or contamination from nearby objects, among others. Furthermore, we take steps to ensure the spectra are not affected by any lingering instrumental artifacts not completely accounted for by the pipeline. The main ones considered here (where relevant) are (i) the scaling of the spectrum to a rest-frame UV flux in order to account for the overall slit loss, and (ii) determining whether wavelength-dependent slit loss effects are present or significant in the prism observations. One limitation we do not account for is any potential difference between the photometry and spectroscopy attributable to spatially varying galaxy properties. For (i), we utilize the extracted HST or NIRCam photometry described in Section \ref{subsec:photo} to account for the overall slit loss by scaling the spectrum to the closest filter (preferentially NIRCam but also HST if the former is lacking) corresponding to a rest-frame wavelength of 1750 \AA, where the photometry probes the rest-frame UV but is unaffected by Lyman-$\alpha$ emission or the Lyman break. For (ii), we use the available NIRCam photometry longward of the Lyman break to evaluate whether wavelength-dependent slit loss is a prevalent effect in prism observations. A detailed discussion of this residual slit loss and the derivation of an average correction function is presented in the Appendix, from which we find minor losses of 1-2\% in $<2 \mu$m NIRCam bands and up to 10-20\% losses (on average) in redder bands. We derive a correction function from these average slit losses and correct all our composite spectra with this function prior to any analysis. As with our composite spectra, we render the correction function publicly available for community use.

\subsection{Magnification Factors of Lensed Galaxies}
\label{subsec:magnification}
\begin{figure*}
\center
 \includegraphics[width=\textwidth]{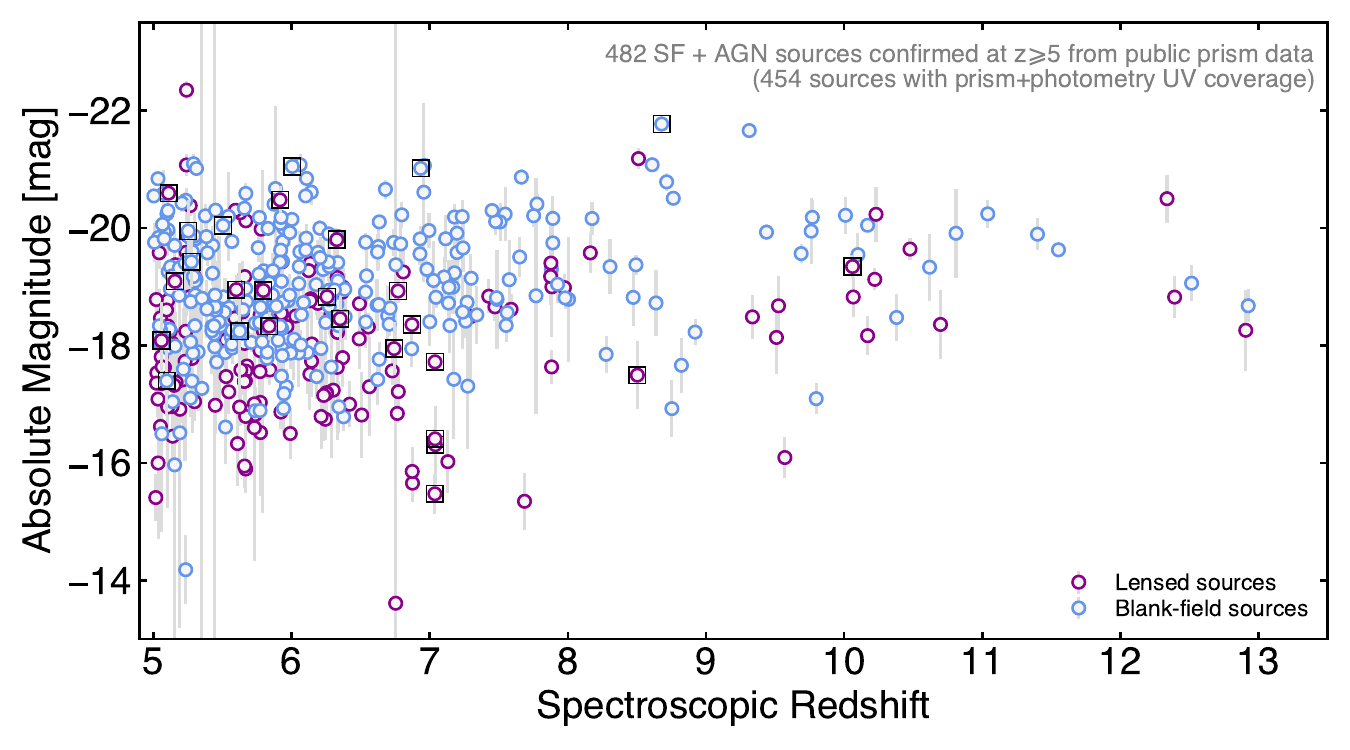}
 \caption{The sample of spectroscopically confirmed sources at $z\geqslant5$ compiled in this work (listed in Tables~\ref{tab:redshifts1}-\ref{tab:redshifts2} in the Appendix), with continuum measurements available for spectrophotometric scaling in the rest-frame UV. Sources identified in blank fields are shown in blue, while lensed sources (approximately 37\% of the parent sample) are plotted in purple. Sources considered to be AGN in this study are indicated with black squares.}\label{fig:zdist}
\end{figure*}

A significant fraction ($\simeq$37\%) of our galaxy sample derives from lensing cluster fields (see Table~\ref{tab:data}) and as such require correction for lensing magnification. For objects behind the A2744 cluster, we use the magnification maps from \citet{bergamini23}, which make use of the recently confirmed $z_{\rm spec.}=9.75$ triply imaged galaxy by \citet{rb23_nature} and 149 multiple images for which a large number include spectroscopic redshifts. For objects behind the RXJ2129 cluster we adopt the magnification maps of \citet{caminha19}, while for objects behind the MACS 0647 cluster we adopt the magnification maps of \citet[][private communication]{meena23}. The spectra and photometry of all lensed sources in our sample of confirmed galaxies are corrected for their adopted magnification factors prior to any analysis, and the $M_{\rm UV}$ values are derived using the measured flux and standard deviation of the resulting spectra in a wavelength window $\lambda_{\rm rest}=1500-2000$ \AA. The distribution of our fiducial sample in the $z-M_{\rm UV}$ plane is shown in Figure~\ref{fig:zdist}.

\subsection{Identification of Active Galactic Nuclei}
\label{subsec:AGNID}
One of the most surprising findings from early JWST results has been the revelation of significant numbers of AGN in the early Universe (e.g., \citealt{harikane23b,larson23,greene23,goulding23,maiolino23,kocevski23}). While optimal identification methods and the nature of some objects has yet to reach consensus, we err on the side of caution and flag all AGN objects identified in the literature (i.e., objects identified as AGN in \citealt{harikane23b}, \citealt{larson23}, \citealt{greene23}, \citealt{goulding23}, \citealt{maiolino23}, \citealt{kocevski23}). Furthermore, for all objects in our sample not included in those analyses, we evaluate via visual inspection whether a two-component (i.e., broad plus narrow) model is clearly required to reproduce the H$\alpha$ profile when the [\oiii]$\lambda\lambda$4960,5008 \AA\ profile is well fit by a single, narrow component (characteristic of Type 1 AGN). We opt not to extend such an analysis to measurements of other Balmer lines (e.g., H$\beta$, H$\gamma$, and H$\delta$) since they often lack sufficient SNR in individual spectra for robust fits. This amounts to 27 AGN candidates (identified here and in the literature), which are flagged (also listed in the bottom half of Table~\ref{tab:redshifts1}) and not utilized in the subsequent analyses. Of course, such a procedure does not preclude some degree of AGN (broad- or narrow-line) contamination remaining in our sample, particularly at redshifts $z>7$ where H$\alpha$ moves beyond the NIRSpec coverage, however these are likely to be weak AGN whose identifications are beyond the typical (per pixel) SNR ratios of the individual spectra used here.

\subsection{A Representative Sample of $z\geqslant5$ Galaxies}
With our fiducial sample of 427 confirmed star-forming galaxies in hand, we assess the degree of our targets as a representative sample of high-redshift sources by evaluating whether they adequately populate well-known planes of global galaxy properties. One of the most ubiquitous correlations in galaxy evolution is the so-called ``main sequence'' \citep{noeske07,speagle14}, i.e., that between galaxy SFRs and stellar masses ($M_{*}$) which is expected to reflect the statistical accumulation of stellar mass from smooth gas accretion and minor mergers. While individual sources may temporarily lie above or below the main sequence through stochastic levels of star formation (e.g., from mergers), generally speaking a source's position on the SFR-$M_{*}$ diagram dictates whether it represents a ``typical'' source at a given epoch.

To illustrate the regions of parameter space populated by our sample, we derive global galaxy properties following the procedure outlined by \citet{morishita23}. In short, we utilize a best-fit continuum model (see Section \ref{sec:method} below) at the redshifts of the photometrically scaled and magnification-corrected NIRSpec spectra to obtain stellar masses and SFRs, the latter derived using the rest-frame UV luminosity ($L_{\rm UV}$) measured at $\sim1600$ \AA\ from the best-fit SED. To account for dust attenuation, each $L_{\rm UV}$ is corrected using the measured UV continuum slope and the $A_{\rm 1600}-\beta$ relation from \citet{meurer99} -- in principle, such a relation is valid only down to slopes of approximately $-2.2$, whereas a large number of our sample display even bluer slopes. In such cases, we assume negligible dust attenuation and do not correct the UV luminosity. The attenuation-corrected luminosity is then converted to an SFR using the \citet{kennicutt98} relation,
scaled to a \citet{chabrier03} initial mass function (IMF):

\begin{equation}
\rm SFR\,[M_{\odot}yr^{-1}] = 8.8\times10^{-29}\,L_{\rm UV}\,[erg\,s^{-1}\,Hz^{-1}]
\end{equation}

\begin{figure}
\center
 \includegraphics[width=\columnwidth]{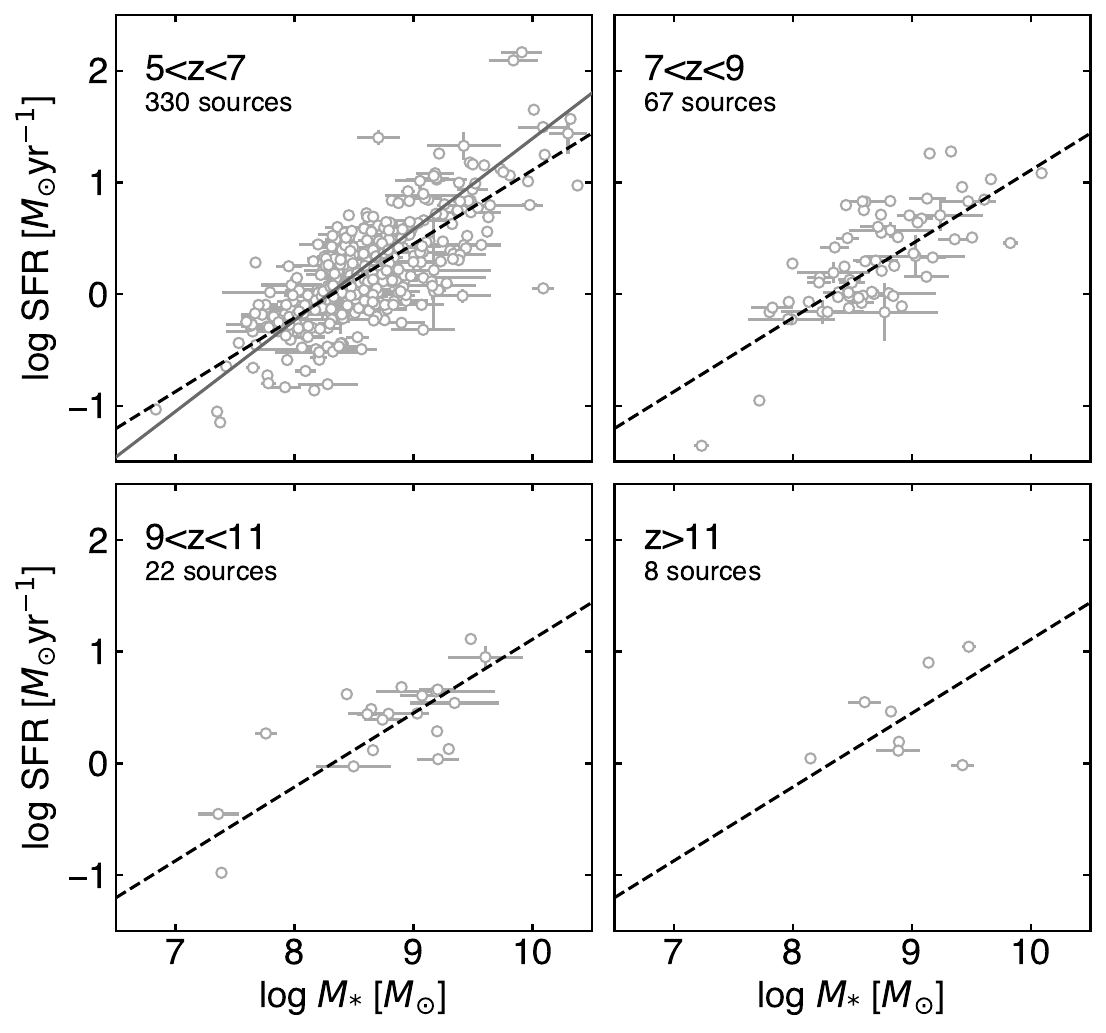}
 \caption{The star-forming main sequence as defined by our sample of confirmed sources, in four redshift bins. Stellar masses are derived from best-fit \texttt{gsf} SED models of the stellar continuum, while SFRs are derived from attenuation-corrected UV luminosities at $\sim1600$ \AA. A best fit straight line to the $5<z<7$ data is shown in each panel, along with the \citet{speagle14} relation at $z\sim6$ in the top left panel.}
 \label{fig:ms}
\end{figure}

We plot the resulting values in four redshift intervals ($5<z<7$, $7<z<9$, $9<z<11$, and $z>11$) in Figure~\ref{fig:ms}, along with a best-fit straight line to the data in the lowest bin where the numbers dominate. We find the sources in each of the four redshift intervals populate expected regions of the parameter space, following a clear trend of increasing SFR with stellar mass. While the number of available data points declines rapidly with redshift, the consistency of the data points in each bin relative to the measured $z\simeq5-7$ main-sequence line (log\,SFR [$M_{\odot}/\rm yr$]$=$0.66\,log\,$M_{*}$ [$M_{\odot}$]$-5.50$) suggests little evolution in the main sequence between the redshifts probed here. We also compare our data points to the \citet{speagle14} $z\sim6$ main-sequence relation, finding good consistency between our points and the slope of their relation, with minor differences in the normalization (the \citealt{speagle14} relation resides approximately $+$0.35 dex above ours, possibly due to potential differences in SFR calibrations, magnification corrections, etc.) which we note are both small and effectively constant with stellar mass, indicating that our points scatter symmetrically (across SFR) around the shifted version of the relation at fixed stellar mass, as expected for a representative sample of star-forming galaxies. The consistency of our inferred galaxy properties with the high-$z$ main sequence, as well as the large range of absolute magnitudes probed in Figure~\ref{fig:zdist}, supports the labeling of this compilation as a representative galaxy sample.

\subsection{A Note Regarding Selection Effects}
The goal of the present paper is to construct representative composite spectra of high-redshift galaxies. Whether a composite spectrum is considered representative of the general galaxy population depends heavily on the variety of objects used in a stack. It is thus important to note that at the time of writing the selection functions for the NIRSpec programs used are extremely heterogenous, often unpublished, and therefore virtually impossible to characterize. This caveat means that one cannot define precisely in what sense our stacks (e.g., especially those at $z>10$) are representative of the general population. They are certainly not flux-limited samples, nor simple color-selected samples. However, a posteriori characterization (Figures~\ref{fig:zdist} and~\ref{fig:ms}) shows that our individual sources lie on the main sequence, and span more or less uniformly a range in $M_{\rm UV}$ between $-$21 mag and $-$17 mag. Furthermore, the properties of individual sources are in excellent agreement with independently derived properties from the literature. Therefore, the stacks presented here should be considered broadly representative of typical star-forming galaxies in the appropriate $M_{\rm UV}$ and redshift ranges, although this caveat should be kept in mind when using them.

Furthermore, we also acknowledge that not all prism spectra have sufficient SNR and spectral resolution to separate, e.g., AGN and Ly$\alpha$ emitters (LAEs; see Sections~\ref{subsec:AGNID} and~\ref{sec:lya}) from their star formation-dominated and/or Ly$\alpha$-attenuated counterparts. Such separations are limited by the data and some contamination may still be present in our stacks (although these are unlikely to be strong AGN or LAEs). Our efforts represent a first step with JWST/NIRSpec, and our stacks will continue to improve in SNR, representation, and diversity of galaxy spectral type as additional data sets become public and selection methods are better understood.

\subsection{Stacking Method and Spectral Modeling}
\label{sec:method}
\begin{figure*}
\center
 \includegraphics[width=\textwidth]{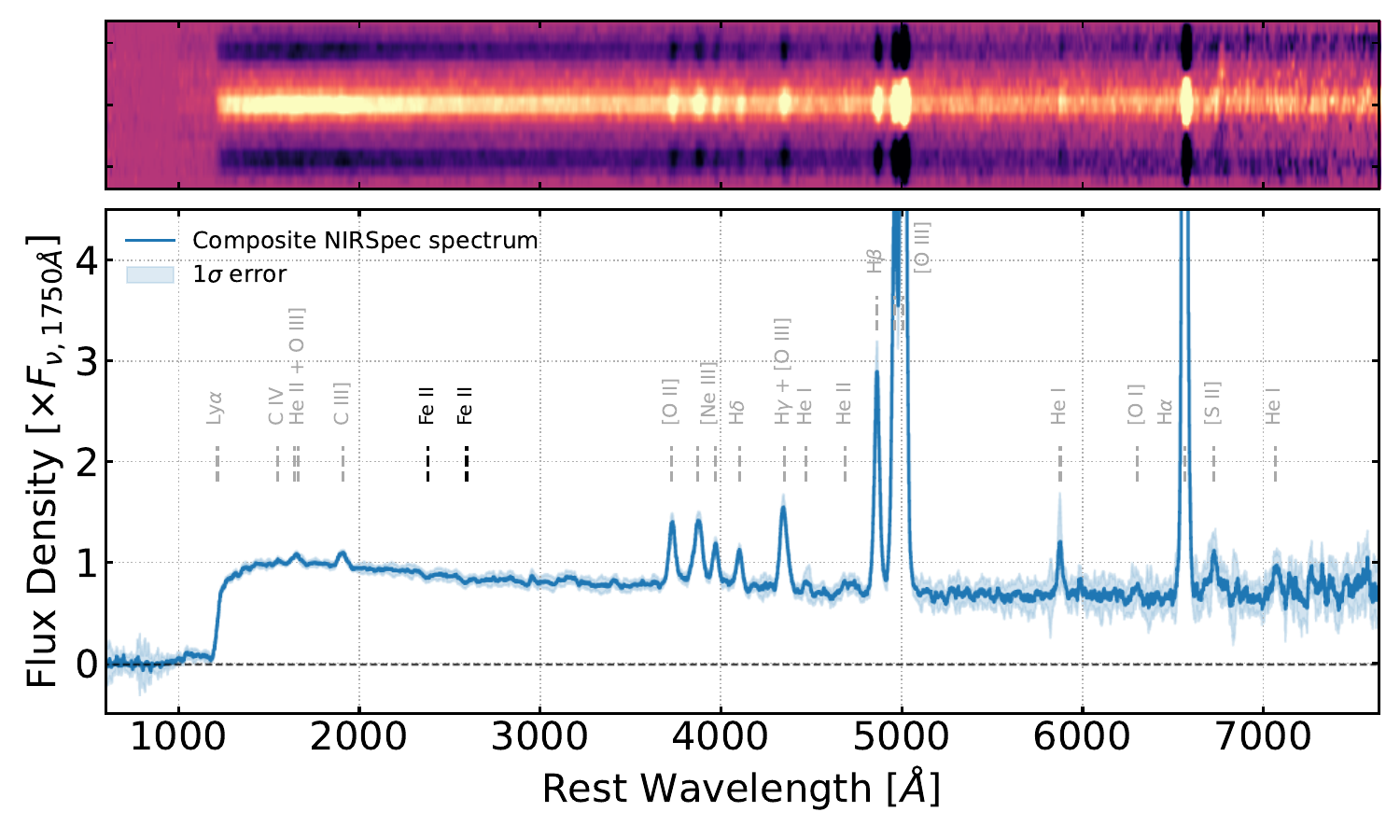}
 \caption{The composite (unscaled) NIRSpec prism spectrum (2D above, 1D below) of all $z\geqslant5$ star-forming galaxies from this study, in the rest frame. Several notable features are seen, from a clear Lyman-$\alpha$ break at $\lambda_{\rm rest}\simeq1216$ \AA\ highlighting the neutral intergalactic medium (IGM) within which most of the objects reside (with some excess UV flux emanating from $z<6$ sources where the IGM becomes ionized), to weak UV lines (the blended \heii$\lambda$1640 \AA\ and [\oiii]$\lambda\lambda$1661,1666 \AA\ doublet, as well as \ciii]$\lambda$1909 \AA\ emission) and both weak and strong optical lines ([\oii]$\lambda\lambda$3727,3730 \AA, [\neiii]$\lambda\lambda$3870,3969 \AA, \hdelta$\lambda$4103 \AA, blended H$\gamma\lambda$4342 \AA\ and [\oiii]$\lambda$4364 \AA, \hei$\lambda$4472 \AA, \heii$\lambda$4687 \AA, \hbeta$\lambda$4863 \AA, [\oiii]$\lambda\lambda$4960,5008 \AA, \hei$\lambda$5877 \AA, [\oi]$\lambda$6302 \AA, \halpha$\lambda$6564 \AA, [\sii]$\lambda\lambda$6718,6732 \AA, and \hei$\lambda$7067 \AA), affording a plethora of emission-line indicators for metallicity, dust, radiation field, and star-formation diagnostics. Furthermore, despite the low resolution of the prism, strong hints of \feii\ absorption can be seen in the rest-frame UV. A clear continuum is also seen at all wavelengths longward of the Lyman and Balmer breaks, indicating the presence of predominantly young, massive stars and longer-lived, lower-mass stars, respectively. The positions of all detected lines are marked by dashed lines (grey for emission lines, black for absorption lines).}
 \label{fig:allstack}
\end{figure*}

We create composite galaxy spectra at high redshift through a stacking of prism data based on a number of galaxy properties. To construct a stack, each galaxy spectrum (already photometrically scaled and magnification corrected) is first de-redshifted then interpolated over a common wavelength grid (with 10 \AA\ intervals) and added to the stack after normalization to unity at its rest-frame UV flux (between $\lambda_{\rm rest}=1500-2000$ \AA). Composite fluxes and associated uncertainties are derived from the nonzero median and the semidifference of the 16th and 84th percentiles of the distribution in each wavelength bin (the latter divided by a square root of $N_{i}$ term, where $N_{i}$ is the number of nonzero fluxes in the stack at a given wavelength element, $i$), respectively, after which the resulting spectrum and uncertainties are shifted and scaled to the median redshift and rest-frame UV magnitude of the stack. We do not apply any weighting to the fluxes in a given stack, in order not to bias toward apparently bright objects or those with longer integration times. An example composite spectrum (1D and 2D) derived using the full sample of $z\geqslant5$ star-forming galaxies (and left unscaled) is shown in Figure~\ref{fig:allstack}, for illustration.

Each composite spectrum is then fit with the \texttt{gsf}\footnote{\url{https://github.com/mtakahiro/gsf}} SED-fitting code \citep{morishita19} to obtain a continuum model. The code fits an assortment of galaxy templates derived from the simple stellar population (SSP) models of \citet{conroy09}, which are convolved to the resolution of the observed prism spectrum. To ensure a representative set of templates able to fit a large variety of spectral types, we generate the templates using a \citet{chabrier03} IMF, a metallicity range of log\,$Z/Z_{\odot}=0.01-1$, and stellar ages of 0.001-1.0 Gyrs. For the fit itself, we adopt a nonparametric star formation history and an SMC-like dust attenuation law (\citealt{gordon03}; see also \citealt{reddy15,reddy20}), leaving $A_{\rm v}$ as a free parameter with $A_{\rm v}=0-3$. Wavelengths corresponding to emission lines strong enough to potentially influence the fit (i.e., [\oii]$\lambda\lambda$3727,3730 \AA, [\neiii]$\lambda\lambda$3969,3870 \AA, H$\delta\lambda$4103 \AA, [\oiii]$\lambda$4364 \AA, H$\beta\lambda$4863 \AA, [\oiii]$\lambda\lambda$4960,5008 \AA, \hei$\lambda$5877 \AA, H$\alpha\lambda$6565 \AA, and [\sii]$\lambda\lambda$6718,6733 \AA) are masked prior to the fit and are measured separately. IGM absorption is modeled via the prescription in \citet{dijkstra14}, however we mask wavelengths between \lya\ and (rest-frame) 1500 \AA\ given the presence of strong line emission and damping wing effects identified by a number of recent studies for sources in our sample (e.g., \citealt{tang23,saxena23,heintz23}). The continuum spectrum and all \texttt{gsf}-derived parameters are taken from the 50th percentile of the posterior distribution, with the 16th to 84th percentile ranges used for uncertainties.

Emission lines are fit using the \texttt{emcee} code \citep{emcee} after subtraction of the best-fit \texttt{gsf} continuum model, the latter of which also accounts for stellar absorption. Given the large wavelength (and thus resolution) range over which each of the aforementioned lines span, we perform the (stellar and emission line) fits in separate wavelength groups and model each of the lines within a group simultaneously -- e.g., [\oii]$\lambda\lambda$3727,3730 $+$ \neiii$\lambda\lambda$3969,3870 $+$ H$\delta\lambda$4103 $+$ [\oiii]$\lambda$4364 \AA, H$\beta\lambda$4863 $+$ [\oiii]$\lambda\lambda$4960,5008, and H$\alpha\lambda$6565 $+$ [\sii]$\lambda\lambda$6718,6733 \AA. For each emission line in a set, we adopt a simple Gaussian profile with a line width ($0.001<\sigma<100$ \AA) and derive a posterior distribution for each of the free parameters (central wavelength, amplitude, and line width). All fluxes are derived using the best-fit parameters and their uncertainties, taken as the 50th percentile and semidifference of the 16th-84th percentiles of their posterior distributions, respectively. EW measurements are derived using those fluxes as well as the underlying \texttt{gsf} continuum model, unless stated otherwise.

\section{Spectral Features of $\lowercase{z}\geqslant5$ Galaxies}
\label{sec:spectralfeatures}
\subsection{Composite Spectra in Bins of Redshift, Absolute Magnitude, and Stellar Mass}
We have already provided a first look at the median spectrum of $z\geqslant5$ sources from the composite spectrum in Figure~\ref{fig:allstack}. The spectrum displays a rich variety of features hinting at a variety of environmental, stellar and nebular effects: a clear Lyman break arising from the absorption of UV photons by an increasingly neutral IGM, a blue rest-frame UV continuum ($\beta=-2.42\pm0.08$) dominated by stellar and nebular light from young OB stars, rest-frame optical emission governed primarily by a mix of A- and OB-type stars, and a host of clear line emission from \heii$\lambda$1640 \AA\ and [\oiii]$\lambda\lambda$1661,1666 \AA\ (all blended), \ciii]$\lambda\lambda$1907,1909 \AA, [\oii]$\lambda\lambda$3727,3729 \AA, [\neiii]$\lambda\lambda$3869,3968 \AA, H$\gamma$$\lambda$4102 \AA, H$\delta$$\lambda$4341 \AA\ and [\oiii]$\lambda$4264 \AA\ (blended), H$\beta\lambda$4862 \AA, [\oiii]$\lambda\lambda$4960,5008 \AA, \hei$\lambda$5877 \AA, [\oi]$\lambda$6302 \AA, H$\alpha$$\lambda$6564 \AA\ and [\nii]$\lambda\lambda$6549,6585 \AA\ (blended), [\sii]$\lambda\lambda$6718,6732 \AA, and \hei$\lambda$7067 \AA\ (as well as a number of additional marginal lines from, e.g., \hei$\lambda$4472 \AA\ and \heii$\lambda$4687 \AA). Additionally, we tentatively detect absorption features from resonant \feii$\lambda\lambda\lambda\lambda$2374,2382,2586,2600 \AA\ lines in the rest-frame UV, which in principle allow for determinations of gas outflows along with estimates of gas covering fraction and column density, among other properties (e.g., \citealt{martin12}).

\begin{figure*}
\center
 \includegraphics[width=\textwidth]{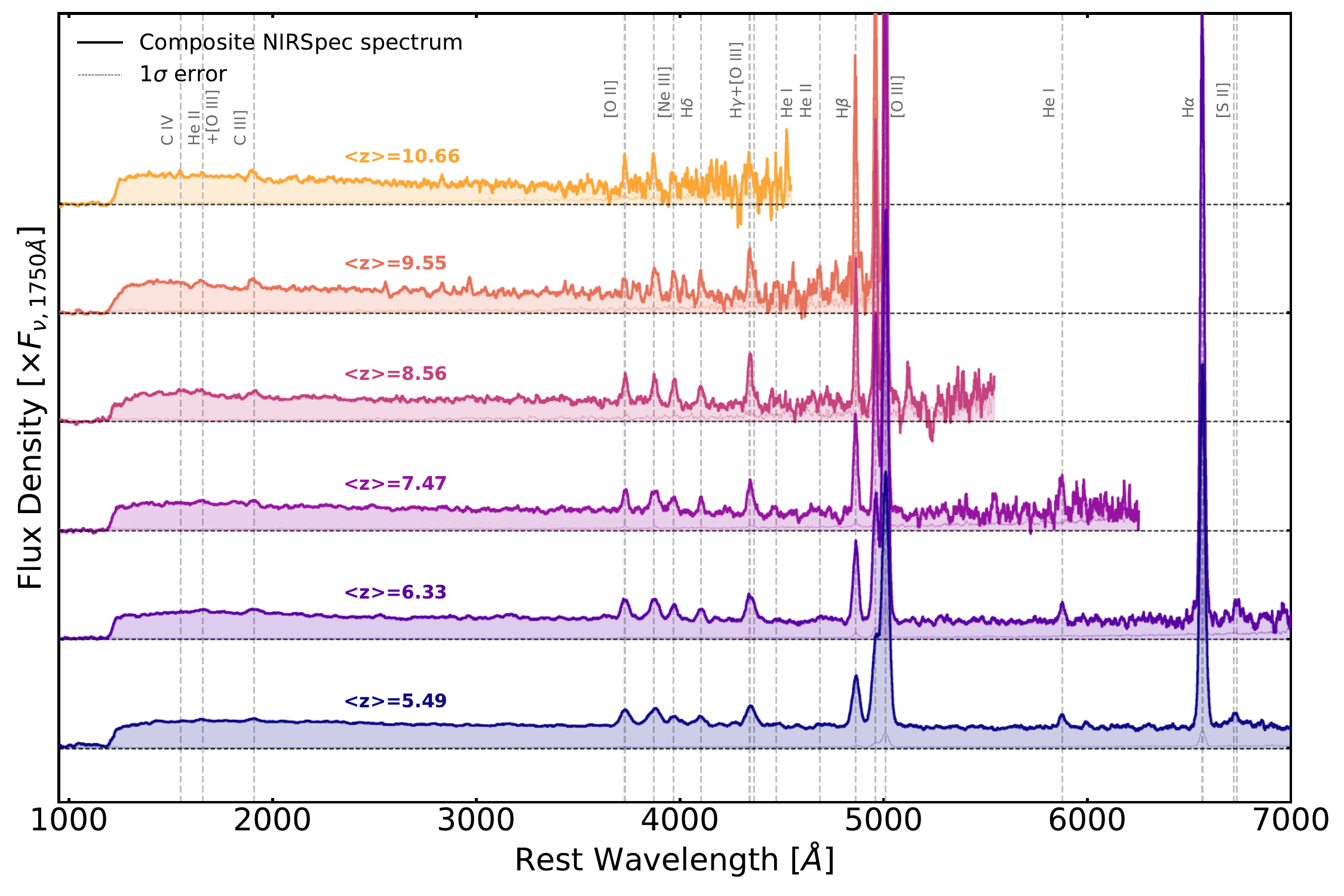}
 \caption{Composite NIRSpec spectra (solid lines and filled) and their $1\sigma$ standard deviations (i.e., dashed line) as a function of spectroscopic redshift. Each composite spectrum is shifted to the rest frame using the median spectroscopic redshift (indicated) of the stack and normalized to its $\lambda\simeq1750$ \AA\ flux. All spectra are vertically offset for clarity, and the positions of several rest-frame optical emission lines are indicated by dashed gray lines.}
 \label{fig:zstack}
\end{figure*}

While remarkably clear, such a spectrum combines galaxies at different epochs, luminosities, and stellar masses, which together yield unphysical results. As such, we construct three sets of more physically motivated stacks, one in bins of redshift only, a second in bins of redshift and $M_{\rm UV}$, and a third in bins of redshift and stellar mass. The redshift ranges for each set are $z=$[5-6, 6-7, 7-8, 8-9, 9-10, 10-13], while the absolute magnitude bins are $-22.00<M_{\rm UV}<-19.25$ or $-18.75<M_{\rm UV}<-16.00$, and the stellar mass bins are log\,$M_{\odot}$/$M_{*}<8.4$ and log\,$M_{\odot}$/$M_{*}>8.6$. The limits of $M_{\rm UV}$ and stellar mass are chosen to cover comparable parameter ranges and completeness levels either side of the divide, while leaving a gap to ensure sufficient contrast in average galaxy properties at fixed redshift (the exact size of the gap is limited by the sample size at the time of writing). The resulting composite spectra for the first set of stacks (i.e., redshift only) are shown in their rest frame in Figure~\ref{fig:zstack} and we adopt these as our fiducial set of templates. The median redshifts, absolute magnitudes, and stellar masses for those spectra, as well as the number of objects in each stack, are listed in Table~\ref{tab:stacks}.

\begin{deluxetable*}{llllllll}
\colnumbers
\tablecaption{Spectral Properties of the Composite Spectra, for Star-forming Galaxies Stacked by Redshift, $M_{\rm UV}$, and Stellar Mass.}
\tablehead{
\colhead{Composite Name} &
\colhead{$N_{\rm gals}$} &
\colhead{$z_{\rm spec}$-Range ($z_{\rm med}$)} &
\colhead{$M_{\rm UV}$} &
\colhead{log\,$M_{*}$} &
\colhead{$\beta$} &
\colhead{$E(B-V)_{\rm SED}$} &
\colhead{Optical Break} \\
\\[-0.6cm]
\colhead{} &
\colhead{} &
\colhead{} &
\colhead{(mag)} &
\colhead{(M$_{\odot}$)} &
\colhead{} &
\colhead{(mag)} &
\colhead{$F_{\nu,\rm 4200}$/$F_{\nu,\rm 3500}$}
}
\startdata
\multicolumn{8}{c}{$z_{\rm spec}$ stacks}\\[0.2cm]
z5\_All & 228 & 5.00-5.99 (5.49$\pm$0.31) & $-$18.44$\pm$0.03 & 8.43$\pm$0.11 & $-$2.29$\pm$0.01 & 0.10$\pm$0.01 & 1.05$\pm$0.01 \\
z6\_All & 107 & 6.00-6.99 (6.33$\pm$0.27) & $-$18.64$\pm$0.05 & 7.99$\pm$0.10 & $-$2.50$\pm$0.10 & 0.07$\pm$0.01 & 0.93$\pm$0.01 \\
z7\_All & 54 & 7.00-7.98 (7.47$\pm$0.29) & $-$19.03$\pm$0.07 & 7.72$\pm$0.11 & $-$2.50$\pm$0.04 & 0.07$\pm$0.01 & 0.87$\pm$0.01 \\
z8\_All & 16 & 8.01-8.92 (8.56$\pm$0.27) & $-$19.30$\pm$0.11 & 7.89$\pm$0.09 & $-$2.62$\pm$0.02 & 0.03$\pm$0.01 & 0.80$\pm$0.02 \\
z9\_All & 10 & 9.31-9.80 (9.55$\pm$0.17) & $-$19.07$\pm$0.12 & 8.00$\pm$0.11 & $-$2.60$\pm$0.01 & 0.04$\pm$0.01 & 0.77$\pm$0.02 \\
z10\_All & 20 & 10.01-12.93 (10.66$\pm$1.00) & $-$19.40$\pm$0.09 & 8.98$\pm$0.17 & $-$2.66$\pm$0.02 & 0.03$\pm$0.01 & -- \\[0.2cm]
\multicolumn{8}{c}{$z_{\rm spec}$-$M_{\rm UV}$ stacks}\\[0.2cm]
z5\_MUVfaint & 126 & 5.02-5.99 (5.55$\pm$0.30) & $-$17.94$\pm$0.04 & 8.16$\pm$0.10 & $-$2.46$\pm$0.01 & 0.08$\pm$0.01 & 1.03$\pm$0.01 \\
z6\_MUVfaint & 51 & 6.02-6.87 (6.33$\pm$0.22) & $-$18.06$\pm$0.08 & 7.79$\pm$0.04 & $-$2.60$\pm$0.01 & 0.07$\pm$0.01 & 0.99$\pm$0.01 \\
z7\_MUVfaint & 16 & 7.00-7.88 (7.27$\pm$0.22) & $-$18.39$\pm$0.11 & 7.43$\pm$0.03 & $-$2.56$\pm$0.01 & 0.08$\pm$0.01 & 0.64$\pm$0.02 \\
z8\_MUVfaint & 5 & 8.28-8.92 (8.75$\pm$0.22) & $-$18.03$\pm$0.18 & 7.86$\pm$0.09 & $-$2.60$\pm$0.01 & 0.03$\pm$0.01 & 0.85$\pm$0.03 \\
z9\_MUVfaint & 5 & 9.34-9.80 (9.53$\pm$0.15) & $-$17.98$\pm$0.21 & 7.63$\pm$0.12 & $-$2.62$\pm$0.02 & 0.07$\pm$0.01 & 0.69$\pm$0.03 \\
z5\_MUVbright & 60 & 5.00-5.99 (5.31$\pm$0.32) & $-$19.80$\pm$0.04 & 9.26$\pm$0.12 & $-$2.08$\pm$0.01 & 0.11$\pm$0.01 & 1.10$\pm$0.01 \\
z6\_MUVbright & 37 & 6.00-6.99 (6.31$\pm$0.33) & $-$19.68$\pm$0.05 & 8.17$\pm$0.05 & $-$2.40$\pm$0.01 & 0.08$\pm$0.01 & 0.89$\pm$0.01 \\
z7\_MUVbright & 18 & 7.12-7.89 (7.53$\pm$0.27) & $-$20.04$\pm$0.08 & 8.28$\pm$0.11 & $-$2.48$\pm$0.01 & 0.09$\pm$0.01 & 1.00$\pm$0.02 \\
z8\_MUVbright & 8 & 8.16-8.76 (8.50$\pm$0.22) & $-$20.36$\pm$0.11 & 8.23$\pm$0.08 & $-$2.50$\pm$0.01 & 0.03$\pm$0.01 & 0.86$\pm$0.01 \\
z9\_MUVbright & 5 & 9.31-9.77 (9.69$\pm$0.19) & $-$20.07$\pm$0.12 & 8.20$\pm$0.12 & $-$2.47$\pm$0.01 & 0.02$\pm$0.01 & 0.87$\pm$0.03 \\[0.2cm]
\multicolumn{8}{c}{$z_{\rm spec}$-$M_{*}$ stacks}\\[0.2cm]
z5\_lowMstar & 55 & 5.00-5.99 (5.44$\pm$0.31) & $-$18.53$\pm$0.04 & 7.50$\pm$0.04 & $-$2.63$\pm$0.01 & 0.11$\pm$0.01 & 1.02$\pm$0.01 \\
z6\_lowMstar & 36 & 6.02-6.99 (6.23$\pm$0.29) & $-$18.63$\pm$0.06 & 7.30$\pm$0.07 & $-$2.70$\pm$0.01 & 0.07$\pm$0.01 & 1.01$\pm$0.02 \\
z7\_lowMstar & 19 & 7.03-7.98 (7.45$\pm$0.27) & $-$18.82$\pm$0.09 & 7.42$\pm$0.02 & $-$2.63$\pm$0.02 & 0.07$\pm$0.01 & 1.03$\pm$0.03 \\
z8\_lowMstar & 8 & 8.30-8.92 (8.70$\pm$0.22) & $-$18.61$\pm$0.14 & 7.45$\pm$0.03 & $-$2.60$\pm$0.01 & 0.01$\pm$0.01 & 0.84$\pm$0.03 \\
z9\_lowMstar & 2 & 9.53-9.77 (9.65$\pm$0.12) & $-$19.38$\pm$0.29 & 8.41$\pm$0.28 & $-$2.67$\pm$0.03 & $<0.01$ & 2.91$\pm$0.02 \\
z5\_highMstar & 111 & 5.02-5.99 (5.54$\pm$0.31) & $-$18.37$\pm$0.04 & 9.12$\pm$0.11 & $-$2.09$\pm$0.01 & 0.10$\pm$0.01 & 1.07$\pm$0.01 \\
z6\_highMstar & 52 & 6.00-6.96 (6.37$\pm$0.25) & $-$18.40$\pm$0.06 & 8.30$\pm$0.08 & $-$2.40$\pm$0.04 & 0.09$\pm$0.01 & 0.89$\pm$0.01 \\
z7\_highMstar & 22 & 7.00-7.91 (7.47$\pm$0.31) & $-$18.82$\pm$0.08 & 8.84$\pm$0.09 & $-$2.39$\pm$0.11 & 0.07$\pm$0.01 & 0.74$\pm$0.02 \\
z8\_highMstar & 9 & 8.01-8.76 (8.28$\pm$0.25) & $-$19.15$\pm$0.13 & 8.11$\pm$0.09 & $-$2.50$\pm$0.01 & 0.06$\pm$0.01 & 0.74$\pm$0.02 \\
z9\_highMstar & 4 & 9.31-9.76 (9.51$\pm$0.20) & $-$19.75$\pm$0.12 & 8.69$\pm$0.12 & $-$2.48$\pm$0.02 & 0.07$\pm$0.01 & 0.99$\pm$0.02 \\[0.2cm]
\multicolumn{8}{c}{Ly$\alpha$ versus non-Ly$\alpha$}\\[0.2cm]
z6\_LAE & 52 & 5.00-7.51 (5.92$\pm$0.60) & $-$18.50$\pm$0.08 & 7.54$\pm$0.06 & $-$2.61$\pm$0.06 & 0.07$\pm$0.01 & 0.76$\pm$0.01 \\
z6\_nonLAE & 249 & 5.02-7.98 (5.92$\pm$0.74) & $-$18.66$\pm$0.03 & 8.36$\pm$0.07 & $-$2.39$\pm$0.02 & 0.08$\pm$0.01 & 1.03$\pm$0.01 \\[0.2cm]
\enddata
\tablecomments{All quoted uncertainties represent $1\sigma$ values. Those quoted for the median redshift refer to the standard deviation of the values in the stack, those for $M_{\rm UV}$ represent the standard deviation of the spectral fluxes within the selected wavelength window, while all others derive from the semidifference of the 16th and 84th percentiles of the posterior distribution from \texttt{emcee} parameter fitting. All values are corrected for magnification when necessary.}
\label{tab:stacks}
\end{deluxetable*}

\subsection{Blue Ultraviolet Continuum Slopes}
\label{subsec:beta}
\begin{figure*}
\center
 \includegraphics[width=\textwidth]{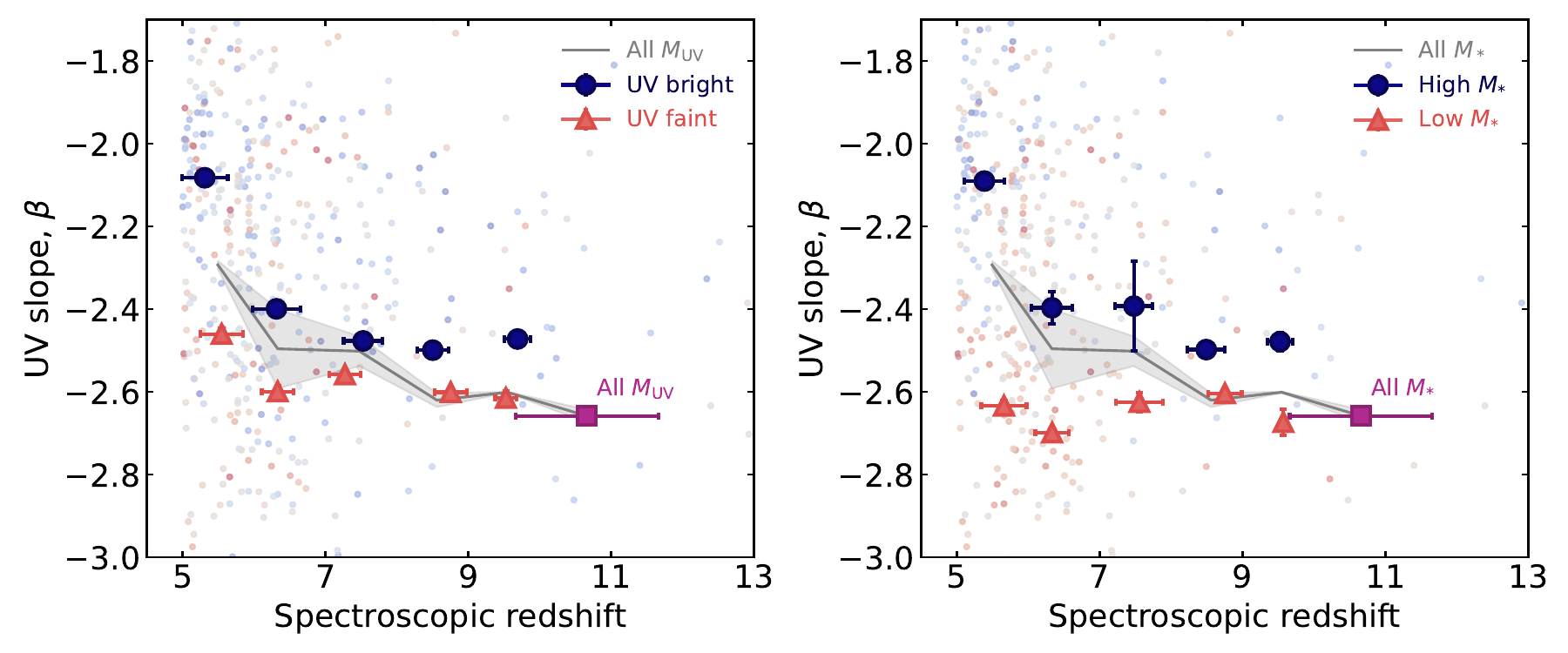}
 \caption{The UV continuum slopes of $z\geqslant5$ galaxies from individual sources (points) and composite JWST/NIRSpec prism spectra, the latter binned as a function of redshift only (gray line and gray shaded uncertainties in both panels), $z-M_{\rm UV}$ (navy and orange points for sources with $-22.00<M_{\rm UV}<-19.25$ mag and $-18.75<M_{\rm UV}<-16.00$ mag, respectively, in the left panel), and $z-M_{*}$ (navy and orange points for sources with log\,$M_{*}/M_{\odot}>8.6$ and log\,$M_{*}/M_{\odot}<8.4$, respectively, in the right panel). Individual points (color coded to their $M_{\rm UV}$ and log\,$M_{*}$ values, with UV-luminous and/or higher-mass sources in bluer colors and the opposite in redder colors) display a large scatter of values (generally between $\beta\sim-1.8$ to $\beta\sim-2.8$, with a mean $1\sigma$ uncertainty of $\simeq0.22$) and an overall declining slope with redshift. Composite points confirm the trend, with clear differences between faint/low-mass systems and their bright/high-mass counterparts and a particularly strong redshift evolution with absolute magnitude but shallower evolution with stellar mass.}
 \label{fig:beta}
\end{figure*}

Similarly to Figure~\ref{fig:allstack}, we find overall blue continuum slopes in each of our high-redshift stacks. Adopting a wavelength window of 1600-2800 \AA, masking the region of the \ciii]$\lambda\lambda$1907,1909 \AA\ doublet, and binning the continuum by a factor of $\times10$ (making it less sensitive to flux outliers), we use \texttt{emcee} with a simple power-law parameterization (i.e., $F_{\lambda}\propto\lambda^{\beta}$) and measure increasingly blue UV slopes of $-$2.29$\pm$0.01, $-$2.50$\pm$0.10, $-$2.50$\pm$0.04, $-$2.62$\pm$0.02, $-$2.60$\pm$0.01, $-$2.66$\pm$0.02 at redshifts $z\simeq5.5$, $z\simeq6.3$, $z\simeq7.5$, $z\simeq8.6$, $z\simeq9.6$, and $z\simeq10.7$, respectively (from the redshift-binned stacks). The results and their uncertainties (from the 50th and semidifference of the 16-84th percentiles of the posterior distribution, respectively) are shown as a gray line and shading in Figure~\ref{fig:beta}, where a clear and steep decline in UV slope at fixed absolute magnitude is seen for $z\simeq5-7$ sources ($\beta\simeq-$2.1 to $-$2.5) before a much more gradual evolution at redshifts beyond ($\beta\simeq-$2.5 to $-$2.6). The trend is suggestive of evolving dust contents and stellar populations, where the lowest-redshift sources are characterized by the most dust- and metal-rich interstellar media and aged stellar populations, while higher-redshift sources reflect more pristine ISM conditions and younger stellar populations.

A secondary dependence of $\beta$ on luminosity is well known (see, e.g., \citealt{bouwens14} and \citealt{cullen23}), and we find that our $z-M_{\rm UV}$ stacks display consistent but systematically offset values compared to the redshift-binned stacks: the most luminous galaxies harbor redder slopes compared to their fainter counterparts (approximately $\Delta\beta=$0.1-0.4 redder, with smaller differences at higher redshifts), likely reflecting more evolved stellar populations and metal- and dust-rich gas reservoirs for the most luminous systems. 

The UV-luminous and UV-faint stacks are characterized by trends mirroring the evolution seen in the redshift-binned stacks, while a similar trend is also found with stellar mass, particularly for the most massive systems, which display a virtually identical set of continuum slopes at fixed redshift compared to the UV-luminous stacks.
The trend is less pronounced for lower-mass galaxies compared to their UV-faint counterparts, however, suggestive of stellar mass playing a more dominant role in regulating $\beta$ with redshift, through the chemical enrichment of the ISM and aging of stellar populations in the most massive galaxies. 

The normalization and evolution of the UV slopes with redshift can first be compared to HST-derived results, which generally display redder slopes at comparable redshifts likely attributable to the modest sampling of near-ultraviolet wavelengths compared to the longer baseline afforded by JWST. For instance, for sources at $z_{\rm phot}\sim5-7$ \citet{wilkins11} report median slopes of $\beta\simeq-1.9$ to $-2.3$, \citet{finkelstein12} report median slopes of $\beta\simeq-1.9$ to $-1.95$ at $z_{\rm phot}\sim5-8$, \citet{dunlop13} report median slopes of $\beta\simeq=-2.1$ to $-1.8$ at $z_{\rm phot}\sim7-9$, and \citet{bouwens14} report median values of $\beta\simeq-2.2$ to $-1.8$ at $z_{\rm phot}\sim5-9$ (for sources at comparable luminosities to here). The $z<8$ values are consistent only with the most luminous of our stacks and, interestingly, also point to a steepening with redshift from $z\simeq5$ to $z\simeq7$ \citep{wilkins11,finkelstein12,bouwens14}. However, beyond $z\simeq7$ where WFC3 photometry samples only up to $\lambda_{\rm rest}\simeq1600$ \AA, the HST photometry lacks the wavelength coverage necessary for accurate constraints, as is evident from the markedly shallower reported slopes ($\Delta\beta\approx0.2-0.7$) compared to our spectroscopic results. 

The extended baseline afforded by the NIRCam imaging alleviates some of the tension, yielding trends and a range of UV slopes similar to those reported here. For instance, \citet{nanayakkara23} report median slopes of $\beta\simeq-2.1$ to $-2.4$ at $z_{\rm phot}\sim5.1-6.5$, while \citet{topping23} report median values of $\beta\simeq-2.3$ to $-2.5$ at $z_{\rm phot}\sim6-12$, and \citet{cullen23} report a median value of $\beta-2.32$ for sources at $z_{\rm phot}\sim6-12$ (with comparable luminosities to our stacks).

Those NIRCam studies all report trends of declining UV slopes with redshift that are similar to what we report here, with some differences in slope and normalization likely reflecting selection effects and possible sample contamination. For instance, a straight-line fit to the redshift trend produced by our fiducial stacks in Figure~\ref{fig:beta} yields a slope of $d\beta/dz=-0.06\pm0.01$, which is moderately steeper than (but consistent with) a fit to the \citet{topping23} median points ($d\beta/dz=-0.03\pm0.01$) over the same redshift range. Conversely, a similar comparison at redshifts of $z\gtrsim8$ with median values from the \citet{cullen23} NIRCam points yields slopes of $d\beta/dz=-0.11\pm0.04$ and $d\beta/dz=-0.02\pm0.02$ for our stacks and their photometric data points, respectively. For the comparisons above, we also find y-axis intercepts of $-2.04\pm0.12$ (these stacks) versus $-2.06\pm0.05$ (fit to the \citealt{topping23} points) and $-2.45\pm0.20$ (these stacks, $z>8$) and $-1.36\pm0.41$ (fit to the derived median \citealt{cullen23} points).

Of particular interest are redshifts beyond which galaxies are deemed to become ``dust-free,'' facilitating the transmission of UV photons from massive stars and enhancing their observed UV luminosities perhaps beyond first-order expectations (e.g., \citealt{castellano22,naidu22,ferrara23,mason23,casey23}). While this analysis and the studies discussed above all associate the bluest slopes to sources at the highest redshifts, none consistently measure dust-free slopes of $\beta\simeq-3$ or bluer. A small minority of individual sources within these samples are reported to show slopes bluer than the canonical $\beta\simeq-2.7$ limit set by nebular continuum contributions \citep{topping23,cullen23}, however single-source measurements from photometry or low SNR spectroscopy are prone to large scatter and uncertainties. Median values consistently measure values redder than or close to this limit and the $z>10$ values presented here confirm the lack of $\beta<-3$ slopes, at least on average.

While a direct comparison with photometric results is challenging, the secure redshifts and high SNR measurements of our NIRSpec composite spectra confirm a clear steepening in UV slopes (i.e., toward bluer values) from lower redshifts to $z\sim7$ across the general galaxy population, and a probable flattening out beyond. The measurements point primarily to a declining dust content, possibly arriving close to relatively pristine systems (on average) by redshifts $z\simeq8-9$. While this does not preclude some exceptionally luminous sources to have accelerated their dust buildup compared to the general population (e.g., \citealt{ferrara22}), such examples are rarely seen at redshifts beyond $z\simeq8-9$ and likely represent the exception rather than the rule. 

\subsection{Continuum Breaks Indicative of Evolving Stellar Populations}
One of the primary discontinuities in the rest-frame optical spectra of galaxies are the so-called Balmer and 4000 \AA\ breaks, both of which serve as indirect tracers for the ages of the underlying stellar populations. While both derive from absorption within the atmospheres of stars and are often treated as a single feature, their origins differ -- the Balmer break at 3646 \AA\ represents the limit of the Balmer series and occurs in stellar populations younger than 1 Gyr, while the 4000 \AA\ break derives from the absorption of stellar continuum by ionized metals in the atmospheres of stars (including \caii\ H and K) and as such increases monotonically with age \citep{kauffmann03}, with some dependence also on metallicity.

\begin{figure}
\center
 \includegraphics[width=\columnwidth]{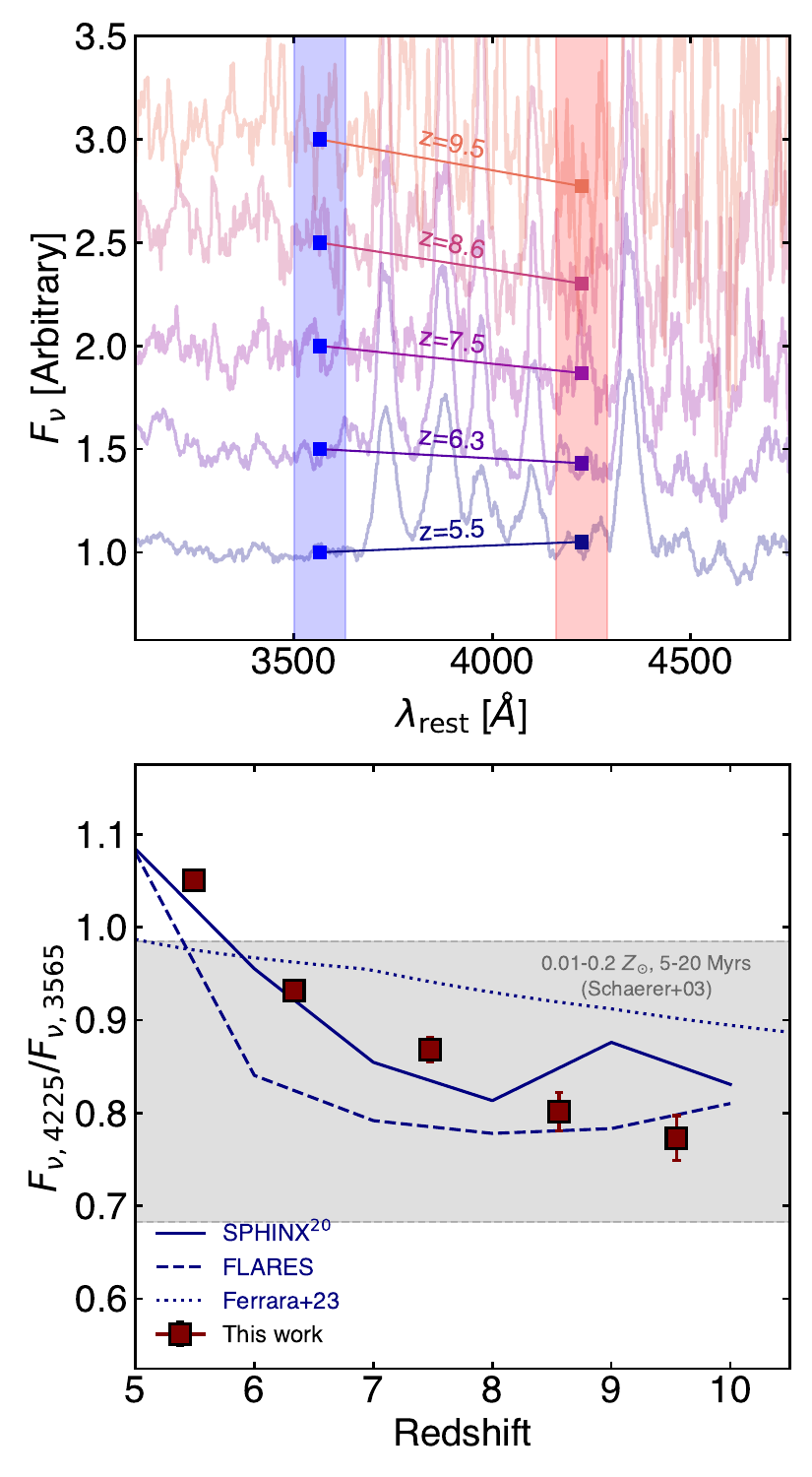}
\caption{The Balmer break as measured directly from our fiducial composite spectra ($F_{\nu \rm 4225}/F_{\nu,\rm 3565}$), as a function of redshift (red squares and uncertainties). For comparison, the expected range of values for low-metallicity (0.01-0.2 $Z_{\odot}$) and young (5-20 Myr) sources \citet{schaerer03} are shown as the gray region. Median values for comparable sources in the SPHINX$^{20}$ \citep{katz23} and FLARES \citep{wilkins24} simulations are also shown as solid and dashed navy lines, respectively, and the model of \citet{ferrara23} as a dotted line. All comparisons show agreement with the range of values measured here, as well as the observed trend with redshift.}
 \label{fig:d4000}
\end{figure}

Until recently, for $z\geqslant5$ sources these diagnostics were accessible only through Spitzer photometry (e.g., \citealt{hashimoto18,rb22,laporte21}). However, in near-identical fashion to wide JWST/NIRCam bands, the wide coverage of photometric bands can lead to significant degeneracies (see, e.g., \citealt{rb22,desprez23}). Thanks to the extensive wavelength coverage of the NIRSpec prism, the Balmer break is covered in each of our stacks, allowing for a direct and consistent measurement of the continuum across $\sim600$ Myr of cosmic time. A number of definitions have been utilized for sources in the local Universe, differing primarily in the central wavelengths and widths of the pseudobandpasses where the red and blue portions of the continuum are measured. Given the smearing of emission-line fluxes across the stellar continuum, traditional definitions using 4050-4250 \AA\ and 3750-3950 \AA\ \citep{bruzual83} or narrower 3850-3950 \AA\ and 4000-4100 \AA\ \citep{balogh99} windows would result in emission-line-contaminated measurements. For a consistent and clean set of measurements, therefore, we take care to define spectral windows available across our entire redshift range that are free of line emission. We adopt wavelength ranges of $\lambda_{\rm rest}=3500-3630$ \AA\ and $4160-4290$ \AA\ and, to mitigate the impact of noise or lingering features, determine the average fluxes from straight-line fits to the fluxes in each of those windows. We refer to the measurements as Balmer ``indices'' and plot the results for our redshift-binned stacks in Figure~\ref{fig:d4000}, where uncertainties are defined as the $1\sigma$ error on the best-fit normalization parameter.

We find a steadily declining Balmer index with redshift. The largest index is found in our $z\sim5.5$ stack, with $F_{\nu, \rm 4225}/F_{\nu, \rm 3565}=1.05\pm0.06$, which rapidly shifts to $F_{\nu, \rm 4225}/F_{\nu, \rm 3565}<1$ by $z\sim6.5$ and declines further with redshift to 0.75$\pm$0.26 by $z\simeq9.6$. For our highest-redshift stack at $z\simeq10.7$, the red continuum is too noisy to determine a reliable break, while for our $z-M_{\rm UV}$ and $z-M_{*}$ stacks the SNR is insufficient to verify whether differences between subsamples exist (however the general decline toward reduced values with redshift remains).

The small values we observe here are unlike those found for the bulk of the general population at $z\sim0$, which show indices of $\sim1-2$ and include effects from both the Balmer and 4000 \AA\ breaks \citep{kauffmann03}. For those sources, even the youngest galaxies are limited to values of $\sim1$. Local analogs, however, typically associated with extremely metal-poor ISM, young stellar ages, and enhanced nebular continuum (all of which are conducive to Balmer jumps; \citealt{mingozzi22,inoue11}), show greater consistency with the results shown here (e.g., \citealt{izotov21}). At $z>5$, only a handful of sources have been studied with spectroscopic measurements, and these show significant scatter in the reported indices (e.g., \citealt{looser23}, \citealt{cameron23_nebular}, \citealt{vikaeus24}, \citealt{witten24}), likely reflective of stochastic star formation histories.

The trend observed here is reminiscent of the declining UV continuum slopes found in the previous section, and is also matched by decreasing mass-weighted stellar ages inferred from our \texttt{gsf} continuum fitting. Those ages begin at values of $\sim95$ Myr for our $z\sim5.5$ stack, decrease to $\sim18$ Myr at $z\sim6.5$, and steady out to ages of $<10$ Myr at redshifts beyond. The inferred breaks and ages are in line with photometric expectations (e.g., \citealt{stefanon23,whitler23b}) and recent spectroscopic measurements of some individual sources \citep{vikaeus24,cameron23_nebular}, which all require young stellar ages, low quantities of metals and dust, and significant contributions from the nebular continuum to explain the observations.

To verify whether our observations agree with expectations from theory and simulations, we overplot in Figure~\ref{fig:d4000} the expected break index ranges measured in identical fashion on the spectral models of \citet{schaerer03}, for a variety of metallicities ($0.01-0.2 Z_{\odot}$) and stellar ages (5-20 Myr), all of which include contributions from the nebular continuum. The models are in excellent agreement with the values measured here. We also overplot (as navy lines) the median values from the SPHINX$^{\rm 20}$ \citep{katz23} and FLARES \citep{lovell20,vijayan20} hydrodynamical simulations, limiting the comparison to sources with metallicities of $<0.01 Z_{\odot}$, stellar masses of $<10^{8.0} M_{\odot}$, and stellar ages comparable to those inferred by our \texttt{gsf} modeling at each redshift interval. For the former, we measure the indices in identical fashion using the simulated spectra, while the later are derived directly from the FLARES code output using the formalism of \citet{wilkins24}. Both simulations are able to match the range of our measured values, and moreover show excellent agreement with the trend toward reduced values at higher redshift (see \citealt{vikaeus24} and their comparison to the FLARES and DELPHI simulations with a smaller number of confirmed JWST sources). Similar conclusions can be derived when comparing with the model of \citet{ferrara23}, which also shows agreement with the range of values we probe and a decline in the Balmer break index with redshift.

The strength of the break is sensitive to a number of physical properties, including stellar age, dust attenuation, metallicity, the escape fraction of ionizing Lyman continuum photons, and the shape of the star formation history \citep{wilkins24}. Determining the precise contributions of each to the breaks measured here is beyond the scope of the present work, however we note that the overall blue UV slopes seen in Section~\ref{subsec:beta} disfavor significant dust contributions and the consistency between the \texttt{gsf} ages and the age-limited simulations suggest stellar age is the main driver.

\subsection{Prominent \ciii] Emission in Early Galaxies}
\begin{figure}
\center
 \includegraphics[width=\columnwidth]{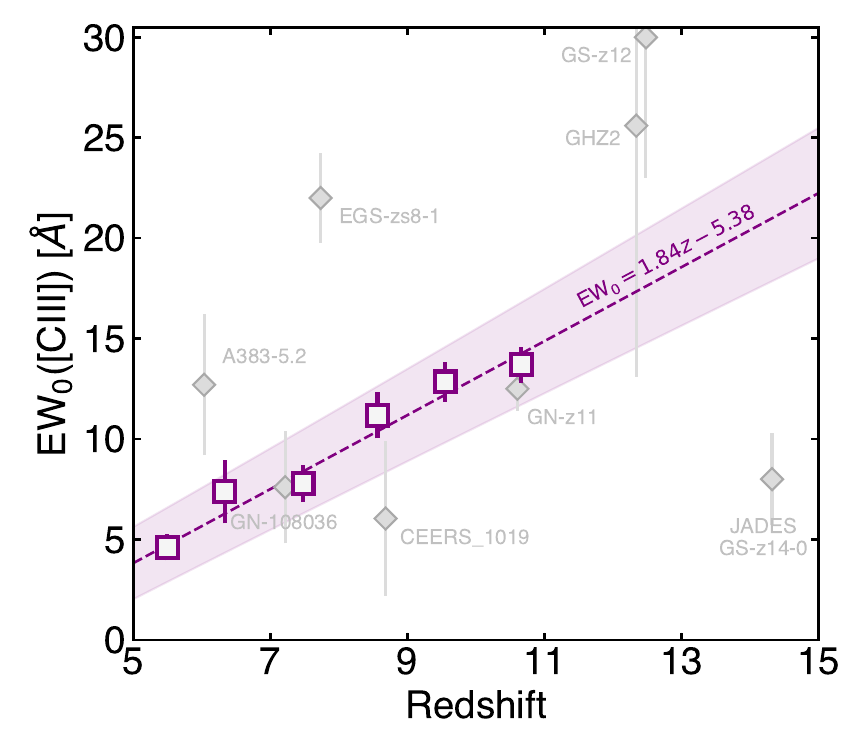}
 \caption{The rest-frame equivalent width of \ciii]$\lambda\lambda$1907,1909 \AA\ line emission as a function of spectroscopic redshift, as measured from our redshift-binned composite spectra. The trend of increasing EW with redshift is well parametrized by a straight line fit (dashed purple line with pink uncertainties). Measurements from the literature \citep{stark15,stark17,larson23,bunker23,deugenio23,castellano24,carniani24} are added as grey points, for comparison.}
 \label{fig:ciii}
\end{figure}

\ciii]$\lambda\lambda$1907,1909 \AA\ emission is a useful indicator for the metal enrichment and electron density of the ISM (e.g., \citealt{jaskot16,deugenio23}), the hardness of the ionizing spectrum (e.g., \citealt{stark17}), and emission-line diagnostics to distinguish between star formation and AGN activity (e.g., \citealt{laporte17b,nakajima18}). Characterizing the presence and evolution of \ciii] thus represents an important step toward understanding the ISM conditions of galaxies out to the earliest times. Our composite spectra reveal significant \ciii] emission at all redshifts, with measured rest-frame EWs (EW$_{0}$) of 4.6$\pm$0.7 \AA, 7.4$\pm$1.6 \AA, 7.8$\pm$0.9 \AA, 11.2$\pm$1.2 \AA, 12.8$\pm$1.0 \AA, and 13.7$\pm$0.9 \AA\ for the redshift-binned stacks at $z\simeq5.5$, $z\simeq6.3$, $z\simeq7.5$, $z\simeq8.6$, $z\simeq9.6$, and $z\simeq10.7$, respectively, marking an increase of a factor of $\sim3$ from $z\sim5$ to $z\sim11$ and a median value of 9.5$\pm$0.7 across the full sample. The values are plotted in Figure~\ref{fig:ciii} and showcase a very clear trend of increasing emission with redshift, with an evolution best described by a straight-line fit of EW$_{0}$(\ciii])=1.84($\pm$0.19)$z-$5.38($\pm$1.53) (indicated by a dashed line and $1\sigma$ uncertainty fill). Furthermore, we find no clear evidence for varying \ciii] strength in our $z-M_{\rm UV}$ stacks, but do find tentative evidence for some enhanced emission in the least massive sources of our $z-M_{\rm *}$ stacks compared to their higher-mass counterparts.

The implied prevalence of \ciii] and the observed correlation are consistent with extrapolations of measurements from lower redshift. High resolution spectroscopy over statistical samples of $z\simeq2-4$ sources find a high incidence of \ciii] emission in both individual sources (e.g., $\sim$24-29\%; \citealt{lefevre19,llerena22,stark14}) and stacked spectra (e.g., \citealt{shapley03,amorin17,lefevre19,llerena22}). Reported EW$_{0}$s for typical sources generally range between a few to 10 \AA\, although most studies also report a small tail in the distribution toward larger values, generally attributed to extreme radiation fields from very metal-poor galaxies or AGN \citep{jaskot16,nakajima18}; for instance, for a statistical sample of 3899 \citep[][see also \citealt{erb10,stark14,amorin17,llerena22}]{lefevre19} report EWs of 3-10 \AA\ for 20\% of their sample and find higher values of $>10$ \AA\ for only $\sim4$\% of the sample. The same authors find median values of 2.0 \AA\ and 2.2 \AA\ for sources at $2<z<3$ and $3<z<4$, respectively, while \citet{shapley03} find a marginally smaller but consistent value of 1.7 \AA\ in stacked spectra of UV-selected $z\sim3$ LBGs. Each of these are in good agreement with the trend we observe in Figure~\ref{fig:ciii}, which if extrapolated suggests EWs of 0.1-2.0 \AA\ at $z=3-4$. At even lower redshifts of $z\sim1$, \citet{du17} find a detection rate of $\sim$22\% and a median EW of 1.3 \AA, consistent with the decline seen in Figure~\ref{fig:ciii}. The ubiquity of our (stacked) detections suggests \ciii] is likely to be at least as prominent in $z\geqslant5$ sources as in $z\sim1-4$ sources, while a well-established link between \ciii] EW and metallicity \citep{jaskot16,nakajima18,llerena22} suggests the latter is the main driver of our trend.

For comparisons at high redshift, in Figure~\ref{fig:ciii} we also plot a number of reported values for known \ciii] emitters in the Epoch of Reionization \citep{stark15,stark17,bunker23,larson23,deugenio23,castellano24}. We find a large scatter of individual objects around the relation (albeit with significant uncertainties), where a number of objects (A383-5.2, EGS-zs8-1, GS-z12, and GHZ2) fall above the relation and by this metric alone could indicate extreme ISM conditions -- e.g., EGS-zs8-1 is a known $z=7.7$ LAE with extreme [\oiii]+H$\beta$ line emission \citep{oesch15,rb16}, while GHZ2 displays remarkably luminous rest-frame UV lines possibly indicative of AGN activity \citep{castellano24}. For the sources instead lying close to the range probed by the general high-$z$ population, these include apparently extreme sources such as GN-z11 and CEERS\_1019, two Ly$\alpha$-emitting sources with atypical UV line emission and nitrogen enhancement \citep{bunker23,rui24,senchyna24}. In the case of GN-z11, we note the source exhibits a measured EW$_{0}$(\ciii]) of 12.5$\pm$1.1 \AA\ \citep{bunker23}, compared to a predicted value of 14.1$\pm$2.5 \AA. Our results imply the classification of ``extreme'' systems for these particular objects are not reflected in their \ciii] emission, based on a comparison to the range of values characteristic of the general galaxy population.

At even higher redshifts of $z\sim14$, JADES-GS-z14-0 appears well below the extrapolation of the trend, suggestive of an exceptionally low metallicity \citep{carniani24}. Should other sources at similarly high redshifts reveal comparable \ciii] EWs, this might imply a turnover in the trend somewhere around $z\sim13$ which is not captured by the extrapolation of the evolution from $z<11$, where sources begin to reveal virtually pristine ISM conditions.

Lastly, although the presence of \ciii] in our stacks is unambiguous, the presence or otherwise of additional rest-frame UV lines is less clear. For instance, although we find detections of \heii+[\oiii] rest-frame UV line emission in Figure~\ref{fig:allstack} and Figure~\ref{fig:zstack}, those detections are lower SNR and do not translate to the smaller samples in any of our $z-M_{\rm UV}$ or $z-M_{*}$ stacks. Similar conclusions can be derived for other rest-frame UV lines such as \nv$\lambda$1243 \AA, \niv$\lambda$1486 \AA, and \civ$\lambda\lambda$1548,1550 \AA, all of which are seen in individual Lyman break or \ciii]-emitting galaxies at $z\sim2-3$ (e.g., \citealt{shapley03,stark14}) and out to higher redshifts of $z>6$ (e.g., \citealt{bunker23}, \citealt{topping24}), but not in our stacks.

\subsection{Rest-frame Optical Line Emission}
By far the most common features found in our stacked spectra are signatures of strong rest-frame optical line emission at virtually all redshifts, UV magnitudes, and/or stellar masses, highlighting the nature by which most of our $z>5$ sample were selected. These include clear signatures of line emission from [\oii]$\lambda\lambda$3727,3729 \AA, [\neiii]$\lambda$3869 \AA, [\neiii]$\lambda$3968 \AA, H$\delta\lambda$4102 \AA, H$\gamma\lambda$4341 \AA+[\oiii]$\lambda$4364 \AA, H$\beta\lambda$4862 \AA, [\oiii]$\lambda\lambda$4960,5008 \AA, \hei$\lambda$5877 \AA, H$\alpha\lambda$6564 \AA+[\nii]$\lambda$6584 \AA, and [\sii]$\lambda\lambda$6718,6732 \AA. Due to the wavelength resolution and limits of the NIRSpec prism, most line doublets and closely spaced pairs (e.g., H$\gamma$+[\oiii]) are blended, while emission from \hei, H$\alpha$, and [\sii] are detectable only out to $z\sim7-7.5$ and strong lines of [\oiii]+H$\beta$ out to $z\simeq9.6$.

The weakest lines across the $\lambda_{\rm rest}\simeq$3750-4500 \AA\ wavelength range are generally the [\neiii]$\lambda$3968 \AA\ and H$\delta$ lines, with rest-frame equivalent widths in the range EW$_{0}\simeq$11-30 \AA\ for the former and EW$_{0}\simeq$16-22 \AA\ for the latter. Next in strength are the [\oii]$\lambda\lambda$3727,3729 \AA\ doublet and H$\gamma$+[\oiii], with EW$_{0}\simeq$14-39 \AA\ and EW$_{0}\simeq$48-76 \AA, respectively.

Prominent [\oiii]+H$\beta$ lines are found across all redshifts (except the highest redshift of $z\sim10.66$ where the lines fall beyond the red limit of the prism) with combined rest-frame EW$_{0}$s of 752$\pm$10 \AA, 1031$\pm$31 \AA, 1292$\pm$68 \AA, 1214$\pm$54 \AA, and 910$\pm$18 \AA\ for redshifts $z\simeq5.5$, $z\simeq6.3$, $z\simeq7.5$, $z\simeq8.6$, and $z\simeq9.6$. An increase of $\sim\times1.7$ is found between the lowest and highest values, which reflect some possible redshift evolution at least between $z\sim5.5-8.5$. The range of values reported here illustrate the strong line nature of the underlying sample.

\begin{figure}
\center
 \includegraphics[width=\columnwidth]{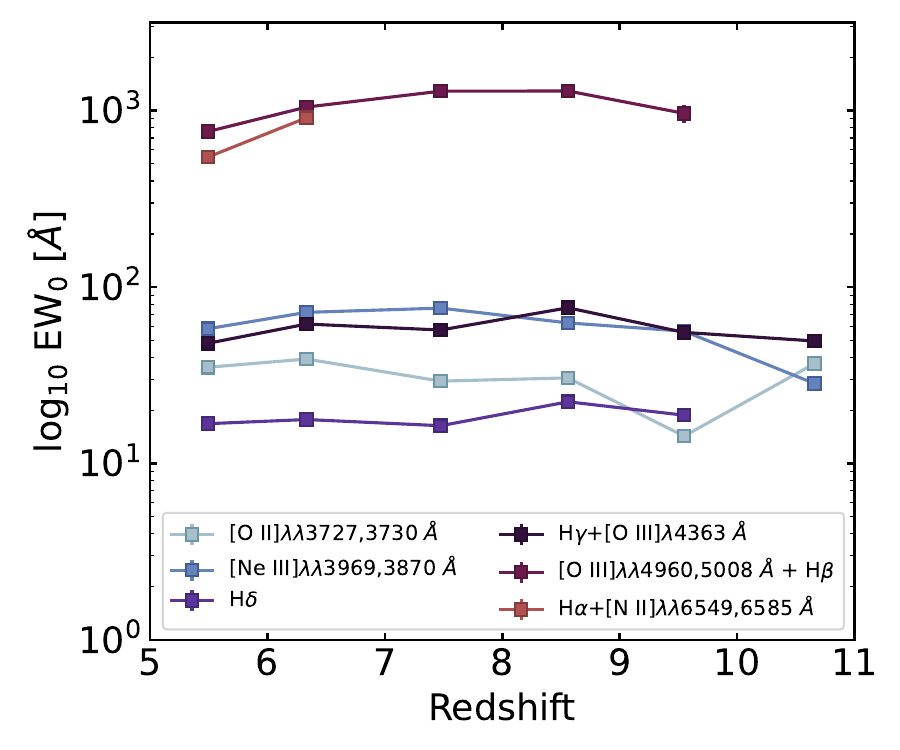}
 \caption{The distribution of EW$_{0}$s for the rest-frame optical emission lines observed in our redshift-stacked spectra, as measured from the Gaussian fits described in Section \ref{sec:method}.}
 \label{fig:lines}
\end{figure}

The median value of the sample is 1031$\pm$31 \AA\ which is in excellent agreement with values reported in the literature from JWST photometry at comparable redshifts (e.g., \citealt{endsley21b}). Similarly to observations of the \ciii] line, we find no significant evidence for stronger line emission in UV-faint galaxies compared to their UV-bright counterparts, but do find evidence for such a trend with stellar mass. This is comparable to the trend found by \citet{reddy18}, who show a clear decrease in virtually all rest-frame optical line EWs at higher stellar masses. Prominent \hei$\lambda$5877 \AA\ and H$\alpha$ emission are also seen in the lowest-redshift stacks (at redshifts of $z\simeq5.5-7.5$ for the former and $z\simeq5.5-6.3$ for the latter) before moving beyond the prism coverage. Their EWs are reported to be 16-32 \AA\ for \hei\ at redshifts $z\simeq5.5-7.5$, and 559-922 \AA\ for H$\alpha$ at $z\simeq5.5-6.3$. The EWs of each of the rest-optical lines are shown in Figure~\ref{fig:lines}, and we list these along with their H$\beta$-normalized fluxes in Table~\ref{tab:line_ratios_redshift} in the Appendix.

\section{Inferred Properties of High-redshift Sources}
\label{sec:spectraprops}
\subsection{Dust Attenuation from Spectral Energy Distribution Fitting and Balmer Decrements}
Constraints on the prevalence and evolution of dust at high redshift have yielded mixed results. Until recently, constraints in Reionization Era galaxies have come from SED fitting of (primarily) rest-frame UV photometry \citep[e.g.,][]{strait20,rb22}, or NIR continuum measurements from Atacama Large Millimeter/submillimeter Array \citep[ALMA; e.g.,][]{laporte17,tamura19,bakx21,bouwens21_rebels,ferrara22}. Although the vast majority of objects have yielded apparently low to negligible dust contents, some marked exceptions from ALMA have challenged expectations with large dust quantities by $z\sim7-8$ \citep{laporte17,ferrara22}. More direct constraints of nebular dust attenuation with JWST spectroscopy of the Balmer decrement (i.e., the ratio of Balmer lines to H$\beta$) have indicated ISM approaching ``dust-free'' states by $z\simeq6$, with a clear correlation of enhanced color excesses ($E(B-V)_{\rm neb}$) with stellar mass (above log\,$M_{*}/M_{\odot}\gtrsim9.5$) that appears invariant at $z\simeq2-6$ \citep{shapley22,shapley23}. Importantly, however, the scatter and uncertainties on both individual measurements and small sample sizes remain large, making general conclusions challenging.

We evaluate the potential contributions of dust to the attenuation of our composite spectra, via both continuum and nebular measurements. The evolution of the UV continuum slope presented in Section \ref{subsec:beta} already provides strong indication of declining dust obscuration toward higher redshift. This interpretation is mirrored by our SED-fitting results, which, assuming the SMC-like dust curve of \citet{gordon03} and assuming $R_{\rm V}=2.74$, yield negligible $E(B-V)_{\rm stel}$ values of 0.03-0.10 and a clear decline in values from lower to higher redshift.

A consistent assessment of the nebular color excess with redshift is more challenging: H$\alpha$ is covered only in our two lowest-redshift stacks and blended with [\nii]$\lambda$6585 \AA, H$\gamma$ is blended with [\oiii]$\lambda$4364 \AA, and the faintness of H$\delta$ makes its measurements extremely sensitive to small fluctuations in the underlying continuum fit. Assuming the Milky Way extinction law of \citet[][see \citealt{reddy20} for its appropriateness to high-$z$ nebular spectra]{cardelli89}, we derive upper limits on the H$\alpha$/H$\beta$ decrement (intrinsic ratio H$\alpha$/H$\beta$=2.860) of 2.65 and 3.38 for our $z\simeq5.5$ and $z\simeq6.3$ spectra, which are consistent with no and low ($E(B-V)_{\rm neb}\simeq0.17$) dust scenarios, respectively. Given those values are derived from our lowest-redshift stacks, they can be considered upper limits on the dust attenuation for our higher-redshift stacks as well. The low $E(B-V)_{\rm stel}$ values inferred from our SED fitting and lowest-redshift spectra suggest little to no dust in our stacks, and thus we do not make additional corrections to our spectra in forthcoming sections.

\subsection{Metallicities and Ionization States}
The plethora of emission lines in our stacks allows us to constrain the chemical enrichment of galaxies to high significance and out to the earliest times. In theory, the presence of the [\oiii]$\lambda$4364 \AA\ and [\oiii]$\lambda$5008 \AA\ lines in our spectra also enables us to determine direct T$_{e}$-based oxygen abundances out to $z\sim9.7$. However, the prism's spectral resolution blends the [\oiii]$\lambda$4364 \AA\ line with H$\gamma$, preventing us from modeling it with high confidence (see \citealt{hu24}). We thus derive gas-phase oxygen abundances using the JWST-derived calibrations in \citet{sanders_metal}, specifically using the so-called ``O32'', ``R23'', and ``Ne3O2'' ratios which yield metallicities of 12+log\,(O/H)$\approx$7.3-7.9, 7.1-7.5, and 7.3-7.6, respectively (approximately 0.03-0.16 solar, assuming 12+log\,(O/H)$_{\odot}$=8.69; \citealt{asplund09}). The most obvious trend comes from the $O32$ calibration, which at face value appears to show a well-defined but small evolution of decreasing metallicity with redshift that is in agreement with single-object and average derivations in the literature \citep{sanders_metal,curti23,nakajima23,hu24}. We verify the significance of the evolution by comparing the difference between stacked points and the uncertainties on those differences. We find the decrease between points to be larger than the associated uncertainties by 2-5$\sigma$ (except for the decrease between the $z\sim7.5$ and $z\sim8.6$ points), adding statistical robustness to the trend. Despite the agreement, suggestive of some evolution in the mass-metallicity relation within the redshift range probed, we consider whether moderate stellar mass differences between our stacks could induce some artificial evolution if one assumes an invariant mass-metallicity relation across those redshifts. In such a case, based on the stellar mass differences between our stacks alone, we would infer a metallicity difference of only 0.09 dex \citep{nakajima23}. This is significantly smaller than the metallicity difference we observe between our $z\sim5.5$ and $z\sim9.5$ stacks, which amounts to 0.58 dex. Furthermore, were the metallicity evolution we observe driven by differences in the stellar masses sampled in each redshift stack, the redshift–metallicity trend would then have to mirror the redshift-$M_{*}$ evolution between our stacks, which it does not -- i.e., the metallicity would have to scatter in the same fashion as the stellar mass between our stacks. As such, while the uncertainties mean we cannot rule out some modest influence by stellar mass variations, the overall trend of decreasing metallicities with redshift appears most consistent with some small evolution in the mass-metallicity relation between $z\sim5$ and 10.

\begin{figure}
\center
\includegraphics[width=\columnwidth]{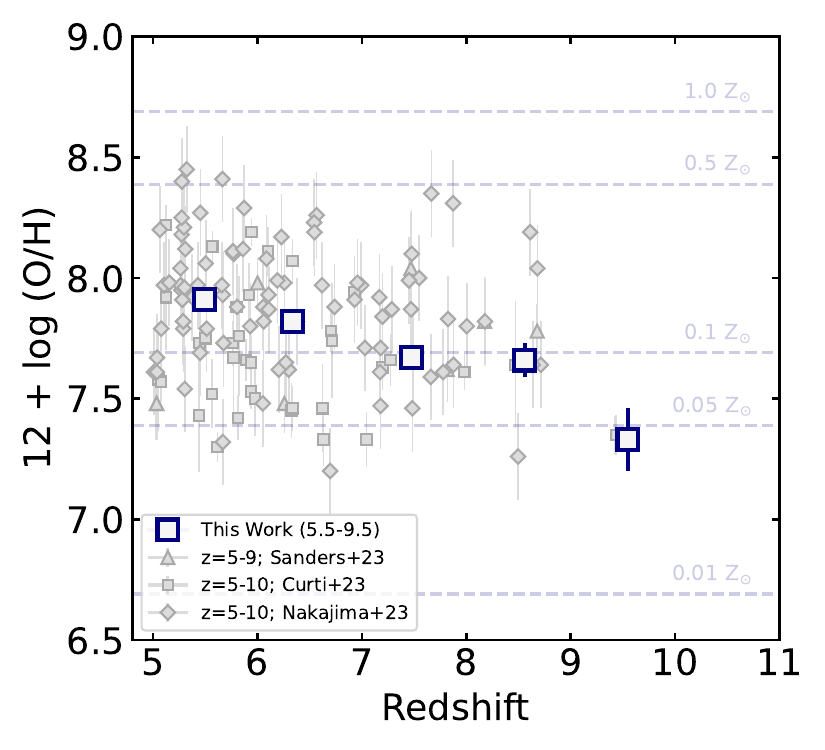}
 \caption{Gas-phase oxygen metallicities as a function of redshift for the stacks derived in this work (blue squares), as estimated from the strong line $O32$ prescription of \citet{sanders_metal}. Literature points from \citet{sanders_metal}, \citet{curti23}, and \citet{nakajima23} are plotted as gray symbols. The dashed lines mark levels of constant metallicity in solar units, assuming 12+log\,(O/H)$_{\odot}$=8.69 \citep{asplund09}. A clear trend of decreasing metallicity toward higher redshifts is evident, likely reflecting larger \hi\ gas fractions of high-$z$ sources.}
 \label{fig:metal}
\end{figure}

\begin{figure*}
\center
\includegraphics[width=\textwidth]{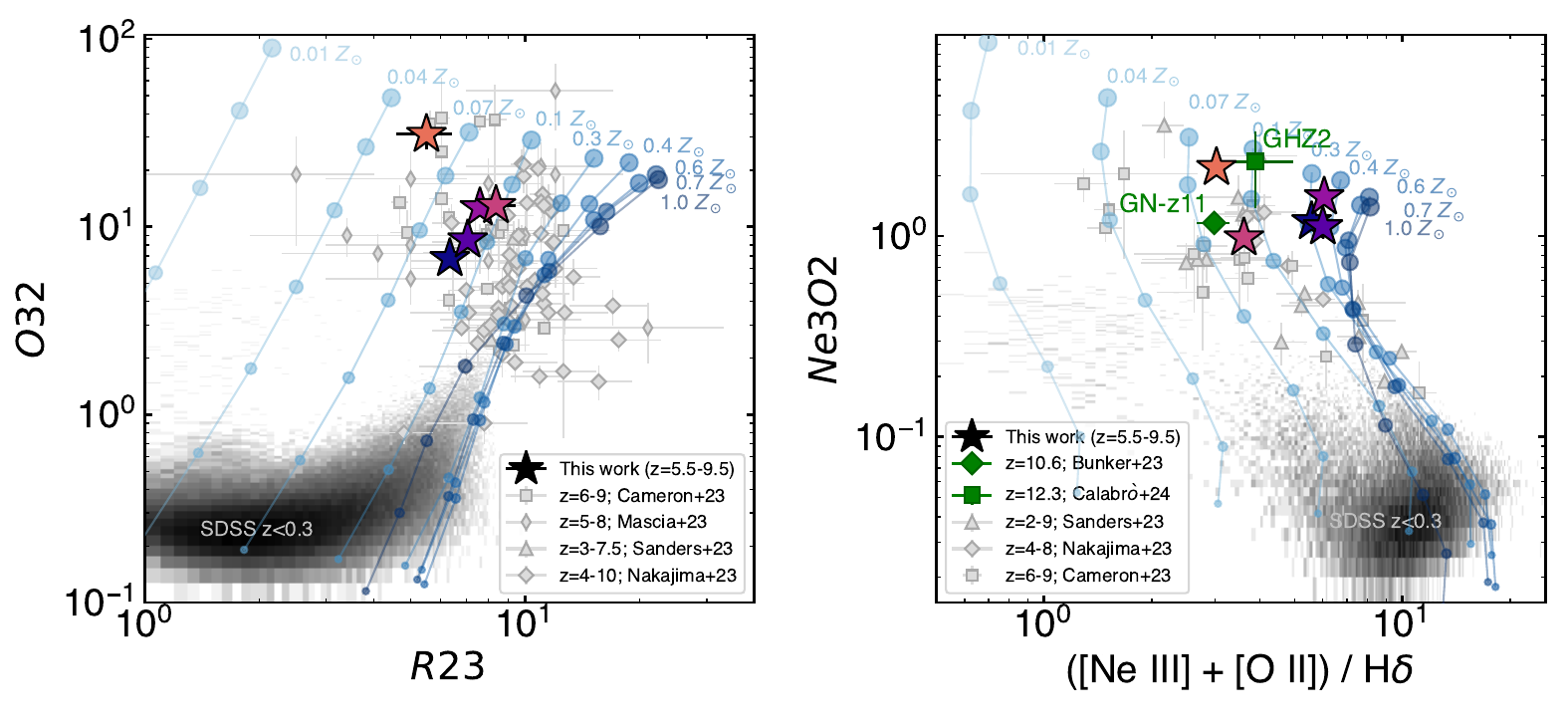}
 \caption{The $R23-O32$ (left) and $Ne3O2-$([\neiii]+[\oii])/H$\delta$ (right) diagrams tracing the ionization and excitation states of galaxy ISM across cosmic time. In each plot, the values and uncertainties (often smaller than the symbols) for our stacked sources are shown as colored stars (color-coded according to Figure~\ref{fig:zstack}), while values from the literature at comparable redshifts (\citealt{sanders23,mascia23,cameron23,nakajima23}) are shown as gray points with varying marker symbols. GN-z11 is shown as a green diamond in the right panel using values from \citet{bunker23}, while GHZ2 is shown as a green square using values from \citet{calabro24}. For reference, the locus of $z<0.3$ points from SDSS are shown as a density plot. Tracks from the photoionization models of \citet{gutkin16} are shown in blue for a variety of metallicities and ionization parameters, where the shade of blue denotes the metallicity (from 0.01 to 0.4 $Z_{\odot}$ going from light shaded to dark shaded) and the size of the connecting points describes the ionization parameter (from log\,$U=-3$ to $-1$ in $-0.5$ intervals, with the largest values represented by the largest circles).}
 \label{fig:r23o32}
\end{figure*}

Ionization versus excitation diagrams also provide a means of interpreting the underlying conditions of the ISM. We show two variations of such a diagram in Figure~\ref{fig:r23o32}, in similar fashion to Figure 8 of \citet{bunker23}. In the left panel we show the $R23-O32$ line ratios, where $R23$ is used as a proxy for gas-phase metallicity while $O32$ is invoked as a proxy for ionization parameter (e.g., \citealt{kewley19}). Low-redshift ($z<0.3$) points from the Sloan Digital Sky Survey (SDSS) MPA-JHU catalogs\footnote{\url{https://www.sdss3.org/dr10/spectro/galaxy_mpajhu.php}} \citep{aihara11} are added to the plots (in black), for reference. A clear trend appears from these diagnostics, whereby the lowest $R23$ and $O32$ values are found in local galaxies and the highest values in the highest-redshift sources, indicating an evolution towards more extreme and ionized ISM with decreased gas-phase metallicities at higher redshift, i.e., the harder ionizing spectrum (traced by an ionization parameter, $U$) associated with high-$z$ sources drags the $R23$ ratio to higher values at fixed $O32$ \citep{steidel16}. Our points, shown as color-coded stars, are in excellent agreement with the single-source measurements at comparable redshifts by \citet{mascia23}, \citet{cameron23}, \citet{sanders23}, and \citet{nakajima23}.

The evolution is seen not only between the locus of SDSS sources and locus of stacked (and literature) high-$z$ points, but possibly between the stacked points themselves, where the spectrum at $z\sim5.5$ displays the lowest values of $R23=6.3\pm0.2$ and $O32=6.8\pm0.1$ and increases to $R23=8.4\pm1.1$ and $O32=13.0\pm1.3$ by $z\sim8.6$ (our redshift $z\sim9.5$ point appears to be somewhat scattered off the main locus of points with $R23=5.5\pm0.9$ and $O32=31.1\pm5.3$, possibly due to the comparatively small numbers in the stack). Despite the limited range in parameter space probed by the stacks, the $z\sim5.5-8.6$ spectra hint at nonnegligible evolution in the ISM conditions of star-forming galaxies within the first $\sim1$ Gyr of the Universe (or $\sim500$ Myr within the redshift range). For comparison, we plot the updated line ratios for star-forming sources from \citet{gutkin16}, as a function of metallicity and ionization parameter, log\,$U$. Our points are consistent, as expected, with subsolar metallicity tracks and are characterized by large log\,$U$ values ($-2.5$ to $-1.5$) compared to $z\sim0$ sources. Lastly, the two locuses of low- and high-redshift sources are also well connected by measurements at intermediate redshifts, namely those of \citet{sanders23} at $z\sim3.0-7.5$ and \citet{nakajima23} at $z\sim4-10$, which sit firmly between the SDSS points and ours.

Analogous to the $R23-O32$ diagram is the $Ne3O2$ line ratio paired with ([\neiii]$\lambda$3869 \AA+[\oii]$\lambda\lambda$3726,3729 \AA) / H$\delta$, neither of which make use of strong [\oiii] and H$\beta$ lines and thus allow for an inspection of ISM conditions with NIRSpec in principle out to $z\sim13.2$. In this case the ([\neiii]+[\oii])/H$\delta$ ratio traces the gas-phase metallicity of galaxies while the $Ne3O2$ ratio traces their ionization strengths. We plot our values and a number of literature values \citep{sanders23,nakajima23,cameron23} in the right panel of Figure~\ref{fig:r23o32}. Similarly to above, we also plot the MPA-JHU values for $z<0.3$ SDSS star-forming galaxies, as well as the photoionization models of \citet{gutkin16}. For reference, we also show the reported values for GN-z11 \citep{bunker23}. Again, we find a clear separation between sources at low redshift, which dominate regions of parameter space associated with low-ionization and high-metallicity systems, and high redshift, which shift toward high values of $Ne3O2$ and low values of ([\neiii]+[\oii])/H$\delta$ where high-ionization and low-metallicity sources dominate. Interestingly, the ratios presented by \citet{bunker23} for GN-z11 are in excellent agreement with those found for our $z\sim9.5$ stack, reflecting ISM conditions comparable to those of ``typical'' sources at higher redshift and a conclusion those authors found when comparing to the individual measurements in $z\sim5.5-9.5$ sources by \citet{cameron23}. A similar conclusion can be derived for GHZ2, whose $Ne3O2$ value (see \citealt{calabro24}) places it at the extreme end of the distribution but within the locus of high-redshift sources clearly distinct from the ISM conditions of $z\sim0$ galaxies. Given its proximity to the measurements of other individual sources at high redshift (e.g., \citealt{sanders_metal,nakajima23}) and our stacked spectra, the result suggests GHZ2 may too be reflective of somewhat typical ISM conditions above $z>5$.

To conclude, the modest change in the $O32$ and $R23$ values from the stacked and literature results shown in the left panel of Figure~\ref{fig:r23o32} point to some possible (albeit modest) evolution in the ionization state and gas-phase metallicities from redshifts $z\simeq9-10$ to $z\simeq3-5$. The combination of both panels clearly indicate a far more drastic change in the ionization state and excitation conditions of galaxies from redshifts $z\simeq3-5$ to $z\sim0$, suggesting the bulk of metal and dust buildup in galaxies occurs between the latter two populations.

\subsection{Do Extreme Properties Lead To Extreme Ly$\alpha$?}
\label{sec:lya}
A significant number of individual sources in our star-forming sample display strong ($\gtrsim30$ \AA) \lya\ emission. Characterizing the properties of such populations and determining whether they exhibit enhanced intrinsic ionizing capabilities compared to their \lya-attenuated counterparts represents a key goal of JWST. Moreover, a thorough understanding of their intrinsic properties is a prerequisite toward using them as tracers of Cosmic Reionization (e.g., through the opacity of \lya; \citealt{Mason2019b}).

We identify a sample of 54 LAEs (12\% of our fiducial sample) via a combination of previous works (specifically \citealt{tang23}, \citealt{saxena23}, \citealt{saxena_faint}, and \citealt{napolitano24}) and visual inspection of our compiled spectra. For reference, 10/54 of the LAEs ($\sim17$\%) fall in the Abell 2744 field, 16/54 ($\sim32$\%) fall in the CANDELS-EGS field, 1/54 ($\sim2$\%) fall in the GOODS-North field, 23/54 ($\sim42$\%) fall in the GOODS-South field, 3/54 ($\sim5$\%) fall in the MACS 0647 field, and 1/54 ($\sim2$\%) fall in the RXJ2129 field. While the main driver of Ly$\alpha$ visibility in each of these fields remains subject to debate, the variability is likely linked to a mixture of intrinsic galaxy properties, large-scale ionized bubbles, and cosmic variance. While the sample of LAEs used here is not sufficiently large to perform stacks in bins of global properties, it is large enough for a comparison to a stacked sample of redshift- and $M_{\rm UV}$-matched non-LAEs.

\begin{figure*}
\center
\includegraphics[width=\textwidth]{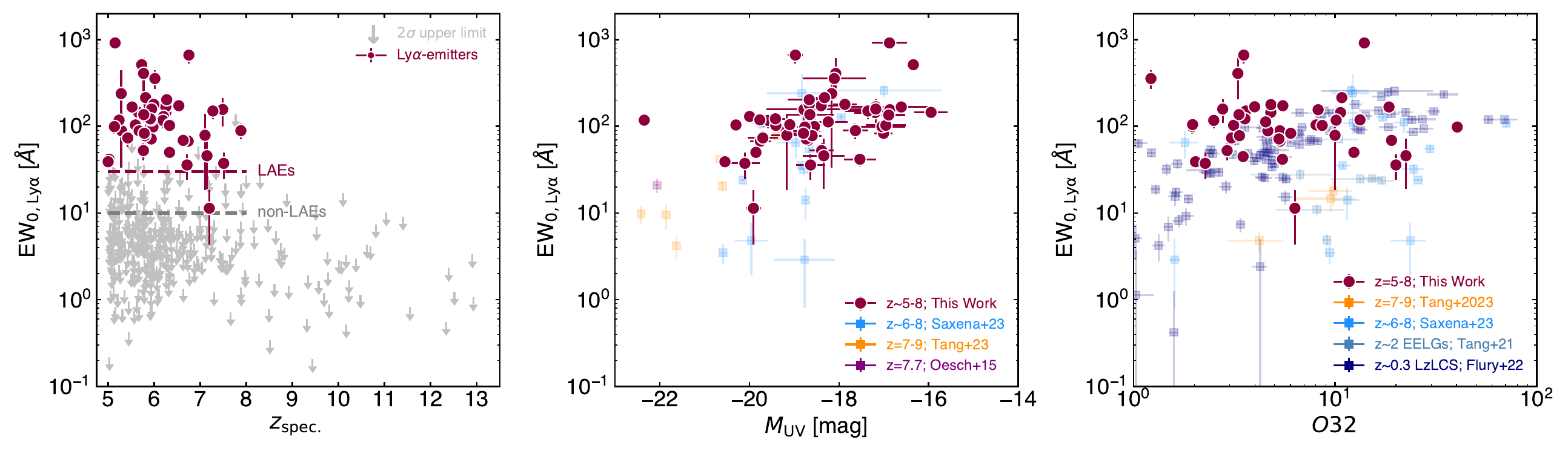}
 \caption{\textit{Left:} the distribution of Ly$\alpha$ rest-frame equivalent widths as a function of spectroscopic redshift for the objects in our fiducial sample. Objects with confirmed Ly$\alpha$ emission from prism observations are highlighted as maroon circles, while undetected sources are represented by $2\sigma$ gray upper limits. Dashed lines represent the EW limits used for the construction of the composite spectra. \textit{Middle:} the same EW measurements from the left panel for Ly$\alpha$-detected sources only, as a function of absolute magnitude. Similar measurements from \citet{oesch15}, \citet{tang23}, and \citet{saxena23} are plotted for comparison. \textit{Right:} the same as the middle plot but as a function of $O32$ parameter. Results for both low- and high-redshift studies using measurements from \citet{flury22}, \citet{tang21,tang23}, and \citet{saxena23} are plotted for comparison.}
 \label{fig:lyasample}
\end{figure*}

To separate the two populations, we first derive \lya\ EWs for each of our LAEs using a Gaussian profile and the median prism continuum flux within a wavelength range $\lambda_{\rm rest}\simeq1250-1750$ \AA. A straight-line fit is included in the Gaussian fit to account for underlying continuum emission (when present). The resulting range of values span EW$_{0,\rm Ly\alpha}\simeq$11-919 \AA\ (with a median value EW$_{0,\rm Ly\alpha}\simeq$155 \AA), consistent with ranges derived by previous works. For all other objects, where the line remains undetected in the prism spectra, we derive $2\sigma$ upper limits using instead the flux measured within a 200 \AA\ window centered at the redshifted position of \lya. The resulting distribution of values, along with the redshifts of their respective hosts, is shown in Figure~\ref{fig:lyasample} where we highlight in maroon our LAE sample and in gray the upper limits, which unsurprisingly span a lower range of values, EW$_{0,\rm non-LAE}\simeq$1-86 \AA.

For comparison, we also include in the middle and right plots of Figure~\ref{fig:lyasample} the \lya\ EWs for the line-detected $z\simeq6-9$ sources in \citet{oesch15}, \citet{tang23}, and \citet{saxena23} and examine potential trends. Firstly, in the middle plot of Figure~\ref{fig:lyasample} we see a clear trend of increasing EW with decreasing luminosity. This is expected for UV-selected sources (see \citealt{schaerer11,schenker14,debarros17,stark17}) where \lya\ is (in general) more easily attenuated in luminous sources dominated by enhanced \hi\ column densities and dust obscuration. However, important selection effects apply: our measurements naturally probe the strongest \lya\ EWs (which correspond to moderately and extremely faint sources with $M_{\rm UV}>-20$ mag) and $R\gtrsim1000$ spectroscopy from the literature probe the weaker emission from more luminous sources, however weak \lya\ (EW$_{0, \rm Ly\alpha}\lesssim$10 \AA) from the intrinsically faint populations ($M_{\rm UV}>-18$ mag) probed by flux-limited samples are beyond the sensitivity of the data sets considered here.

Secondly, we verify whether at fixed $O32$ ratio LAEs at higher redshifts display lower EWs. Although the dynamical range where such comparisons are viable is limited, we do not find any significant evidence for decreasing \lya\ EW at fixed $O32$ ratio across the combined samples plotted in the figure (i.e., this work; \citealt{tang21}, \citealt{tang23}, \citealt{saxena23}, and \citealt{flury22}), which together span a redshift range $z\sim0.3-9.0$. We caution, however, that such an interpretation is sensitive to selection effects, where our prism sample probes only the strongest LAEs and the higher resolution spectroscopy (sensitive to lower EW measurements) is limited in sample size. An enlarged sample of $R>1000$ measurements at $z>6$ is needed for a more complete comparison with which to derive conclusive results.

For our composite spectra we perform two stacks, one of all sources with EW$_{0,\rm Ly\alpha}>30$ \AA\ (which corresponds to the approximate lower limit seen in the LAEs in Figure~\ref{fig:lyasample}) and $z_{\rm spec}<8$, and one with EW$_{0,\rm Ly\alpha}<10$ \AA\ and $z_{\rm spec}<8$. In total, this results in 53 sources for the LAE stack and 249 sources for the non-LAE stack. The composites are an excellent match in redshift ($z_{\rm spec, LAE}=5.92\pm0.60$ and $z_{\rm spec, non-LAE}=5.92\pm0.74$, respectively) and $M_{\rm UV}$ ($M_{\rm UV, LAE}=-18.50\pm0.08$ and $M_{\rm UV, non-LAE}=-18.70\pm0.03$, respectively). Furthermore, while approximately 75\% of compiled LAEs fall in either the GOODS-North or CANDELS-EGS field (where the presence of large-scale ionized bubbles are suspected; \citealt{tilvi20,larson22,endsley22_cosmos}), the sampling of sources from six independent fields should, to a degree, mitigate cosmic variance biases (see \citealt{rb24_borg} for examples of spectroscopy over independent sight lines). The combination of the matched redshifts and $M_{\rm UV}$'s, as well as the sampling from independent fields, allows for a close comparison of the populations' intrinsic properties independently of clustering effects or IGM opacity (see \citealt{rb23}).

\begin{figure*}
\center
 \includegraphics[width=\textwidth]{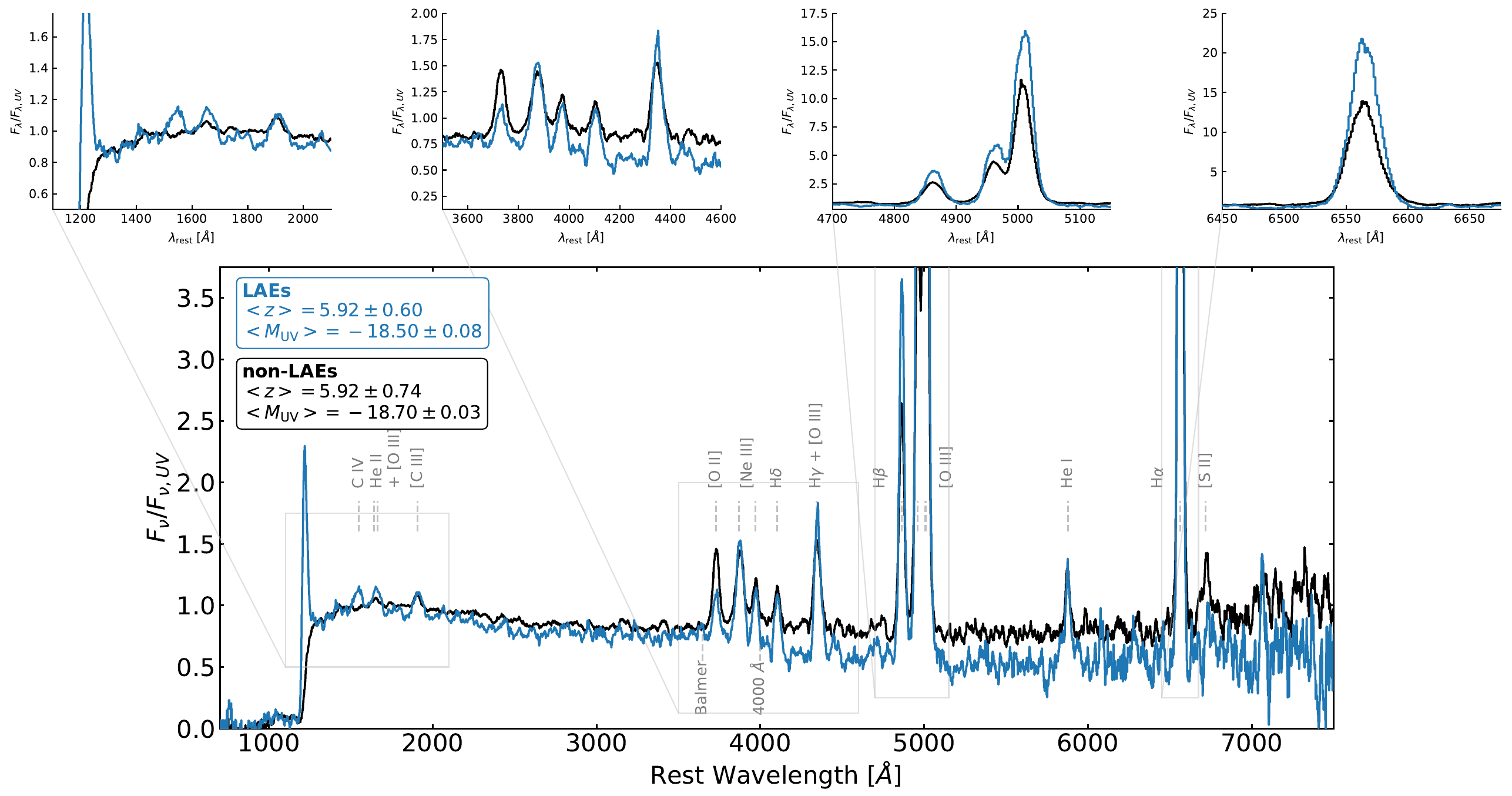}
 \caption{A comparison between the median spectrum of sources with EW$_{0,\rm Ly\alpha}>30$ \AA\ (LAEs; blue) and those with EW$_{0,\rm Ly\alpha}<10$ \AA\ (non-LAEs; black). The two spectra are well matched in median redshift and absolute magnitude, allowing for a close comparison of intrinsic properties: the two spectra reveal modest differences in UV slope, but significant differences in line intensity and ratios (e.g., the presence or otherwise of \civ, \ciii], and \heii+[\oiii] in the rest-frame UV), as well as in the rest-frame optical continuum. Overall, the LAE stack appears to represent sources with more extreme ionizing fields, lower metallicities and stellar masses, and young stellar populations, suggesting intrinsic properties -- in addition to environmental effects -- play some role in regulating the visibility of strong Ly$\alpha$.}
 \label{fig:lyastack}
\end{figure*}

We note modest differences in the UV continuum slopes of the two stacks, with $\beta=-2.59\pm0.05$ for the LAE spectrum and $\beta=-2.40\pm0.02$ for the non-LAE spectrum, both within the range of values probed by the stacks in Figure~\ref{fig:beta} and suggestive of dust-poor systems. Remarkably, we find striking differences in the presence and intensity of rest-frame UV line emission. Our LAE spectrum displays a host of line emission from \civ$\lambda\lambda$1548,1550 \AA\ (EW$_{0}=2.4\pm0.1$ \AA), \heii$\lambda$1640 + [\oiii]$\lambda\lambda$1660,1666 \AA\ (EW$_{0}=3.8\pm0.1$ \AA), and \ciii]$\lambda\lambda$1907,1909 \AA\ (EW$_{0}=6.9\pm0.1$ \AA), while the non-LAE spectrum displays only reduced \ciii]$\lambda\lambda$1907,1909 \AA\ emission (EW$_{0}=4.5\pm0.1$ \AA) and a hint of \heii$\lambda$1640+[\oiii]$\lambda\lambda$1660,1666 \AA (EW$_{0}<1.2$ \AA\ at $2\sigma$), without any obvious \civ\ emission (EW$_{0}<1.8$ \AA\ at $2\sigma$). The \ciii] EWs of the two stacks are consistent with the upper and lower limits of the relation presented in Figure~\ref{fig:ciii} for their respective redshifts, while the enhanced \ciii] emission in the LAE composite relative to the non-LAE composite is not unexpected given results from individual and stacked spectra at $z\sim2-4$ almost always show larger \ciii] EWs with larger \lya\ EWs \citep{shapley03,amorin17,stark17,marchi18}.

The presence of \civ\ in our LAE spectrum alone is also suggestive of particularly extreme ionizing properties, requiring 48 eV to convert \ciii] to \civ\ and most typically seen in extremely metal-poor systems or AGN \citep{feltre16,senchyna19}. \civ\ has already been identified in a number of objects in the Reionization Era (e.g., \citealt{bunker23,topping24}) and their lower-redshift analogs (e.g., \citealt{erb10,stark14,rigby15,berg19,du20}), with a significant fraction of those also displaying \lya\ emission (e.g., \citealt{schenker12,schmidt17,amorin17,naidu22b,bunker23}) which hints at a possible link between \civ\ and high \lya\ escape. This latter consideration has been indirectly proposed in a number of recent studies, which link high \civ/\ciii] ratios to a nonzero escape fraction of Lyman continuum photons \citep{berg19,schaerer22,mascia23b,kramarenko24} and thus by extension \lya. Whether such systems are representative of the global population at $z>6$ remains unclear, although our stacked results suggest they likely are, at least in Ly$\alpha$-emitting systems.

We also find notable difference in the strengths of the rest-frame optical line emission, which we list (fluxes and EWs) in Table~\ref{tab:lyaflux} in the Appendix. The LAE spectrum displays line strengths $\times$1.4-1.7 larger than its non-LAE counterpart for [\neiii]$\lambda\lambda$3969,3870 \AA, H$\delta$, and H$\gamma$+[\oiii], and moderately higher factors of $\sim\times2$ in its \hei$\lambda$4473 \AA, [\oiii]+H$\beta$, and H$\alpha$ emission. The largest difference is seen in the [\oiii]+H$\beta$ line strength, with EWs of 1573$\pm$34 \AA\ and 734$\pm$11 \AA\ for the LAE and non-LAE stacks, respectively. The two measurements are in excellent agreement with the differences observed by \citet{castellano17} for LAEs and non-LAEs through stacked Spitzer/IRAC photometry, as well as more general measurements inferred from both SED fitting (e.g., \citealt{rb16,endsley21b,endsley23,laporte23}) or NIRSpec spectroscopy (e.g., \citealt{williams22,tang23,saxena23}). The more extreme environments inferred from enhanced [\oiii]+H$\beta$ EWs are reflected in the $R23$-$O32$ ratios of the two stacks, with $R23_{\rm LAE}=6.12\pm0.23$ and $O32_{\rm LAE}=21.00\pm0.57$ compared to $R23_{\rm non-LAE}=6.54\pm0.22$ and $O32_{\rm non-LAE}=6.69\pm0.30$. While the $R23$ ratios of the two stacks are consistent with each other, the LAE stack displays a $O32$ ratio more than $\times3$ that of its non-LAE counterpart and $\sim\times1.6-3.0$ larger than our redshift-binned stacks, clearly indicating a higher ionization parameter likely paired with lower metallicity. The inferred metallicities for the two stacks using the $O32$ prescription of \citet{sanders_metal} are log\,(O/H)$_{\rm LAE}=7.48\pm0.01$ and log\,(O/H)$_{\rm non-LAE}=7.91\pm0.02$, supporting the interpretation (see \citealt{cullen20}).

Finally, we also observe significant differences in the rest-frame optical continuum. Our LAE spectrum displays a Balmer index of $F_{\rm red}/F_{\rm blue}=0.79\pm0.07$, similar to the values observed at $z\sim8.5$ in our redshift-binned stacks (see Figure~\ref{fig:d4000}) which suggests young stellar ages and enhanced contributions from the nebular continuum. Our non-LAE stack instead displays a larger Balmer index of $F_{\rm red}/F_{\rm blue}=1.03\pm0.04$, more typical of aged stars and enhanced stellar contributions to the spectrum. The interpretation is corroborated by contrasting stellar mass and mass-weighted ages, where \texttt{gsf}-derived estimates point to 
log\,$M_{*,\rm LAE}/M_{\odot}\sim7.54$ and $\sim4$ Myr for the LAE composite, and log\,$M_{*,\rm non-LAE}/M_{\odot}\sim8.36$ with $\sim$110 Myr for the non-LAE composite. The significant contrast in line and continuum features between the two stacks is demonstrative of intrinsic properties playing some role in regulating the visibility of Ly$\alpha$, and possibly the recent star formation histories of the hosts. This is likely in combination with the presence of overdense regions and/or large ionized bubbles in sources at $z\geqslant6$ (c.f., \citealt{rb23}), and/or with effects from a largely neutral medium for sources at $z\lesssim5.3-5.8$ (see \citealt{eilers18,bosman22,Zhu2023} for discussions on the end of reionization).

The SED fitting of the LAE composite also indicates an effectively dust-free system ($E(B-V)\simeq0.07$), therefore as a final inspection we can use the measured H$\beta$, H$\alpha$, and rest-frame UV continuum fluxes to infer a number of fundamental quantities related to the \lya\ visibility. One of the most informative is the production rate of ionizing continuum photons, $\xi_{\rm ion}$. Given $\xi_{\rm ion}$ is related to H$\alpha$ via:

\begin{equation}
\begin{split}
\xi_{\rm ion}&=\frac{N(H^{0})}{L_{\rm 1500}} \\
&=\frac{L({\rm H\alpha})}{(1-f_{\rm esc})}\frac{1}{L_{\rm 1500}}\,\rm [7.4\times10^{11}\,\,s^{-1}\,erg^{-1}\,s^{-1}\,Hz^{-1}]
\end{split}
\end{equation}

where $N(H^{0})$ is the production rate of ionizing continuum photons, $L(\rm H\alpha)$ is the luminosity of H$\alpha$, $L_{\rm 1500}$ is the continuum luminosity at approximately $\lambda_{\rm rest}\simeq1500$ \AA, and $f_{\rm esc}$ is the escape fraction of ionizing continuum photons (assumed to be zero here). From the measured H$\alpha$ flux ($f(\rm H\alpha)=1.1\times10^{-18}$ erg\,s$^{-1}$\,cm$^{-2}$), we derive log\,$\xi_{\rm ion, \rm LAE}=25.49\pm0.01$ Hz\,erg$^{-1}$, which is consistent with canonical values derived from SED fitting of high-redshift sources and/or low-redshift populations (e.g., \citealt{bouwens16,matthee17,shivae18,simmonds24}). Applying the same procedure to the non-LAE stack, we derive a similarly consistent value of log\,$\xi_{\rm ion, \rm non-LAE}=25.30\pm0.01$ Hz\,erg$^{-1}$. We note those values increase by only 0.04 dex for modest escape fractions of 10\%.

Additionally, we evaluate the \lya\ escape fraction, $f^{\rm Ly\alpha}_{\rm esc}$ of our composite LAE spectrum. From Case B recombination theory, we adopt an \lya/H$\beta$ ratio of 24.273 (assuming \lya/H$\alpha$=8.7 and H$\alpha$/H$\beta$=2.790) and, multiplied by our measured H$\beta$ flux, $f(\rm H\beta)=4.0\times10^{-19}$ erg\,s$^{-1}$\,cm$^{-2}$, derive an intrinsic \lya\ flux of $f^{\rm Ly\alpha}_{\rm int}=9.8\times10^{-18}$ erg\,s$^{-1}$\,cm$^{-2}$. From a simple Gaussian fit to the continuum-subtracted \lya\ profile, we derive an observed \lya\ flux of $f^{\rm Ly\alpha}_{\rm obs}=3.5\times10^{-18}$ erg\,s$^{-1}$\,cm$^{-2}$ which equates to an escape fraction of approximately $f^{\rm Ly\alpha}_{\rm esc}\simeq36$\%, on the high end of but consistent with measurements of individual objects at $z\geqslant5$ reported in the literature (see \citealt{saxena23,napolitano24}).

A more in-depth interpretation of such values and their implications for the contributions of galaxies to the reionization process requires a thorough investigation into the underlying properties of the sample going into our stack (which in this case probes the strongest LAEs), which we leave to future papers. Nevertheless, our comparison of the average spectra of LAEs and non-LAEs very clearly illustrates important intrinsic differences between the two populations. Considering the identical median redshifts of the stacks ($z\sim5.9$), those differences are likely related to the physics within the galaxies themselves and their immediate surroundings as opposed to the neutrality of the intervening IGM. These differences will have to be taken into account when using, e.g., galaxies and their \lya\ emission as a probe of the neutral density of the Universe. For example, the high \lya\ detection fraction in $z>7$ samples selected for their extreme optical emission lines \citep{rb16} is easier to reproduce if they have intrinsically higher \lya\ emission than UV-continuum-selected galaxies \citep{rb23}. In order to further characterize these differences, higher spectral resolution spectra are needed, so as to increase the sensitivity to \lya\ and determine its velocity profile in comparison to the rest-frame optical Balmer lines. Our stacks, and average line ratios, are therefore intended as an initial benchmark for reference in future studies.

\section{Summary and Conclusions}
\label{sec:concl}
We have performed the largest compilation to date of $z\geqslant5$ sources from public JWST/NIRSpec prism spectra. We identify 482 sources with secure redshifts primarily from the [\oiii]+H$\beta$ emission-line complex at $z\simeq5.0-9.7$ and the Lyman-$\alpha$ break at higher redshifts, with 356 of those sources representing novel confirmations. After removal of candidate unobscured AGN sources, we verify that our sample is commensurate with the star-forming main sequence at comparable redshifts, and use our statistical sample to construct representative and high SNR composite spectra as a function of redshift, absolute magnitude, stellar mass, and \lya\ EW. We report on their average properties while making those templates available to the community. Our main findings can be summarized as follows.

\begin{itemize}
    \item We provide the tightest constraints yet on the UV continuum slopes of high-redshift galaxies and their evolution between $z\simeq5-11$, finding blue slopes of $\beta\simeq-2.3$ to $-2.7$, indicative of overall dust-poor ISM and young stellar populations. Sources at $z\sim5.5$ display the reddest slopes, followed by a steep decline to bluer slopes for sources at $z\sim6-7$, and a flattening off in the evolution between $z\sim7-11$. UV-luminous and higher-mass sources display systematically redder UV slopes compared to their UV-fainter and lower-mass counterparts, likely reflecting more evolved ISM conditions. On average, sources at $z>10$ do not display slopes of $\beta\lesssim-2.7$ indicative of dust-free systems.

    \item We report a steadily declining rest-frame optical continuum break with redshift, indicative of declining stellar ages and metallicities. The largest breaks ($F_{\nu, \rm 4200}/F_{\nu, \rm 3500}>1$) are attributed to our lowest-redshift ($z\sim5.5$) stack, while our higher-redshift composites are characterized by optical jumps of $F_{\nu, \rm 4200}/F_{\nu, \rm 3500}<1$ attributable to extremely young and metal-poor stellar populations, and a possible enhancement of the nebular continuum.

    \item All of our composite spectra display unambiguous detections of \ciii]1907,1909 \AA\ emission, with equivalent widths of EW$_{0}\simeq5-14$ \AA\ and a clear increase of the line strength with redshift. The evolving EW of the line is consistent with an extrapolation of results at $z\sim0-3$, while its prevalence and strength in our stacks suggests a high cadence among individual spectra. A comparison to confirmed objects in the literature suggests a large number classified as ``extreme'' in fact reside close to or on the relation, reflecting \ciii] strengths comparable to the general star-forming population. Prominent \heii+[\oiii] line emission is also consistently detected, but at comparatively low strengths.
    
    \item Ratios of strong rest-optical line emission ($O32$, $R23$, $Ne3O2$, and (\neiii+[\oiii])/H$\delta$) indicate subsolar gas-phase metallicities (log\,(O/H)$\simeq7.2-7.9$; $\sim$0.03-0.16 $Z_{\odot}$) which decline with redshift. Comparisons to photoionization models further reveal that $z\geqslant5$ galaxies populate regions of ionization versus excitation diagrams typically associated with especially strong ionization parameters, indicative of extreme ionizing radiation fields not observed at $z\sim0$.

    \item Lastly, a direct comparison between the average (redshift- and $M_{\rm UV}$-matched) spectra of LAEs and non-LAEs reveals important differences in their intrinsic properties. The spectrum of the former displays enhanced UV emission (from \ciii], \civ, and \heii+[\oiii]), bluer UV continuum slopes, an inverse Balmer jump, and stronger line emission with enhanced ratios compared to the latter. Inferences from H$\alpha$ and \lya\ emission in the LAE stack suggest a production rate of ionizing continuum photons of log\,$\xi_{\rm ion}=25.49\pm0.01$ Hz\,erg$^{-1}$ and \lya\ escape fraction of $f^{\rm Ly\alpha}_{\rm esc}\simeq35$\%, consistent with canonical values and measurements of individual high-$z$ sources in the literature. The observed intrinsic differences suggest LAEs are governed by more extreme ionizing fields, young stellar populations, dust-poor ISM, and low metallicities, the combination of which -- combined with environmental effects -- may be more conducive to higher escape fractions of \lya\ photons compared to non-LAEs regardless of environmental effects.
\end{itemize}

The combination of JWST/NIRSpec's unprecedented sensitivity and wavelength coverage is opening a new window with which to characterize galaxies over the first $\sim1.2$ Gyr of the Universe. The plethora of available spectroscopic data sets from only $\sim1.5$ years of operations now allow us to move from single-object characterizations to statistical analyses generally attributed to photometric studies, placing novel and direct constraints on the evolution of galaxy properties across entire populations and out to the earliest times. This study represents one of the first (and largest) examples to do so, mirroring spectroscopic studies at $z\sim2-4$ and extending them to star-forming sources representative of the populations dominating the aftermath, the conclusion, and the thick of the reionization process. The findings and templates derived in this study represent a new and observational benchmark for future studies of high-redshift galaxies, allowing for a more direct comparison of galaxy evolution across cosmic time and acting as a spectroscopic reference point for the interpretation of individual sources out to the highest redshifts.

\acknowledgments
We are grateful and extend our thanks to a number of people. First, we thank Adi Zitrin and Asheesh Meena for providing the magnification factors of confirmed sources behind the MACS 0647 cluster. We also thank Anna Feltre, St\'ephane Charlot, and Alex Cameron for providing and discussing the photoionization models used in this study. G.R.B. also thanks Gabe Brammer for helpful discussions on the reduction of the JWST spectra, and for his extensive efforts in developing and maintaining the \texttt{msaexp} code from which the high-$z$ community has greatly benefited. Last but not least, we also thank Daniel Schaerer, Rui Marques-Chaves, and Emma Giovinazzo for additional models used in this study and for useful discussions on the science described here. Some/all of the data presented in this article were obtained from the Mikulski Archive for Space Telescopes (MAST) at the Space Telescope Science Institute. The specific observations analyzed can be accessed via DOIs \dataset[10.17909/j4rg-3f57]{https://doi.org/10.17909/j4rg-3f57} and \dataset[10.17909/8tdj-8n28]{https://doi.org/10.17909/8tdj-8n28}.

\bibliography{sample63}{}

\begin{thebibliography}{}
\expandafter\ifx\csname natexlab\endcsname\relax\def\natexlab#1{#1}\fi
\providecommand{\url}[1]{\href{#1}{#1}}
\providecommand{\dodoi}[1]{doi:~\href{http://doi.org/#1}{\nolinkurl{#1}}}
\providecommand{\doeprint}[1]{\href{http://ascl.net/#1}{\nolinkurl{http://ascl.net/#1}}}
\providecommand{\doarXiv}[1]{\href{https://arxiv.org/abs/#1}{\nolinkurl{https://arxiv.org/abs/#1}}}

\bibitem[{{Aihara} {et~al.}(2011){Aihara}, {Allende Prieto}, {An}, {Anderson}, {Aubourg}, {Balbinot}, {Beers}, {Berlind}, {Bickerton}, {Bizyaev}, {Blanton}, {Bochanski}, {Bolton}, {Bovy}, {Brandt}, {Brinkmann}, {Brown}, {Brownstein}, {Busca}, {Campbell}, {Carr}, {Chen}, {Chiappini}, {Comparat}, {Connolly}, {Cortes}, {Croft}, {Cuesta}, {da Costa}, {Davenport}, {Dawson}, {Dhital}, {Ealet}, {Ebelke}, {Edmondson}, {Eisenstein}, {Escoffier}, {Esposito}, {Evans}, {Fan}, {Femen{\'\i}a Castell{\'a}}, {Font-Ribera}, {Frinchaboy}, {Ge}, {Gillespie}, {Gilmore}, {Gonz{\'a}lez Hern{\'a}ndez}, {Gott}, {Gould}, {Grebel}, {Gunn}, {Hamilton}, {Harding}, {Harris}, {Hawley}, {Hearty}, {Ho}, {Hogg}, {Holtzman}, {Honscheid}, {Inada}, {Ivans}, {Jiang}, {Johnson}, {Jordan}, {Jordan}, {Kazin}, {Kirkby}, {Klaene}, {Knapp}, {Kneib}, {Kochanek}, {Koesterke}, {Kollmeier}, {Kron}, {Lampeitl}, {Lang}, {Le Goff}, {Lee}, {Lin}, {Long}, {Loomis}, {Lucatello}, {Lundgren}, {Lupton}, {Ma}, {MacDonald}, {Mahadevan}, {Maia}, {Makler},
  {Malanushenko}, {Malanushenko}, {Mandelbaum}, {Maraston}, {Margala}, {Masters}, {McBride}, {McGehee}, {McGreer}, {M{\'e}nard}, {Miralda-Escud{\'e}}, {Morrison}, {Mullally}, {Muna}, {Munn}, {Murayama}, {Myers}, {Naugle}, {Neto}, {Nguyen}, {Nichol}, {O'Connell}, {Ogando}, {Olmstead}, {Oravetz}, {Padmanabhan}, {Palanque-Delabrouille}, {Pan}, {Pandey}, {P{\^a}ris}, {Percival}, {Petitjean}, {Pfaffenberger}, {Pforr}, {Phleps}, {Pichon}, {Pieri}, {Prada}, {Price-Whelan}, {Raddick}, {Ramos}, {Reyl{\'e}}, {Rich}, {Richards}, {Rix}, {Robin}, {Rocha-Pinto}, {Rockosi}, {Roe}, {Rollinde}, {Ross}, {Ross}, {Rossetto}, {S{\'a}nchez}, {Sayres}, {Schlegel}, {Schlesinger}, {Schmidt}, {Schneider}, {Sheldon}, {Shu}, {Simmerer}, {Simmons}, {Sivarani}, {Snedden}, {Sobeck}, {Steinmetz}, {Strauss}, {Szalay}, {Tanaka}, {Thakar}, {Thomas}, {Tinker}, {Tofflemire}, {Tojeiro}, {Tremonti}, {Vandenberg}, {Vargas Maga{\~n}a}, {Verde}, {Vogt}, {Wake}, {Wang}, {Weaver}, {Weinberg}, {White}, {White}, {Yanny}, {Yasuda}, {Yeche}, \&
  {Zehavi}}]{aihara11}
{Aihara}, H., {Allende Prieto}, C., {An}, D., {et~al.} 2011, \apjs, 193, 29, \dodoi{10.1088/0067-0049/193/2/29}

\bibitem[{{Amor{\'\i}n} {et~al.}(2017){Amor{\'\i}n}, {Fontana}, {P{\'e}rez-Montero}, {Castellano}, {Guaita}, {Grazian}, {Le F{\`e}vre}, {Ribeiro}, {Schaerer}, {Tasca}, {Thomas}, {Bardelli}, {Cassar{\`a}}, {Cassata}, {Cimatti}, {Contini}, {de Barros}, {Garilli}, {Giavalisco}, {Hathi}, {Koekemoer}, {Le Brun}, {Lemaux}, {Maccagni}, {Pentericci}, {Pforr}, {Talia}, {Tresse}, {Vanzella}, {Vergani}, {Zamorani}, {Zucca}, \& {Merlin}}]{amorin17}
{Amor{\'\i}n}, R., {Fontana}, A., {P{\'e}rez-Montero}, E., {et~al.} 2017, Nature Astronomy, 1, 0052, \dodoi{10.1038/s41550-017-0052}

\bibitem[{{Arrabal Haro} {et~al.}(2023{\natexlab{a}}){Arrabal Haro}, {Dickinson}, {Finkelstein}, {Fujimoto}, {Fern{\'a}ndez}, {Kartaltepe}, {Jung}, {Cole}, {Burgarella}, {Chworowsky}, {Hutchison}, {Morales}, {Papovich}, {Simons}, {Amor{\'\i}n}, {Backhaus}, {Bagley}, {Bisigello}, {Calabr{\`o}}, {Castellano}, {Cleri}, {Dav{\'e}}, {Dekel}, {Ferguson}, {Fontana}, {Gawiser}, {Giavalisco}, {Harish}, {Hathi}, {Hirschmann}, {Holwerda}, {Huertas-Company}, {Koekemoer}, {Larson}, {Lucas}, {Mobasher}, {P{\'e}rez-Gonz{\'a}lez}, {Pirzkal}, {Rose}, {Santini}, {Trump}, {de la Vega}, {Wang}, {Weiner}, {Wilkins}, {Yang}, {Yung}, \& {Zavala}}]{arrabal23b}
{Arrabal Haro}, P., {Dickinson}, M., {Finkelstein}, S.~L., {et~al.} 2023{\natexlab{a}}, arXiv e-prints, arXiv:2304.05378, \dodoi{10.48550/arXiv.2304.05378}

\bibitem[{{Arrabal Haro} {et~al.}(2023{\natexlab{b}}){Arrabal Haro}, {Dickinson}, {Finkelstein}, {Kartaltepe}, {Donnan}, {Burgarella}, {Carnall}, {Cullen}, {Dunlop}, {Fern{\'a}ndez}, {Fujimoto}, {Jung}, {Krips}, {Larson}, {Papovich}, {P{\'e}rez-Gonz{\'a}lez}, {Amor{\'\i}n}, {Bagley}, {Buat}, {Casey}, {Chworowsky}, {Cohen}, {Ferguson}, {Giavalisco}, {Huertas-Company}, {Hutchison}, {Kocevski}, {Koekemoer}, {Lucas}, {McLeod}, {McLure}, {Pirzkal}, {Trump}, {Weiner}, {Wilkins}, \& {Zavala}}]{arrabal23}
---. 2023{\natexlab{b}}, arXiv e-prints, arXiv:2303.15431, \dodoi{10.48550/arXiv.2303.15431}

\bibitem[{{Asplund} {et~al.}(2009){Asplund}, {Grevesse}, {Sauval}, \& {Scott}}]{asplund09}
{Asplund}, M., {Grevesse}, N., {Sauval}, A.~J., \& {Scott}, P. 2009, \araa, 47, 481, \dodoi{10.1146/annurev.astro.46.060407.145222}

\bibitem[{{Bagley} {et~al.}(2023){Bagley}, {Finkelstein}, {Koekemoer}, {Ferguson}, {Arrabal Haro}, {Dickinson}, {Kartaltepe}, {Papovich}, {P{\'e}rez-Gonz{\'a}lez}, {Pirzkal}, {Somerville}, {Willmer}, {Yang}, {Yung}, {Fontana}, {Grazian}, {Grogin}, {Hirschmann}, {Kewley}, {Kirkpatrick}, {Kocevski}, {Lotz}, {Medrano}, {Morales}, {Pentericci}, {Ravindranath}, {Trump}, {Wilkins}, {Calabr{\`o}}, {Cooper}, {Costantin}, {de la Vega}, {Hilbert}, {Hutchison}, {Larson}, {Lucas}, {McGrath}, {Ryan}, {Wang}, \& {Wuyts}}]{bagley23}
{Bagley}, M.~B., {Finkelstein}, S.~L., {Koekemoer}, A.~M., {et~al.} 2023, \apjl, 946, L12, \dodoi{10.3847/2041-8213/acbb08}

\bibitem[{{Bakx} {et~al.}(2021){Bakx}, {Sommovigo}, {Carniani}, {Ferrara}, {Akins}, {Fujimoto}, {Hagimoto}, {Knudsen}, {Pallottini}, {Tamura}, \& {Watson}}]{bakx21}
{Bakx}, T. J.~L.~C., {Sommovigo}, L., {Carniani}, S., {et~al.} 2021, \mnras, 508, L58, \dodoi{10.1093/mnrasl/slab104}

\bibitem[{{Balogh} {et~al.}(1999){Balogh}, {Morris}, {Yee}, {Carlberg}, \& {Ellingson}}]{balogh99}
{Balogh}, M.~L., {Morris}, S.~L., {Yee}, H.~K.~C., {Carlberg}, R.~G., \& {Ellingson}, E. 1999, \apj, 527, 54, \dodoi{10.1086/308056}

\bibitem[{Barbary(2016)}]{barbary2016}
Barbary, K. 2016, Journal of Open Source Software, 1, 58, \dodoi{10.21105/joss.00058}

\bibitem[{{Berg} {et~al.}(2019){Berg}, {Chisholm}, {Erb}, {Pogge}, {Henry}, \& {Olivier}}]{berg19}
{Berg}, D.~A., {Chisholm}, J., {Erb}, D.~K., {et~al.} 2019, \apjl, 878, L3, \dodoi{10.3847/2041-8213/ab21dc}

\bibitem[{{Bergamini} {et~al.}(2023){Bergamini}, {Acebron}, {Grillo}, {Rosati}, {Bartosch Caminha}, {Mercurio}, {Vanzella}, {Mason}, {Treu}, {Angora}, {Brammer}, {Meneghetti}, {Nonino}, {Boyett}, {Bradac}, {Castellano}, {Fontana}, {Morishita}, {Paris}, {Prieto-Lyon}, {Roberts-Borsani}, {Roy}, {Santini}, {Vulcani}, {Wang}, \& {Yang}}]{bergamini23}
{Bergamini}, P., {Acebron}, A., {Grillo}, C., {et~al.} 2023, arXiv e-prints, arXiv:2303.10210, \dodoi{10.48550/arXiv.2303.10210}

\bibitem[{{Bertin} \& {Arnouts}(1996)}]{bertin96}
{Bertin}, E., \& {Arnouts}, S. 1996, \aaps, 117, 393, \dodoi{10.1051/aas:1996164}

\bibitem[{{Bezanson} {et~al.}(2022){Bezanson}, {Labbe}, {Whitaker}, {Leja}, {Price}, {Franx}, {Brammer}, {Marchesini}, {Zitrin}, {Wang}, {Weaver}, {Furtak}, {Atek}, {Coe}, {Cutler}, {Dayal}, {van Dokkum}, {Feldmann}, {Forster Schreiber}, {Fujimoto}, {Geha}, {Glazebrook}, {de Graaff}, {Greene}, {Juneau}, {Kassin}, {Kriek}, {Khullar}, {Maseda}, {Mowla}, {Muzzin}, {Nanayakkara}, {Nelson}, {Oesch}, {Pacifici}, {Pan}, {Papovich}, {Setton}, {Shapley}, {Smit}, {Stefanon}, {Taylor}, \& {Williams}}]{bezanson22}
{Bezanson}, R., {Labbe}, I., {Whitaker}, K.~E., {et~al.} 2022, arXiv e-prints, arXiv:2212.04026, \dodoi{10.48550/arXiv.2212.04026}

\bibitem[{{Bosman} {et~al.}(2022){Bosman}, {Davies}, {Becker}, {Keating}, {Davies}, {Zhu}, {Eilers}, {D'Odorico}, {Bian}, {Bischetti}, {Cristiani}, {Fan}, {Farina}, {Haehnelt}, {Hennawi}, {Kulkarni}, {Mesinger}, {Meyer}, {Onoue}, {Pallottini}, {Qin}, {Ryan-Weber}, {Schindler}, {Walter}, {Wang}, \& {Yang}}]{bosman22}
{Bosman}, S. E.~I., {Davies}, F.~B., {Becker}, G.~D., {et~al.} 2022, \mnras, 514, 55, \dodoi{10.1093/mnras/stac1046}

\bibitem[{{Bouwens} {et~al.}(2016){Bouwens}, {Smit}, {Labb{\'e}}, {Franx}, {Caruana}, {Oesch}, {Stefanon}, \& {Rasappu}}]{bouwens16}
{Bouwens}, R.~J., {Smit}, R., {Labb{\'e}}, I., {et~al.} 2016, \apj, 831, 176, \dodoi{10.3847/0004-637X/831/2/176}

\bibitem[{{Bouwens} {et~al.}(2014){Bouwens}, {Illingworth}, {Oesch}, {Labb{\'e}}, {van Dokkum}, {Trenti}, {Franx}, {Smit}, {Gonzalez}, \& {Magee}}]{bouwens14}
{Bouwens}, R.~J., {Illingworth}, G.~D., {Oesch}, P.~A., {et~al.} 2014, \apj, 793, 115, \dodoi{10.1088/0004-637X/793/2/115}

\bibitem[{{Bouwens} {et~al.}(2015){Bouwens}, {Illingworth}, {Oesch}, {Trenti}, {Labb{\'e}}, {Bradley}, {Carollo}, {van Dokkum}, {Gonzalez}, {Holwerda}, {Franx}, {Spitler}, {Smit}, \& {Magee}}]{bouwens15}
---. 2015, \apj, 803, 34, \dodoi{10.1088/0004-637X/803/1/34}

\bibitem[{{Bouwens} {et~al.}(2021){Bouwens}, {Smit}, {Schouws}, {Stefanon}, {Bowler}, {Endsley}, {Gonzalez}, {Inami}, {Stark}, {Oesch}, {Hodge}, {Aravena}, {da Cunha}, {Dayal}, {de Looze}, {Ferrara}, {Fudamoto}, {Graziani}, {Li}, {Nanayakkara}, {Pallotini}, {Schneider}, {Sommovigo}, {Topping}, {van der Werf}, {Barrufet}, {Hygate}, {Labbe}, {Riechers}, \& {Witstok}}]{bouwens21_rebels}
{Bouwens}, R.~J., {Smit}, R., {Schouws}, S., {et~al.} 2021, arXiv e-prints, arXiv:2106.13719.
\newblock \doarXiv{2106.13719}

\bibitem[{Brammer(2021)}]{brammer_eazy}
Brammer, G. 2021, {eazy-py}, 0.5.2, \dodoi{10.5281/zenodo.5012704}

\bibitem[{Brammer(2023)}]{grizli}
---. 2023, {grizli}, \dodoi{10.5281/zenodo.1146904}

\bibitem[{{Bruzual A.}(1983)}]{bruzual83}
{Bruzual A.}, G. 1983, \apj, 273, 105, \dodoi{10.1086/161352}

\bibitem[{{Bunker} {et~al.}(2023{\natexlab{a}}){Bunker}, {Saxena}, {Cameron}, {Willott}, {Curtis-Lake}, {Jakobsen}, {Carniani}, {Smit}, {Maiolino}, {Witstok}, {Curti}, {D'Eugenio}, {Jones}, {Ferruit}, {Arribas}, {Charlot}, {Chevallard}, {Giardino}, {de Graaff}, {Looser}, {Luetzgendorf}, {Maseda}, {Rawle}, {Rix}, {Rodriguez Del Pino}, {Alberts}, {Egami}, {Eisenstein}, {Endsley}, {Hainline}, {Hausen}, {Johnson}, {Rieke}, {Rieke}, {Robertson}, {Shivaei}, {Stark}, {Sun}, {Tacchella}, {Tang}, {Williams}, {Willmer}, {Baker}, {Baum}, {Bhatawdekar}, {Bowler}, {Boyett}, {Chen}, {Circosta}, {Helton}, {Ji}, {Lyu}, {Nelson}, {Parlanti}, {Perna}, {Sandles}, {Scholtz}, {Suess}, {Topping}, {Uebler}, {Wallace}, \& {Whitler}}]{bunker23}
{Bunker}, A.~J., {Saxena}, A., {Cameron}, A.~J., {et~al.} 2023{\natexlab{a}}, arXiv e-prints, arXiv:2302.07256, \dodoi{10.48550/arXiv.2302.07256}

\bibitem[{{Bunker} {et~al.}(2023{\natexlab{b}}){Bunker}, {Cameron}, {Curtis-Lake}, {Jakobsen}, {Carniani}, {Curti}, {Witstok}, {Maiolino}, {D'Eugenio}, {Looser}, {Willott}, {Bonaventura}, {Hainline}, {Uebler}, {Willmer}, {Saxena}, {Smit}, {Alberts}, {Arribas}, {Baker}, {Baum}, {Bhatawdekar}, {Bowler}, {Boyett}, {Charlot}, {Chen}, {Chevallard}, {Circosta}, {DeCoursey}, {de Graaff}, {Egami}, {Eisenstein}, {Endsley}, {Ferruit}, {Giardino}, {Hausen}, {Helton}, {Hviding}, {Ji}, {Johnson}, {Jones}, {Kumari}, {Laseter}, {Luetzgendorf}, {Maseda}, {Nelson}, {Parlanti}, {Perna}, {Rawle}, {Rix}, {Rieke}, {Robertson}, {Rodriguez Del Pino}, {Sandles}, {Scholtz}, {Sharpe}, {Skarbinski}, {Stark}, {Sun}, {Tacchella}, {Topping}, {Villanueva}, {Wallace}, {Williams}, \& {Woodrum}}]{bunker_jades}
{Bunker}, A.~J., {Cameron}, A.~J., {Curtis-Lake}, E., {et~al.} 2023{\natexlab{b}}, arXiv e-prints, arXiv:2306.02467, \dodoi{10.48550/arXiv.2306.02467}

\bibitem[{{Calabro} {et~al.}(2024){Calabro}, {Castellano}, {Zavala}, {Pentericci}, {Arrabal Haro}, {Bakx}, {Burgarella}, {Casey}, {Dickinson}, {Finkelstein}, {Fontana}, {Llerena}, {Mascia}, {Merlin}, {Mitsuhashi}, {Napolitano}, {Paris}, {Perez-Gonzalez}, {Roberts-Borsani}, {Santini}, {Treu}, \& {Vanzella}}]{calabro24}
{Calabro}, A., {Castellano}, M., {Zavala}, J.~A., {et~al.} 2024, arXiv e-prints, arXiv:2403.12683, \dodoi{10.48550/arXiv.2403.12683}

\bibitem[{{Cameron} {et~al.}(2023{\natexlab{a}}){Cameron}, {Katz}, {Witten}, {Saxena}, {Laporte}, \& {Bunker}}]{cameron23_nebular}
{Cameron}, A.~J., {Katz}, H., {Witten}, C., {et~al.} 2023{\natexlab{a}}, arXiv e-prints, arXiv:2311.02051, \dodoi{10.48550/arXiv.2311.02051}

\bibitem[{{Cameron} {et~al.}(2023{\natexlab{b}}){Cameron}, {Saxena}, {Bunker}, {D'Eugenio}, {Carniani}, {Maiolino}, {Curtis-Lake}, {Ferruit}, {Jakobsen}, {Arribas}, {Bonaventura}, {Charlot}, {Chevallard}, {Curti}, {Looser}, {Maseda}, {Rawle}, {Rodr{\'\i}guez Del Pino}, {Smit}, {{\"U}bler}, {Willott}, {Witstok}, {Egami}, {Eisenstein}, {Johnson}, {Hainline}, {Rieke}, {Robertson}, {Stark}, {Tacchella}, {Williams}, {Bhatawdekar}, {Bowler}, {Boyett}, {Circosta}, {Helton}, {Jones}, {Kumari}, {Ji}, {Nelson}, {Parlanti}, {Sandles}, {Scholtz}, \& {Sun}}]{cameron23}
{Cameron}, A.~J., {Saxena}, A., {Bunker}, A.~J., {et~al.} 2023{\natexlab{b}}, arXiv e-prints, arXiv:2302.04298, \dodoi{10.48550/arXiv.2302.04298}

\bibitem[{{Caminha} {et~al.}(2019){Caminha}, {Rosati}, {Grillo}, {Rosani}, {Caputi}, {Meneghetti}, {Mercurio}, {Balestra}, {Bergamini}, {Biviano}, {Nonino}, {Umetsu}, {Vanzella}, {Annunziatella}, {Broadhurst}, {Delgado-Correal}, {Demarco}, {Koekemoer}, {Lombardi}, {Maier}, {Verdugo}, \& {Zitrin}}]{caminha19}
{Caminha}, G.~B., {Rosati}, P., {Grillo}, C., {et~al.} 2019, \aap, 632, A36, \dodoi{10.1051/0004-6361/201935454}

\bibitem[{{Cardelli} {et~al.}(1989){Cardelli}, {Clayton}, \& {Mathis}}]{cardelli89}
{Cardelli}, J.~A., {Clayton}, G.~C., \& {Mathis}, J.~S. 1989, \apj, 345, 245, \dodoi{10.1086/167900}

\bibitem[{{Carniani} {et~al.}(2024){Carniani}, {Hainline}, {D'Eugenio}, {Eisenstein}, {Jakobsen}, {Witstok}, {Johnson}, {Chevallard}, {Maiolino}, {Helton}, {Willott}, {Robertson}, {Alberts}, {Arribas}, {Baker}, {Bhatawdekar}, {Boyett}, {Bunker}, {Cameron}, {Cargile}, {Charlot}, {Curti}, {Curtis-Lake}, {Egami}, {Giardino}, {Isaak}, {Ji}, {Jones}, {Maseda}, {Parlanti}, {Rawle}, {Rieke}, {Rieke}, {Rodr{\'\i}guez Del Pino}, {Saxena}, {Scholtz}, {Smit}, {Sun}, {Tacchella}, {{\"U}bler}, {Venturi}, {Williams}, \& {Willmer}}]{carniani24}
{Carniani}, S., {Hainline}, K., {D'Eugenio}, F., {et~al.} 2024, arXiv e-prints, arXiv:2405.18485, \dodoi{10.48550/arXiv.2405.18485}

\bibitem[{{Casey} {et~al.}(2023){Casey}, {Akins}, {Shuntov}, {Ilbert}, {Paquereau}, {Franco}, {Hayward}, {Finkelstein}, {Boylan-Kolchin}, {Robertson}, {Allen}, {Brinch}, {Cooper}, {Ding}, {Drakos}, {Faisst}, {Fujimoto}, {Gillman}, {Harish}, {Hirschmann}, {Jin}, {Kartaltepe}, {Koekemoer}, {Kokorev}, {Liu}, {Long}, {Magdis}, {Maraston}, {Martin}, {McCracken}, {McKinney}, {Mobasher}, {Rhodes}, {Rich}, {Sanders}, {Silverman}, {Toft}, {Vijayan}, {Weaver}, {Wilkins}, {Yang}, \& {Zavala}}]{casey23}
{Casey}, C.~M., {Akins}, H.~B., {Shuntov}, M., {et~al.} 2023, arXiv e-prints, arXiv:2308.10932, \dodoi{10.48550/arXiv.2308.10932}

\bibitem[{{Castellano} {et~al.}(2017){Castellano}, {Pentericci}, {Fontana}, {Vanzella}, {Merlin}, {De Barros}, {Amorin}, {Caputi}, {Cristiani}, {Finkelstein}, {Giallongo}, {Grazian}, {Koekemoer}, {Maiolino}, {Paris}, {Pilo}, {Santini}, \& {Yan}}]{castellano17}
{Castellano}, M., {Pentericci}, L., {Fontana}, A., {et~al.} 2017, \apj, 839, 73, \dodoi{10.3847/1538-4357/aa696e}

\bibitem[{{Castellano} {et~al.}(2022){Castellano}, {Fontana}, {Treu}, {Santini}, {Merlin}, {Leethochawalit}, {Trenti}, {Vanzella}, {Mestric}, {Bonchi}, {Belfiori}, {Nonino}, {Paris}, {Polenta}, {Roberts-Borsani}, {Boyett}, {Brada{\v{c}}}, {Calabr{\`o}}, {Glazebrook}, {Grillo}, {Mascia}, {Mason}, {Mercurio}, {Morishita}, {Nanayakkara}, {Pentericci}, {Rosati}, {Vulcani}, {Wang}, \& {Yang}}]{castellano22}
{Castellano}, M., {Fontana}, A., {Treu}, T., {et~al.} 2022, \apjl, 938, L15, \dodoi{10.3847/2041-8213/ac94d0}

\bibitem[{{Castellano} {et~al.}(2024){Castellano}, {Napolitano}, {Fontana}, {Roberts-Borsani}, {Treu}, {Vanzella}, {Zavala}, {Arrabal Haro}, {Calabr{\`o}}, {Llerena}, {Mascia}, {Merlin}, {Paris}, {Pentericci}, {Santini}, {Bakx}, {Bergamini}, {Cupani}, {Dickinson}, {Filippenko}, {Glazebrook}, {Grillo}, {Kelly}, {Malkan}, {Mason}, {Morishita}, {Nanayakkara}, {Rosati}, {Sani}, {Wang}, \& {Yoon}}]{castellano24}
{Castellano}, M., {Napolitano}, L., {Fontana}, A., {et~al.} 2024, arXiv e-prints, arXiv:2403.10238, \dodoi{10.48550/arXiv.2403.10238}

\bibitem[{{Chabrier}(2003)}]{chabrier03}
{Chabrier}, G. 2003, \pasp, 115, 763, \dodoi{10.1086/376392}

\bibitem[{{Coe} {et~al.}(2019){Coe}, {Salmon}, {Brada{\v{c}}}, {Bradley}, {Sharon}, {Zitrin}, {Acebron}, {Cerny}, {Cibirka}, {Strait}, {Paterno-Mahler}, {Mahler}, {Avila}, {Ogaz}, {Huang}, {Pelliccia}, {Stark}, {Mainali}, {Oesch}, {Trenti}, {Carrasco}, {Dawson}, {Rodney}, {Strolger}, {Riess}, {Jones}, {Frye}, {Czakon}, {Umetsu}, {Vulcani}, {Graur}, {Jha}, {Graham}, {Molino}, {Nonino}, {Hjorth}, {Selsing}, {Christensen}, {Kikuchihara}, {Ouchi}, {Oguri}, {Welch}, {Lemaux}, {Andrade-Santos}, {Hoag}, {Johnson}, {Peterson}, {Past}, {Fox}, {Agulli}, {Livermore}, {Ryan}, {Lam}, {Sendra-Server}, {Toft}, {Lovisari}, \& {Su}}]{coe19}
{Coe}, D., {Salmon}, B., {Brada{\v{c}}}, M., {et~al.} 2019, \apj, 884, 85, \dodoi{10.3847/1538-4357/ab412b}

\bibitem[{{Conroy} {et~al.}(2009){Conroy}, {Gunn}, \& {White}}]{conroy09}
{Conroy}, C., {Gunn}, J.~E., \& {White}, M. 2009, \apj, 699, 486, \dodoi{10.1088/0004-637X/699/1/486}

\bibitem[{{Cullen} {et~al.}(2020){Cullen}, {McLure}, {Dunlop}, {Carnall}, {McLeod}, {Shapley}, {Amor{\'\i}n}, {Bolzonella}, {Castellano}, {Cimatti}, {Cirasuolo}, {Cucciati}, {Fontana}, {Fontanot}, {Garilli}, {Guaita}, {Jarvis}, {Pentericci}, {Pozzetti}, {Talia}, {Zamorani}, {Calabr{\`o}}, {Cresci}, {Fynbo}, {Hathi}, {Giavalisco}, {Koekemoer}, {Mannucci}, \& {Saxena}}]{cullen20}
{Cullen}, F., {McLure}, R.~J., {Dunlop}, J.~S., {et~al.} 2020, \mnras, 495, 1501, \dodoi{10.1093/mnras/staa1260}

\bibitem[{{Cullen} {et~al.}(2023){Cullen}, {McLure}, {McLeod}, {Dunlop}, {Donnan}, {Carnall}, {Bowler}, {Begley}, {Hamadouche}, \& {Stanton}}]{cullen23}
{Cullen}, F., {McLure}, R.~J., {McLeod}, D.~J., {et~al.} 2023, \mnras, 520, 14, \dodoi{10.1093/mnras/stad073}

\bibitem[{{Curti} {et~al.}(2023){Curti}, {Maiolino}, {Curtis-Lake}, {Chevallard}, {Carniani}, {D'Eugenio}, {Looser}, {Scholtz}, {Charlot}, {Cameron}, {{\"U}bler}, {Witstok}, {Boyett}, {Laseter}, {Sandles}, {Arribas}, {Bunker}, {Giardino}, {Maseda}, {Rawle}, {Rodr{\'\i}guez Del Pino}, {Smit}, {Willott}, {Eisenstein}, {Hausen}, {Johnson}, {Rieke}, {Robertson}, {Tacchella}, {Williams}, {Willmer}, {Baker}, {Bhatawdekar}, {Egami}, {Helton}, {Ji}, {Kumari}, {Perna}, {Shivaei}, \& {Sun}}]{curti23}
{Curti}, M., {Maiolino}, R., {Curtis-Lake}, E., {et~al.} 2023, arXiv e-prints, arXiv:2304.08516, \dodoi{10.48550/arXiv.2304.08516}

\bibitem[{{Curtis-Lake} {et~al.}(2023){Curtis-Lake}, {Carniani}, {Cameron}, {Charlot}, {Jakobsen}, {Maiolino}, {Bunker}, {Witstok}, {Smit}, {Chevallard}, {Willott}, {Ferruit}, {Arribas}, {Bonaventura}, {Curti}, {D'Eugenio}, {Franx}, {Giardino}, {Looser}, {L{\"u}tzgendorf}, {Maseda}, {Rawle}, {Rix}, {Rodr{\'\i}guez del Pino}, {{\"U}bler}, {Sirianni}, {Dressler}, {Egami}, {Eisenstein}, {Endsley}, {Hainline}, {Hausen}, {Johnson}, {Rieke}, {Robertson}, {Shivaei}, {Stark}, {Tacchella}, {Williams}, {Willmer}, {Bhatawdekar}, {Bowler}, {Boyett}, {Chen}, {de Graaff}, {Helton}, {Hviding}, {Jones}, {Kumari}, {Lyu}, {Nelson}, {Perna}, {Sandles}, {Saxena}, {Suess}, {Sun}, {Topping}, {Wallace}, \& {Whitler}}]{curtislake23}
{Curtis-Lake}, E., {Carniani}, S., {Cameron}, A., {et~al.} 2023, Nature Astronomy, 7, 622, \dodoi{10.1038/s41550-023-01918-w}

\bibitem[{{De Barros} {et~al.}(2017){De Barros}, {Pentericci}, {Vanzella}, {Castellano}, {Fontana}, {Grazian}, {Conselice}, {Yan}, {Koekemoer}, {Cristiani}, {Dickinson}, {Finkelstein}, \& {Maiolino}}]{debarros17}
{De Barros}, S., {Pentericci}, L., {Vanzella}, E., {et~al.} 2017, \aap, 608, A123, \dodoi{10.1051/0004-6361/201731476}

\bibitem[{{Desprez} {et~al.}(2023){Desprez}, {Martis}, {Asada}, {Sawicki}, {Willott}, {Muzzin}, {Abraham}, {Brada{\v{c}}}, {Brammer}, {Estrada-Carpenter}, {Iyer}, {Matharu}, {Mowla}, {Noirot}, {Sarrouh}, {Strait}, {Gledhill}, \& {Rihtar{\v{s}}i{\v{c}}}}]{desprez23}
{Desprez}, G., {Martis}, N.~S., {Asada}, Y., {et~al.} 2023, arXiv e-prints, arXiv:2310.03063, \dodoi{10.48550/arXiv.2310.03063}

\bibitem[{{D'Eugenio} {et~al.}(2023){D'Eugenio}, {Maiolino}, {Carniani}, {Curtis-Lake}, {Witstok}, {Chevallard}, {Charlot}, {Baker}, {Arribas}, {Boyett}, {Bunker}, {Curti}, {Eisenstein}, {Hainline}, {Ji}, {Johnson}, {Looser}, {Nakajima}, {Nelson}, {Rieke}, {Robertson}, {Scholtz}, {Smit}, {Venturi}, {Tacchella}, {Uebler}, {Willmer}, \& {Willott}}]{deugenio23}
{D'Eugenio}, F., {Maiolino}, R., {Carniani}, S., {et~al.} 2023, arXiv e-prints, arXiv:2311.09908, \dodoi{10.48550/arXiv.2311.09908}

\bibitem[{{Dijkstra}(2014)}]{dijkstra14}
{Dijkstra}, M. 2014, \pasa, 31, e040, \dodoi{10.1017/pasa.2014.33}

\bibitem[{{Du} {et~al.}(2017){Du}, {Shapley}, {Martin}, \& {Coil}}]{du17}
{Du}, X., {Shapley}, A.~E., {Martin}, C.~L., \& {Coil}, A.~L. 2017, \apj, 838, 63, \dodoi{10.3847/1538-4357/aa64cf}

\bibitem[{{Du} {et~al.}(2020){Du}, {Shapley}, {Tang}, {Stark}, {Martin}, {Mobasher}, {Topping}, \& {Chevallard}}]{du20}
{Du}, X., {Shapley}, A.~E., {Tang}, M., {et~al.} 2020, \apj, 890, 65, \dodoi{10.3847/1538-4357/ab67b8}

\bibitem[{{Dunlop} {et~al.}(2013){Dunlop}, {Rogers}, {McLure}, {Ellis}, {Robertson}, {Koekemoer}, {Dayal}, {Curtis-Lake}, {Wild}, {Charlot}, {Bowler}, {Schenker}, {Ouchi}, {Ono}, {Cirasuolo}, {Furlanetto}, {Stark}, {Targett}, \& {Schneider}}]{dunlop13}
{Dunlop}, J.~S., {Rogers}, A.~B., {McLure}, R.~J., {et~al.} 2013, \mnras, 432, 3520, \dodoi{10.1093/mnras/stt702}

\bibitem[{{Eilers} {et~al.}(2018){Eilers}, {Davies}, \& {Hennawi}}]{eilers18}
{Eilers}, A.-C., {Davies}, F.~B., \& {Hennawi}, J.~F. 2018, \apj, 864, 53, \dodoi{10.3847/1538-4357/aad4fd}

\bibitem[{{Eisenstein} {et~al.}(2023){Eisenstein}, {Johnson}, {Robertson}, {Tacchella}, {Hainline}, {Jakobsen}, {Maiolino}, {Bonaventura}, {Bunker}, {Cameron}, {Cargile}, {Curtis-Lake}, {Hausen}, {Pusk{\'a}s}, {Rieke}, {Sun}, {Willmer}, {Willott}, {Alberts}, {Arribas}, {Baker}, {Baum}, {Bhatawdekar}, {Carniani}, {Charlot}, {Chen}, {Chevallard}, {Curti}, {DeCoursey}, {D'Eugenio}, {de Graaff}, {Egami}, {Helton}, {Ji}, {Jones}, {Kumari}, {L{\"u}tzgendorf}, {Laseter}, {Looser}, {Lyu}, {Maseda}, {Nelson}, {Parlanti}, {Rauscher}, {Rawle}, {Rieke}, {Rix}, {Rujopakarn}, {Sandles}, {Saxena}, {Scholtz}, {Sharpe}, {Shivaei}, {Simmonds}, {Smit}, {Topping}, {{\"U}bler}, {Venturi}, {Williams}, {Witstok}, \& {Woodrum}}]{eisenstein23}
{Eisenstein}, D.~J., {Johnson}, B.~D., {Robertson}, B., {et~al.} 2023, arXiv e-prints, arXiv:2310.12340, \dodoi{10.48550/arXiv.2310.12340}

\bibitem[{{Endsley} \& {Stark}(2022)}]{endsley22_cosmos}
{Endsley}, R., \& {Stark}, D.~P. 2022, \mnras, 511, 6042, \dodoi{10.1093/mnras/stac524}

\bibitem[{{Endsley} {et~al.}(2021{\natexlab{a}}){Endsley}, {Stark}, {Charlot}, {Chevallard}, {Robertson}, {Bouwens}, \& {Stefanon}}]{endsley21}
{Endsley}, R., {Stark}, D.~P., {Charlot}, S., {et~al.} 2021{\natexlab{a}}, \mnras, 502, 6044, \dodoi{10.1093/mnras/stab432}

\bibitem[{{Endsley} {et~al.}(2021{\natexlab{b}}){Endsley}, {Stark}, {Chevallard}, \& {Charlot}}]{endsley21b}
{Endsley}, R., {Stark}, D.~P., {Chevallard}, J., \& {Charlot}, S. 2021{\natexlab{b}}, \mnras, 500, 5229, \dodoi{10.1093/mnras/staa3370}

\bibitem[{{Endsley} {et~al.}(2023){Endsley}, {Stark}, {Whitler}, {Topping}, {Chen}, {Plat}, {Chisholm}, \& {Charlot}}]{endsley23}
{Endsley}, R., {Stark}, D.~P., {Whitler}, L., {et~al.} 2023, \mnras, 524, 2312, \dodoi{10.1093/mnras/stad1919}

\bibitem[{{Erb} {et~al.}(2010){Erb}, {Pettini}, {Shapley}, {Steidel}, {Law}, \& {Reddy}}]{erb10}
{Erb}, D.~K., {Pettini}, M., {Shapley}, A.~E., {et~al.} 2010, \apj, 719, 1168, \dodoi{10.1088/0004-637X/719/2/1168}

\bibitem[{{Feltre} {et~al.}(2016){Feltre}, {Charlot}, \& {Gutkin}}]{feltre16}
{Feltre}, A., {Charlot}, S., \& {Gutkin}, J. 2016, \mnras, 456, 3354, \dodoi{10.1093/mnras/stv2794}

\bibitem[{{Ferrara} {et~al.}(2023){Ferrara}, {Pallottini}, \& {Dayal}}]{ferrara23}
{Ferrara}, A., {Pallottini}, A., \& {Dayal}, P. 2023, \mnras, 522, 3986, \dodoi{10.1093/mnras/stad1095}

\bibitem[{{Ferrara} {et~al.}(2022){Ferrara}, {Sommovigo}, {Dayal}, {Pallottini}, {Bouwens}, {Gonzalez}, {Inami}, {Smit}, {Bowler}, {Endsley}, {Oesch}, {Schouws}, {Stark}, {Stefanon}, {Aravena}, {da Cunha}, {De Looze}, {Fudamoto}, {Graziani}, {Hodge}, {Riechers}, {Schneider}, {Algera}, {Barrufet}, {Hygate}, {Labb{\'e}}, {Li}, {Nanayakkara}, {Topping}, \& {van der Werf}}]{ferrara22}
{Ferrara}, A., {Sommovigo}, L., {Dayal}, P., {et~al.} 2022, \mnras, 512, 58, \dodoi{10.1093/mnras/stac460}

\bibitem[{{Finkelstein} {et~al.}(2012){Finkelstein}, {Papovich}, {Salmon}, {Finlator}, {Dickinson}, {Ferguson}, {Giavalisco}, {Koekemoer}, {Reddy}, {Bassett}, {Conselice}, {Dunlop}, {Faber}, {Grogin}, {Hathi}, {Kocevski}, {Lai}, {Lee}, {McLure}, {Mobasher}, \& {Newman}}]{finkelstein12}
{Finkelstein}, S.~L., {Papovich}, C., {Salmon}, B., {et~al.} 2012, \apj, 756, 164, \dodoi{10.1088/0004-637X/756/2/164}

\bibitem[{{Finkelstein} {et~al.}(2023){Finkelstein}, {Bagley}, {Ferguson}, {Wilkins}, {Kartaltepe}, {Papovich}, {Yung}, {Haro}, {Behroozi}, {Dickinson}, {Kocevski}, {Koekemoer}, {Larson}, {Le Bail}, {Morales}, {P{\'e}rez-Gonz{\'a}lez}, {Burgarella}, {Dav{\'e}}, {Hirschmann}, {Somerville}, {Wuyts}, {Bromm}, {Casey}, {Fontana}, {Fujimoto}, {Gardner}, {Giavalisco}, {Grazian}, {Grogin}, {Hathi}, {Hutchison}, {Jha}, {Jogee}, {Kewley}, {Kirkpatrick}, {Long}, {Lotz}, {Pentericci}, {Pierel}, {Pirzkal}, {Ravindranath}, {Ryan}, {Trump}, {Yang}, {Bhatawdekar}, {Bisigello}, {Buat}, {Calabr{\`o}}, {Castellano}, {Cleri}, {Cooper}, {Croton}, {Daddi}, {Dekel}, {Elbaz}, {Franco}, {Gawiser}, {Holwerda}, {Huertas-Company}, {Jaskot}, {Leung}, {Lucas}, {Mobasher}, {Pandya}, {Tacchella}, {Weiner}, \& {Zavala}}]{finkelstein23}
{Finkelstein}, S.~L., {Bagley}, M.~B., {Ferguson}, H.~C., {et~al.} 2023, \apjl, 946, L13, \dodoi{10.3847/2041-8213/acade4}

\bibitem[{{Flury} {et~al.}(2022){Flury}, {Jaskot}, {Ferguson}, {Worseck}, {Makan}, {Chisholm}, {Saldana-Lopez}, {Schaerer}, {McCandliss}, {Wang}, {Ford}, {Heckman}, {Ji}, {Giavalisco}, {Amorin}, {Atek}, {Blaizot}, {Borthakur}, {Carr}, {Castellano}, {Cristiani}, {De Barros}, {Dickinson}, {Finkelstein}, {Fleming}, {Fontanot}, {Garel}, {Grazian}, {Hayes}, {Henry}, {Mauerhofer}, {Micheva}, {Oey}, {Ostlin}, {Papovich}, {Pentericci}, {Ravindranath}, {Rosdahl}, {Rutkowski}, {Santini}, {Scarlata}, {Teplitz}, {Thuan}, {Trebitsch}, {Vanzella}, {Verhamme}, \& {Xu}}]{flury22}
{Flury}, S.~R., {Jaskot}, A.~E., {Ferguson}, H.~C., {et~al.} 2022, \apjs, 260, 1, \dodoi{10.3847/1538-4365/ac5331}

\bibitem[{{Foreman-Mackey} {et~al.}(2013){Foreman-Mackey}, {Hogg}, {Lang}, \& {Goodman}}]{emcee}
{Foreman-Mackey}, D., {Hogg}, D.~W., {Lang}, D., \& {Goodman}, J. 2013, \pasp, 125, 306, \dodoi{10.1086/670067}

\bibitem[{{Fujimoto} {et~al.}(2023){Fujimoto}, {Arrabal Haro}, {Dickinson}, {Finkelstein}, {Kartaltepe}, {Larson}, {Burgarella}, {Bagley}, {Behroozi}, {Chworowsky}, {Hirschmann}, {Trump}, {Wilkins}, {Yung}, {Koekemoer}, {Papovich}, {Pirzkal}, {Ferguson}, {Fontana}, {Grogin}, {Grazian}, {Kewley}, {Kocevski}, {Lotz}, {Pentericci}, {Ravindranath}, {Somerville}, {Amorin}, {Backhaus}, {Calabro}, {Casey}, {Cooper}, {Franco}, {Giavalisco}, {Hathi}, {Harish}, {Hutchison}, {Iyer}, {Jung}, {Lucas}, \& {Zavala}}]{fujimoto23}
{Fujimoto}, S., {Arrabal Haro}, P., {Dickinson}, M., {et~al.} 2023, arXiv e-prints, arXiv:2301.09482, \dodoi{10.48550/arXiv.2301.09482}

\bibitem[{{Gordon} {et~al.}(2003){Gordon}, {Clayton}, {Misselt}, {Landolt}, \& {Wolff}}]{gordon03}
{Gordon}, K.~D., {Clayton}, G.~C., {Misselt}, K.~A., {Landolt}, A.~U., \& {Wolff}, M.~J. 2003, \apj, 594, 279, \dodoi{10.1086/376774}

\bibitem[{{Goulding} {et~al.}(2023){Goulding}, {Greene}, {Setton}, {Labbe}, {Bezanson}, {Miller}, {Atek}, {Bogdan}, {Brammer}, {Chemerynska}, {Cutler}, {Dayal}, {Fudamoto}, {Fujimoto}, {Furtak}, {Kokorev}, {Khullar}, {Leja}, {Marchesini}, {Natarajan}, {Nelson}, {Oesch}, {Pan}, {Papovich}, {Price}, {van Dokkum}, {Wang}, {Weaver}, {Whitaker}, \& {Zitrin}}]{goulding23}
{Goulding}, A.~D., {Greene}, J.~E., {Setton}, D.~J., {et~al.} 2023, arXiv e-prints, arXiv:2308.02750.
\newblock \doarXiv{2308.02750}

\bibitem[{{Greene} {et~al.}(2023){Greene}, {Labbe}, {Goulding}, {Furtak}, {Chemerynska}, {Kokorev}, {Dayal}, {Williams}, {Wang}, {Setton}, {Burgasser}, {Bezanson}, {Atek}, {Brammer}, {Cutler}, {Feldmann}, {Fujimoto}, {Glazebrook}, {de Graaff}, {Leja}, {Marchesini}, {Maseda}, {Matthee}, {Miller}, {Naidu}, {Nanayakkara}, {Oesch}, {Pan}, {Papovich}, {Price}, {van Dokkum}, {Weaver}, {Whitaker}, \& {Zitrin}}]{greene23}
{Greene}, J.~E., {Labbe}, I., {Goulding}, A.~D., {et~al.} 2023, arXiv e-prints, arXiv:2309.05714, \dodoi{10.48550/arXiv.2309.05714}

\bibitem[{{Grogin} {et~al.}(2011){Grogin}, {Kocevski}, {Faber}, {Ferguson}, {Koekemoer}, {Riess}, {Acquaviva}, {Alexander}, {Almaini}, {Ashby}, {Barden}, {Bell}, {Bournaud}, {Brown}, {Caputi}, {Casertano}, {Cassata}, {Castellano}, {Challis}, {Chary}, {Cheung}, {Cirasuolo}, {Conselice}, {Roshan Cooray}, {Croton}, {Daddi}, {Dahlen}, {Dav{\'e}}, {de Mello}, {Dekel}, {Dickinson}, {Dolch}, {Donley}, {Dunlop}, {Dutton}, {Elbaz}, {Fazio}, {Filippenko}, {Finkelstein}, {Fontana}, {Gardner}, {Garnavich}, {Gawiser}, {Giavalisco}, {Grazian}, {Guo}, {Hathi}, {H{\"a}ussler}, {Hopkins}, {Huang}, {Huang}, {Jha}, {Kartaltepe}, {Kirshner}, {Koo}, {Lai}, {Lee}, {Li}, {Lotz}, {Lucas}, {Madau}, {McCarthy}, {McGrath}, {McIntosh}, {McLure}, {Mobasher}, {Moustakas}, {Mozena}, {Nandra}, {Newman}, {Niemi}, {Noeske}, {Papovich}, {Pentericci}, {Pope}, {Primack}, {Rajan}, {Ravindranath}, {Reddy}, {Renzini}, {Rix}, {Robaina}, {Rodney}, {Rosario}, {Rosati}, {Salimbeni}, {Scarlata}, {Siana}, {Simard}, {Smidt}, {Somerville}, {Spinrad},
  {Straughn}, {Strolger}, {Telford}, {Teplitz}, {Trump}, {van der Wel}, {Villforth}, {Wechsler}, {Weiner}, {Wiklind}, {Wild}, {Wilson}, {Wuyts}, {Yan}, \& {Yun}}]{grogin11}
{Grogin}, N.~A., {Kocevski}, D.~D., {Faber}, S.~M., {et~al.} 2011, \apjs, 197, 35, \dodoi{10.1088/0067-0049/197/2/35}

\bibitem[{{Gutkin} {et~al.}(2016){Gutkin}, {Charlot}, \& {Bruzual}}]{gutkin16}
{Gutkin}, J., {Charlot}, S., \& {Bruzual}, G. 2016, \mnras, 462, 1757, \dodoi{10.1093/mnras/stw1716}

\bibitem[{{Harikane} {et~al.}(2023{\natexlab{a}}){Harikane}, {Nakajima}, {Ouchi}, {Umeda}, {Isobe}, {Ono}, {Xu}, \& {Zhang}}]{harikane23}
{Harikane}, Y., {Nakajima}, K., {Ouchi}, M., {et~al.} 2023{\natexlab{a}}, arXiv e-prints, arXiv:2304.06658, \dodoi{10.48550/arXiv.2304.06658}

\bibitem[{{Harikane} {et~al.}(2023{\natexlab{b}}){Harikane}, {Zhang}, {Nakajima}, {Ouchi}, {Isobe}, {Ono}, {Hatano}, {Xu}, \& {Umeda}}]{harikane23b}
{Harikane}, Y., {Zhang}, Y., {Nakajima}, K., {et~al.} 2023{\natexlab{b}}, arXiv e-prints, arXiv:2303.11946, \dodoi{10.48550/arXiv.2303.11946}

\bibitem[{{Hashimoto} {et~al.}(2018){Hashimoto}, {Laporte}, {Mawatari}, {Ellis}, {Inoue}, {Zackrisson}, {Roberts-Borsani}, {Zheng}, {Tamura}, {Bauer}, {Fletcher}, {Harikane}, {Hatsukade}, {Hayatsu}, {Matsuda}, {Matsuo}, {Okamoto}, {Ouchi}, {Pell{\'o}}, {Rydberg}, {Shimizu}, {Taniguchi}, {Umehata}, \& {Yoshida}}]{hashimoto18}
{Hashimoto}, T., {Laporte}, N., {Mawatari}, K., {et~al.} 2018, \nat, 557, 392, \dodoi{10.1038/s41586-018-0117-z}

\bibitem[{{Heintz} {et~al.}(2023){Heintz}, {Watson}, {Brammer}, {Vejlgaard}, {Hutter}, {Strait}, {Matthee}, {Oesch}, {Jakobsson}, {Tanvir}, {Laursen}, {Naidu}, {Mason}, {Killi}, {Jung}, {Hsiao}, {Abdurro'uf}, {Coe}, {Arrabal Haro}, {Finkelstein}, \& {Toft}}]{heintz23}
{Heintz}, K.~E., {Watson}, D., {Brammer}, G., {et~al.} 2023, arXiv e-prints, arXiv:2306.00647, \dodoi{10.48550/arXiv.2306.00647}

\bibitem[{{Horne}(1986)}]{horne86}
{Horne}, K. 1986, \pasp, 98, 609, \dodoi{10.1086/131801}

\bibitem[{{Hsiao} {et~al.}(2023){Hsiao}, {Abdurro'uf}, {Coe}, {Larson}, {Jung}, {Mingozzi}, {Dayal}, {Kumari}, {Kokorev}, {Vikaeus}, {Brammer}, {Furtak}, {Adamo}, {Andrade-Santos}, {Antwi-Danso}, {Bradac}, {Bradley}, {Broadhurst}, {Carnall}, {Conselice}, {Diego}, {Donahue}, {Eldridge}, {Fujimoto}, {Henry}, {Hernandez}, {Hutchison}, {James}, {Norman}, {Park}, {Pirzkal}, {Postman}, {Ricotti}, {Rigby}, {Vanzella}, {Welch}, {Wilkins}, {Windhorst}, {Xu}, {Zackrisson}, \& {Zitrin}}]{hsiao23}
{Hsiao}, T. Y.-Y., {Abdurro'uf}, {Coe}, D., {et~al.} 2023, arXiv e-prints, arXiv:2305.03042, \dodoi{10.48550/arXiv.2305.03042}

\bibitem[{{Hu} {et~al.}(2024){Hu}, {Papovich}, {Dickinson}, {Kennicutt}, {Shen}, {Amor{\'\i}n}, {Arrabal Haro}, {Bagley}, {Bhatawdekar}, {Cleri}, {Cole}, {Dekel}, {de la Vega}, {Finkelstein}, {Grogin}, {Hathi}, {Hirschmann}, {Holwerda}, {Hutchison}, {Jung}, {Koekemoer}, {Kartaltepe}, {Lucas}, {Llerena}, {Mascia}, {Mobasher}, {Napolitano}, {Newman}, {Pentericci}, {P{\'e}rez-Gonz{\'a}lez}, {Trump}, {Wilkins}, \& {Yung}}]{hu24}
{Hu}, W., {Papovich}, C., {Dickinson}, M., {et~al.} 2024, \apj, 971, 21, \dodoi{10.3847/1538-4357/ad5015}

\bibitem[{{Inoue}(2011)}]{inoue11}
{Inoue}, A.~K. 2011, \mnras, 415, 2920, \dodoi{10.1111/j.1365-2966.2011.18906.x}

\bibitem[{{Izotov} {et~al.}(2021){Izotov}, {Thuan}, \& {Guseva}}]{izotov21}
{Izotov}, Y.~I., {Thuan}, T.~X., \& {Guseva}, N.~G. 2021, \mnras, 504, 3996, \dodoi{10.1093/mnras/stab1099}

\bibitem[{{Jaskot} \& {Ravindranath}(2016)}]{jaskot16}
{Jaskot}, A.~E., \& {Ravindranath}, S. 2016, \apj, 833, 136, \dodoi{10.3847/1538-4357/833/2/136}

\bibitem[{{Jung} {et~al.}(2018){Jung}, {Finkelstein}, {Livermore}, {Dickinson}, {Larson}, {Papovich}, {Song}, {Tilvi}, \& {Wold}}]{jung18}
{Jung}, I., {Finkelstein}, S.~L., {Livermore}, R.~C., {et~al.} 2018, \apj, 864, 103, \dodoi{10.3847/1538-4357/aad686}

\bibitem[{{Katz} {et~al.}(2023){Katz}, {Rosdahl}, {Kimm}, {Blaizot}, {Choustikov}, {Farcy}, {Garel}, {Haehnelt}, {Michel-Dansac}, \& {Ocvirk}}]{katz23}
{Katz}, H., {Rosdahl}, J., {Kimm}, T., {et~al.} 2023, The Open Journal of Astrophysics, 6, 44, \dodoi{10.21105/astro.2309.03269}

\bibitem[{{Kauffmann} {et~al.}(2003){Kauffmann}, {Heckman}, {White}, {Charlot}, {Tremonti}, {Brinchmann}, {Bruzual}, {Peng}, {Seibert}, {Bernardi}, {Blanton}, {Brinkmann}, {Castander}, {Cs{\'a}bai}, {Fukugita}, {Ivezic}, {Munn}, {Nichol}, {Padmanabhan}, {Thakar}, {Weinberg}, \& {York}}]{kauffmann03}
{Kauffmann}, G., {Heckman}, T.~M., {White}, S. D.~M., {et~al.} 2003, \mnras, 341, 33, \dodoi{10.1046/j.1365-8711.2003.06291.x}

\bibitem[{{Kennicutt}(1998)}]{kennicutt98}
{Kennicutt}, Robert~C., J. 1998, \araa, 36, 189, \dodoi{10.1146/annurev.astro.36.1.189}

\bibitem[{{Kewley} {et~al.}(2019){Kewley}, {Nicholls}, \& {Sutherland}}]{kewley19}
{Kewley}, L.~J., {Nicholls}, D.~C., \& {Sutherland}, R.~S. 2019, \araa, 57, 511, \dodoi{10.1146/annurev-astro-081817-051832}

\bibitem[{{Kocevski} {et~al.}(2023){Kocevski}, {Onoue}, {Inayoshi}, {Trump}, {Arrabal Haro}, {Grazian}, {Dickinson}, {Finkelstein}, {Kartaltepe}, {Hirschmann}, {Aird}, {Holwerda}, {Fujimoto}, {Juneau}, {Amor{\'\i}n}, {Backhaus}, {Bagley}, {Barro}, {Bell}, {Bisigello}, {Calabr{\`o}}, {Cleri}, {Cooper}, {Ding}, {Grogin}, {Ho}, {Hutchison}, {Inoue}, {Jiang}, {Jones}, {Koekemoer}, {Li}, {Li}, {McGrath}, {Molina}, {Papovich}, {P{\'e}rez-Gonz{\'a}lez}, {Pirzkal}, {Wilkins}, {Yang}, \& {Yung}}]{kocevski23}
{Kocevski}, D.~D., {Onoue}, M., {Inayoshi}, K., {et~al.} 2023, \apjl, 954, L4, \dodoi{10.3847/2041-8213/ace5a0}

\bibitem[{{Kramarenko} {et~al.}(2024){Kramarenko}, {Kerutt}, {Verhamme}, {Oesch}, {Barrufet}, {Matthee}, {Kusakabe}, {Goovaerts}, \& {Thai}}]{kramarenko24}
{Kramarenko}, I.~G., {Kerutt}, J., {Verhamme}, A., {et~al.} 2024, \mnras, 527, 9853, \dodoi{10.1093/mnras/stad3853}

\bibitem[{{Kriek} {et~al.}(2015){Kriek}, {Shapley}, {Reddy}, {Siana}, {Coil}, {Mobasher}, {Freeman}, {de Groot}, {Price}, {Sanders}, {Shivaei}, {Brammer}, {Momcheva}, {Skelton}, {van Dokkum}, {Whitaker}, {Aird}, {Azadi}, {Kassis}, {Bullock}, {Conroy}, {Dav{\'e}}, {Kere{\v{s}}}, \& {Krumholz}}]{kriek15}
{Kriek}, M., {Shapley}, A.~E., {Reddy}, N.~A., {et~al.} 2015, \apjs, 218, 15, \dodoi{10.1088/0067-0049/218/2/15}

\bibitem[{{Labb{\'e}} {et~al.}(2013){Labb{\'e}}, {Oesch}, {Bouwens}, {Illingworth}, {Magee}, {Gonz{\'a}lez}, {Carollo}, {Franx}, {Trenti}, {van Dokkum}, \& {Stiavelli}}]{labbe13}
{Labb{\'e}}, I., {Oesch}, P.~A., {Bouwens}, R.~J., {et~al.} 2013, \apjl, 777, L19, \dodoi{10.1088/2041-8205/777/2/L19}

\bibitem[{{Laporte} {et~al.}(2023){Laporte}, {Ellis}, {Witten}, \& {Roberts-Borsani}}]{laporte23}
{Laporte}, N., {Ellis}, R.~S., {Witten}, C.~E.~C., \& {Roberts-Borsani}, G. 2023, \mnras, 523, 3018, \dodoi{10.1093/mnras/stad1597}

\bibitem[{{Laporte} {et~al.}(2021){Laporte}, {Meyer}, {Ellis}, {Robertson}, {Chisholm}, \& {Roberts-Borsani}}]{laporte21}
{Laporte}, N., {Meyer}, R.~A., {Ellis}, R.~S., {et~al.} 2021, \mnras, 505, 3336, \dodoi{10.1093/mnras/stab1239}

\bibitem[{{Laporte} {et~al.}(2017{\natexlab{a}}){Laporte}, {Nakajima}, {Ellis}, {Zitrin}, {Stark}, {Mainali}, \& {Roberts-Borsani}}]{laporte17b}
{Laporte}, N., {Nakajima}, K., {Ellis}, R.~S., {et~al.} 2017{\natexlab{a}}, \apj, 851, 40, \dodoi{10.3847/1538-4357/aa96a8}

\bibitem[{{Laporte} {et~al.}(2017{\natexlab{b}}){Laporte}, {Ellis}, {Boone}, {Bauer}, {Qu{\'e}nard}, {Roberts-Borsani}, {Pell{\'o}}, {P{\'e}rez-Fournon}, \& {Streblyanska}}]{laporte17}
{Laporte}, N., {Ellis}, R.~S., {Boone}, F., {et~al.} 2017{\natexlab{b}}, \apjl, 837, L21, \dodoi{10.3847/2041-8213/aa62aa}

\bibitem[{{Larson} {et~al.}(2022){Larson}, {Finkelstein}, {Hutchison}, {Papovich}, {Bagley}, {Dickinson}, {Rojas-Ruiz}, {Ferguson}, {Jung}, {Giavalisco}, {Grazian}, {Pentericci}, \& {Tacchella}}]{larson22}
{Larson}, R.~L., {Finkelstein}, S.~L., {Hutchison}, T.~A., {et~al.} 2022, \apj, 930, 104, \dodoi{10.3847/1538-4357/ac5dbd}

\bibitem[{{Larson} {et~al.}(2023){Larson}, {Finkelstein}, {Kocevski}, {Hutchison}, {Trump}, {Arrabal Haro}, {Bromm}, {Cleri}, {Dickinson}, {Fujimoto}, {Kartaltepe}, {Koekemoer}, {Papovich}, {Pirzkal}, {Tacchella}, {Zavala}, {Bagley}, {Behroozi}, {Champagne}, {Cole}, {Jung}, {Morales}, {Yang}, {Zhang}, {Zitrin}, {Amor{\'\i}n}, {Burgarella}, {Casey}, {Ch{\'a}vez Ortiz}, {Cox}, {Chworowsky}, {Fontana}, {Gawiser}, {Grazian}, {Grogin}, {Harish}, {Hathi}, {Hirschmann}, {Holwerda}, {Juneau}, {Leung}, {Lucas}, {McGrath}, {P{\'e}rez-Gonz{\'a}lez}, {Rigby}, {Seill{\'e}}, {Simons}, {Weiner}, {Wilkins}, {Yung}, \& {The CEERS Team}}]{larson23}
{Larson}, R.~L., {Finkelstein}, S.~L., {Kocevski}, D.~D., {et~al.} 2023, arXiv e-prints, arXiv:2303.08918, \dodoi{10.48550/arXiv.2303.08918}

\bibitem[{{Le F{\`e}vre} {et~al.}(2019){Le F{\`e}vre}, {Lemaux}, {Nakajima}, {Schaerer}, {Talia}, {Zamorani}, {Cassata}, {Garilli}, {Maccagni}, {Pentericci}, {Tasca}, {Zucca}, {Amorin}, {Bardelli}, {Cimatti}, {Giavalisco}, {Guaita}, {Hathi}, {Marchi}, {Vanzella}, {Vergani}, \& {Dunlop}}]{lefevre19}
{Le F{\`e}vre}, O., {Lemaux}, B.~C., {Nakajima}, K., {et~al.} 2019, \aap, 625, A51, \dodoi{10.1051/0004-6361/201732197}

\bibitem[{{Leethochawalit} {et~al.}(2023){Leethochawalit}, {Trenti}, {Santini}, {Yang}, {Merlin}, {Castellano}, {Fontana}, {Treu}, {Mason}, {Glazebrook}, {Jones}, {Vulcani}, {Nanayakkara}, {Marchesini}, {Mascia}, {Morishita}, {Roberts-Borsani}, {Bonchi}, {Paris}, {Boyett}, {Strait}, {Calabr{\`o}}, {Pentericci}, {Bradac}, {Wang}, \& {Scarlata}}]{leethochawalit23a}
{Leethochawalit}, N., {Trenti}, M., {Santini}, P., {et~al.} 2023, \apjl, 942, L26, \dodoi{10.3847/2041-8213/ac959b}

\bibitem[{{Llerena} {et~al.}(2022){Llerena}, {Amor{\'\i}n}, {Cullen}, {Pentericci}, {Calabr{\`o}}, {McLure}, {Carnall}, {P{\'e}rez-Montero}, {Marchi}, {Bongiorno}, {Castellano}, {Fontana}, {McLeod}, {Talia}, {Hathi}, {Hibon}, {Mannucci}, {Saxena}, {Schaerer}, \& {Zamorani}}]{llerena22}
{Llerena}, M., {Amor{\'\i}n}, R., {Cullen}, F., {et~al.} 2022, \aap, 659, A16, \dodoi{10.1051/0004-6361/202141651}

\bibitem[{{Looser} {et~al.}(2023){Looser}, {D'Eugenio}, {Maiolino}, {Witstok}, {Sandles}, {Curtis-Lake}, {Chevallard}, {Tacchella}, {Johnson}, {Baker}, {Suess}, {Carniani}, {Ferruit}, {Arribas}, {Bonaventura}, {Bunker}, {Cameron}, {Charlot}, {Curti}, {de Graaff}, {Maseda}, {Rawle}, {Rix}, {Rodriguez Del Pino}, {Smit}, {{\"U}bler}, {Willott}, {Alberts}, {Egami}, {Eisenstein}, {Endsley}, {Hausen}, {Rieke}, {Robertson}, {Shivaei}, {Williams}, {Boyett}, {Chen}, {Ji}, {Jones}, {Kumari}, {Nelson}, {Perna}, {Saxena}, \& {Scholtz}}]{looser23}
{Looser}, T.~J., {D'Eugenio}, F., {Maiolino}, R., {et~al.} 2023, arXiv e-prints, arXiv:2302.14155, \dodoi{10.48550/arXiv.2302.14155}

\bibitem[{{Lotz} {et~al.}(2017){Lotz}, {Koekemoer}, {Coe}, {Grogin}, {Capak}, {Mack}, {Anderson}, {Avila}, {Barker}, {Borncamp}, {Brammer}, {Durbin}, {Gunning}, {Hilbert}, {Jenkner}, {Khandrika}, {Levay}, {Lucas}, {MacKenty}, {Ogaz}, {Porterfield}, {Reid}, {Robberto}, {Royle}, {Smith}, {Storrie-Lombardi}, {Sunnquist}, {Surace}, {Taylor}, {Williams}, {Bullock}, {Dickinson}, {Finkelstein}, {Natarajan}, {Richard}, {Robertson}, {Tumlinson}, {Zitrin}, {Flanagan}, {Sembach}, {Soifer}, \& {Mountain}}]{lotz17}
{Lotz}, J.~M., {Koekemoer}, A., {Coe}, D., {et~al.} 2017, \apj, 837, 97, \dodoi{10.3847/1538-4357/837/1/97}

\bibitem[{Lovell {et~al.}(2020)Lovell, Vijayan, Thomas, Wilkins, Barnes, Irodotou, \& Roper}]{lovell20}
Lovell, C.~C., Vijayan, A.~P., Thomas, P.~A., {et~al.} 2020, Monthly Notices of the Royal Astronomical Society, 500, 2127, \dodoi{10.1093/mnras/staa3360}

\bibitem[{{Mainali} {et~al.}(2018){Mainali}, {Zitrin}, {Stark}, {Ellis}, {Richard}, {Tang}, {Laporte}, {Oesch}, \& {McGreer}}]{mainali18}
{Mainali}, R., {Zitrin}, A., {Stark}, D.~P., {et~al.} 2018, \mnras, 479, 1180, \dodoi{10.1093/mnras/sty1640}

\bibitem[{{Maiolino} {et~al.}(2023){Maiolino}, {Scholtz}, {Curtis-Lake}, {Carniani}, {Baker}, {de Graaff}, {Tacchella}, {{\"U}bler}, {D'Eugenio}, {Witstok}, {Curti}, {Arribas}, {Bunker}, {Charlot}, {Chevallard}, {Eisenstein}, {Egami}, {Ji}, {Jones}, {Lyu}, {Rawle}, {Robertson}, {Rujopakarn}, {Perna}, {Sun}, {Venturi}, {Williams}, \& {Willott}}]{maiolino23}
{Maiolino}, R., {Scholtz}, J., {Curtis-Lake}, E., {et~al.} 2023, arXiv e-prints, arXiv:2308.01230, \dodoi{10.48550/arXiv.2308.01230}

\bibitem[{{Marchi} {et~al.}(2018){Marchi}, {Pentericci}, {Guaita}, {Schaerer}, {Verhamme}, {Castellano}, {Ribeiro}, {Garilli}, {Le F{\`e}vre}, {Amorin}, {Bardelli}, {Cassata}, {Durkalec}, {Grazian}, {Hathi}, {Lemaux}, {Maccagni}, {Vanzella}, \& {Zucca}}]{marchi18}
{Marchi}, F., {Pentericci}, L., {Guaita}, L., {et~al.} 2018, \aap, 614, A11, \dodoi{10.1051/0004-6361/201732133}

\bibitem[{{Marques-Chaves} {et~al.}(2024){Marques-Chaves}, {Schaerer}, {Kuruvanthodi}, {Korber}, {Prantzos}, {Charbonnel}, {Weibel}, {Izotov}, {Messa}, {Brammer}, {Dessauges-Zavadsky}, \& {Oesch}}]{rui24}
{Marques-Chaves}, R., {Schaerer}, D., {Kuruvanthodi}, A., {et~al.} 2024, \aap, 681, A30, \dodoi{10.1051/0004-6361/202347411}

\bibitem[{{Martin} {et~al.}(2012){Martin}, {Shapley}, {Coil}, {Kornei}, {Bundy}, {Weiner}, {Noeske}, \& {Schiminovich}}]{martin12}
{Martin}, C.~L., {Shapley}, A.~E., {Coil}, A.~L., {et~al.} 2012, \apj, 760, 127, \dodoi{10.1088/0004-637X/760/2/127}

\bibitem[{{Mascia} {et~al.}(2023{\natexlab{a}}){Mascia}, {Pentericci}, {Calabro'}, {Treu}, {Santini}, {Yang}, {Napolitano}, {Roberts-Borsani}, {Bergamini}, {Grillo}, {Rosati}, {Vulcani}, {Castellano}, {Boyett}, {Fontana}, {Glazebrook}, {Henry}, {Mason}, {Merlin}, {Morishita}, {Nanayakkara}, {Paris}, {Roy}, {Williams}, {Wang}, {Brammer}, {Bradac}, {Chen}, {Kelly}, {Koekemoer}, {Trenti}, \& {Windhorst}}]{mascia23}
{Mascia}, S., {Pentericci}, L., {Calabro'}, A., {et~al.} 2023{\natexlab{a}}, arXiv e-prints, arXiv:2301.02816, \dodoi{10.48550/arXiv.2301.02816}

\bibitem[{{Mascia} {et~al.}(2023{\natexlab{b}}){Mascia}, {Pentericci}, {Saxena}, {Belfiori}, {Calabr{\`o}}, {Castellano}, {Saldana-Lopez}, {Talia}, {Amor{\'\i}n}, {Cullen}, {Garilli}, {Guaita}, {LLerena}, {McLure}, {Moresco}, {Santini}, \& {Schaerer}}]{mascia23b}
{Mascia}, S., {Pentericci}, L., {Saxena}, A., {et~al.} 2023{\natexlab{b}}, \aap, 674, A221, \dodoi{10.1051/0004-6361/202245152}

\bibitem[{{Mason} {et~al.}(2019{\natexlab{a}}){Mason}, {Naidu}, {Tacchella}, \& {Leja}}]{Mason2019b}
{Mason}, C.~A., {Naidu}, R.~P., {Tacchella}, S., \& {Leja}, J. 2019{\natexlab{a}}, \mnras, 489, 2669, \dodoi{10.1093/mnras/stz2291}

\bibitem[{{Mason} {et~al.}(2023){Mason}, {Trenti}, \& {Treu}}]{mason23}
{Mason}, C.~A., {Trenti}, M., \& {Treu}, T. 2023, \mnras, 521, 497, \dodoi{10.1093/mnras/stad035}

\bibitem[{{Mason} {et~al.}(2019{\natexlab{b}}){Mason}, {Fontana}, {Treu}, {Schmidt}, {Hoag}, {Abramson}, {Amorin}, {Brada{\v{c}}}, {Guaita}, {Jones}, {Henry}, {Malkan}, {Pentericci}, {Trenti}, \& {Vanzella}}]{mason19}
{Mason}, C.~A., {Fontana}, A., {Treu}, T., {et~al.} 2019{\natexlab{b}}, \mnras, 485, 3947, \dodoi{10.1093/mnras/stz632}

\bibitem[{{Matthee} {et~al.}(2017){Matthee}, {Sobral}, {Best}, {Khostovan}, {Oteo}, {Bouwens}, \& {R{\"o}ttgering}}]{matthee17}
{Matthee}, J., {Sobral}, D., {Best}, P., {et~al.} 2017, \mnras, 465, 3637, \dodoi{10.1093/mnras/stw2973}

\bibitem[{{McLure} {et~al.}(2018){McLure}, {Pentericci}, {Cimatti}, {Dunlop}, {Elbaz}, {Fontana}, {Nandra}, {Amorin}, {Bolzonella}, {Bongiorno}, {Carnall}, {Castellano}, {Cirasuolo}, {Cucciati}, {Cullen}, {De Barros}, {Finkelstein}, {Fontanot}, {Franzetti}, {Fumana}, {Gargiulo}, {Garilli}, {Guaita}, {Hartley}, {Iovino}, {Jarvis}, {Juneau}, {Karman}, {Maccagni}, {Marchi}, {M{\'a}rmol-Queralt{\'o}}, {Pompei}, {Pozzetti}, {Scodeggio}, {Sommariva}, {Talia}, {Almaini}, {Balestra}, {Bardelli}, {Bell}, {Bourne}, {Bowler}, {Brusa}, {Buitrago}, {Caputi}, {Cassata}, {Charlot}, {Citro}, {Cresci}, {Cristiani}, {Curtis-Lake}, {Dickinson}, {Fazio}, {Ferguson}, {Fiore}, {Franco}, {Fynbo}, {Galametz}, {Georgakakis}, {Giavalisco}, {Grazian}, {Hathi}, {Jung}, {Kim}, {Koekemoer}, {Khusanova}, {Le F{\`e}vre}, {Lotz}, {Mannucci}, {Maltby}, {Matsuoka}, {McLeod}, {Mendez-Hernandez}, {Mendez-Abreu}, {Mignoli}, {Moresco}, {Mortlock}, {Nonino}, {Pannella}, {Papovich}, {Popesso}, {Rosario}, {Salvato}, {Santini}, {Schaerer},
  {Schreiber}, {Stark}, {Tasca}, {Thomas}, {Treu}, {Vanzella}, {Wild}, {Williams}, {Zamorani}, \& {Zucca}}]{mclure18}
{McLure}, R.~J., {Pentericci}, L., {Cimatti}, A., {et~al.} 2018, \mnras, 479, 25, \dodoi{10.1093/mnras/sty1213}

\bibitem[{{Meena} {et~al.}(2023){Meena}, {Zitrin}, {Jim{\'e}nez-Teja}, {Zackrisson}, {Chen}, {Coe}, {Diego}, {Dimauro}, {Furtak}, {Kelly}, {Oguri}, {Welch}, {Abdurro'uf}, {Andrade-Santos}, {Adamo}, {Bhatawdekar}, {Brada{\v{c}}}, {Bradley}, {Broadhurst}, {Conselice}, {Dayal}, {Donahue}, {Frye}, {Fujimoto}, {Hsiao}, {Kokorev}, {Mahler}, {Vanzella}, \& {Windhorst}}]{meena23}
{Meena}, A.~K., {Zitrin}, A., {Jim{\'e}nez-Teja}, Y., {et~al.} 2023, \apjl, 944, L6, \dodoi{10.3847/2041-8213/acb645}

\bibitem[{{Meurer} {et~al.}(1999){Meurer}, {Heckman}, \& {Calzetti}}]{meurer99}
{Meurer}, G.~R., {Heckman}, T.~M., \& {Calzetti}, D. 1999, \apj, 521, 64, \dodoi{10.1086/307523}

\bibitem[{{Mingozzi} {et~al.}(2022){Mingozzi}, {James}, {Arellano-C{\'o}rdova}, {Berg}, {Senchyna}, {Chisholm}, {Brinchmann}, {Aloisi}, {Amor{\'\i}n}, {Charlot}, {Feltre}, {Hayes}, {Heckman}, {Henry}, {Hernandez}, {Kumari}, {Leitherer}, {Llerena}, {Martin}, {Nanayakkara}, {Ravindranath}, {Skillman}, {Sugahara}, {Wofford}, \& {Xu}}]{mingozzi22}
{Mingozzi}, M., {James}, B.~L., {Arellano-C{\'o}rdova}, K.~Z., {et~al.} 2022, \apj, 939, 110, \dodoi{10.3847/1538-4357/ac952c}

\bibitem[{{Morishita} {et~al.}(2019){Morishita}, {Abramson}, {Treu}, {Brammer}, {Jones}, {Kelly}, {Stiavelli}, {Trenti}, {Vulcani}, \& {Wang}}]{morishita19}
{Morishita}, T., {Abramson}, L.~E., {Treu}, T., {et~al.} 2019, \apj, 877, 141, \dodoi{10.3847/1538-4357/ab1d53}

\bibitem[{{Morishita} {et~al.}(2023{\natexlab{a}}){Morishita}, {Roberts-Borsani}, {Treu}, {Brammer}, {Mason}, {Trenti}, {Vulcani}, {Wang}, {Acebron}, {Bah{\'e}}, {Bergamini}, {Boyett}, {Bradac}, {Calabr{\`o}}, {Castellano}, {Chen}, {De Lucia}, {Filippenko}, {Fontana}, {Glazebrook}, {Grillo}, {Henry}, {Jones}, {Kelly}, {Koekemoer}, {Leethochawalit}, {Lu}, {Marchesini}, {Mascia}, {Mercurio}, {Merlin}, {Metha}, {Nanayakkara}, {Nonino}, {Paris}, {Pentericci}, {Rosati}, {Santini}, {Strait}, {Vanzella}, {Windhorst}, \& {Xie}}]{morishita22}
{Morishita}, T., {Roberts-Borsani}, G., {Treu}, T., {et~al.} 2023{\natexlab{a}}, \apjl, 947, L24, \dodoi{10.3847/2041-8213/acb99e}

\bibitem[{{Morishita} {et~al.}(2023{\natexlab{b}}){Morishita}, {Stiavelli}, {Chary}, {Trenti}, {Bergamini}, {Chiaberge}, {Leethochawalit}, {Roberts-Borsani}, {Shen}, \& {Treu}}]{morishita23}
{Morishita}, T., {Stiavelli}, M., {Chary}, R.-R., {et~al.} 2023{\natexlab{b}}, arXiv e-prints, arXiv:2308.05018, \dodoi{10.48550/arXiv.2308.05018}

\bibitem[{{Naidu} {et~al.}(2022{\natexlab{a}}){Naidu}, {Oesch}, {van Dokkum}, {Nelson}, {Suess}, {Brammer}, {Whitaker}, {Illingworth}, {Bouwens}, {Tacchella}, {Matthee}, {Allen}, {Bezanson}, {Conroy}, {Labbe}, {Leja}, {Leonova}, {Magee}, {Price}, {Setton}, {Strait}, {Stefanon}, {Toft}, {Weaver}, \& {Weibel}}]{naidu22}
{Naidu}, R.~P., {Oesch}, P.~A., {van Dokkum}, P., {et~al.} 2022{\natexlab{a}}, \apjl, 940, L14, \dodoi{10.3847/2041-8213/ac9b22}

\bibitem[{{Naidu} {et~al.}(2022{\natexlab{b}}){Naidu}, {Matthee}, {Oesch}, {Conroy}, {Sobral}, {Pezzulli}, {Hayes}, {Erb}, {Amor{\'\i}n}, {Gronke}, {Schaerer}, {Tacchella}, {Kerutt}, {Paulino-Afonso}, {Calhau}, {Llerena}, \& {R{\"o}ttgering}}]{naidu22b}
{Naidu}, R.~P., {Matthee}, J., {Oesch}, P.~A., {et~al.} 2022{\natexlab{b}}, \mnras, 510, 4582, \dodoi{10.1093/mnras/stab3601}

\bibitem[{{Nakajima} {et~al.}(2023){Nakajima}, {Ouchi}, {Isobe}, {Harikane}, {Zhang}, {Ono}, {Umeda}, \& {Oguri}}]{nakajima23}
{Nakajima}, K., {Ouchi}, M., {Isobe}, Y., {et~al.} 2023, arXiv e-prints, arXiv:2301.12825, \dodoi{10.48550/arXiv.2301.12825}

\bibitem[{{Nakajima} {et~al.}(2018){Nakajima}, {Schaerer}, {Le F{\`e}vre}, {Amor{\'\i}n}, {Talia}, {Lemaux}, {Tasca}, {Vanzella}, {Zamorani}, {Bardelli}, {Grazian}, {Guaita}, {Hathi}, {Pentericci}, \& {Zucca}}]{nakajima18}
{Nakajima}, K., {Schaerer}, D., {Le F{\`e}vre}, O., {et~al.} 2018, \aap, 612, A94, \dodoi{10.1051/0004-6361/201731935}

\bibitem[{{Nanayakkara} {et~al.}(2023){Nanayakkara}, {Glazebrook}, {Jacobs}, {Bonchi}, {Castellano}, {Fontana}, {Mason}, {Merlin}, {Morishita}, {Paris}, {Trenti}, {Treu}, {Calabr{\`o}}, {Boyett}, {Bradac}, {Leethochawalit}, {Marchesini}, {Santini}, {Strait}, {Vanzella}, {Vulcani}, {Wang}, \& {Yang}}]{nanayakkara23}
{Nanayakkara}, T., {Glazebrook}, K., {Jacobs}, C., {et~al.} 2023, \apjl, 947, L26, \dodoi{10.3847/2041-8213/acbfb9}

\bibitem[{{Napolitano} {et~al.}(2024){Napolitano}, {Pentericci}, {Santini}, {Calabr{\`o}}, {Mascia}, {Llerena}, {Castellano}, {Dickinson}, {Finkelstein}, {Amorin}, {Arrabal Haro}, {Bagley}, {Bhatawdekar}, {Cleri}, {Davis}, {Gardner}, {Gawiser}, {Giavalisco}, {Hathi}, {Hu}, {Jung}, {Kartaltepe}, {Koekemoer}, {Merlin}, {Mobasher}, {Papovich}, {Park}, {Pirzkal}, {Trump}, {Wilkins}, \& {Yung}}]{napolitano24}
{Napolitano}, L., {Pentericci}, L., {Santini}, P., {et~al.} 2024, arXiv e-prints, arXiv:2402.11220, \dodoi{10.48550/arXiv.2402.11220}

\bibitem[{{Noeske} {et~al.}(2007){Noeske}, {Weiner}, {Faber}, {Papovich}, {Koo}, {Somerville}, {Bundy}, {Conselice}, {Newman}, {Schiminovich}, {Le Floc'h}, {Coil}, {Rieke}, {Lotz}, {Primack}, {Barmby}, {Cooper}, {Davis}, {Ellis}, {Fazio}, {Guhathakurta}, {Huang}, {Kassin}, {Martin}, {Phillips}, {Rich}, {Small}, {Willmer}, \& {Wilson}}]{noeske07}
{Noeske}, K.~G., {Weiner}, B.~J., {Faber}, S.~M., {et~al.} 2007, \apjl, 660, L43, \dodoi{10.1086/517926}

\bibitem[{{Oesch} {et~al.}(2015){Oesch}, {van Dokkum}, {Illingworth}, {Bouwens}, {Momcheva}, {Holden}, {Roberts-Borsani}, {Smit}, {Franx}, {Labb{\'e}}, {Gonz{\'a}lez}, \& {Magee}}]{oesch15}
{Oesch}, P.~A., {van Dokkum}, P.~G., {Illingworth}, G.~D., {et~al.} 2015, \apjl, 804, L30, \dodoi{10.1088/2041-8205/804/2/L30}

\bibitem[{{Ono} {et~al.}(2023){Ono}, {Harikane}, {Ouchi}, {Yajima}, {Abe}, {Isobe}, {Shibuya}, {Wise}, {Zhang}, {Nakajima}, \& {Umeda}}]{ono23}
{Ono}, Y., {Harikane}, Y., {Ouchi}, M., {et~al.} 2023, \apj, 951, 72, \dodoi{10.3847/1538-4357/acd44a}

\bibitem[{{Pentericci} {et~al.}(2018){Pentericci}, {Vanzella}, {Castellano}, {Fontana}, {De Barros}, {Grazian}, {Marchi}, {Bradac}, {Conselice}, {Cristiani}, {Dickinson}, {Finkelstein}, {Giallongo}, {Guaita}, {Koekemoer}, {Maiolino}, {Santini}, \& {Tilvi}}]{pentericci18}
{Pentericci}, L., {Vanzella}, E., {Castellano}, M., {et~al.} 2018, \aap, 619, A147, \dodoi{10.1051/0004-6361/201732465}

\bibitem[{{Reddy} {et~al.}(2015){Reddy}, {Kriek}, {Shapley}, {Freeman}, {Siana}, {Coil}, {Mobasher}, {Price}, {Sanders}, \& {Shivaei}}]{reddy15}
{Reddy}, N.~A., {Kriek}, M., {Shapley}, A.~E., {et~al.} 2015, \apj, 806, 259, \dodoi{10.1088/0004-637X/806/2/259}

\bibitem[{{Reddy} {et~al.}(2018){Reddy}, {Shapley}, {Sanders}, {Kriek}, {Coil}, {Shivaei}, {Freeman}, {Mobasher}, {Siana}, {Azadi}, {Fetherolf}, {Fornasini}, {Leung}, {Price}, {Zick}, \& {Barro}}]{reddy18}
{Reddy}, N.~A., {Shapley}, A.~E., {Sanders}, R.~L., {et~al.} 2018, \apj, 869, 92, \dodoi{10.3847/1538-4357/aaed1e}

\bibitem[{{Reddy} {et~al.}(2020){Reddy}, {Shapley}, {Kriek}, {Steidel}, {Shivaei}, {Sanders}, {Mobasher}, {Coil}, {Siana}, {Freeman}, {Azadi}, {Fetherolf}, {Leung}, {Price}, \& {Zick}}]{reddy20}
{Reddy}, N.~A., {Shapley}, A.~E., {Kriek}, M., {et~al.} 2020, \apj, 902, 123, \dodoi{10.3847/1538-4357/abb674}

\bibitem[{{Rieke} \& {the JADES Collaboration}(2023)}]{rieke23}
{Rieke}, M., \& {the JADES Collaboration}. 2023, arXiv e-prints, arXiv:2306.02466, \dodoi{10.48550/arXiv.2306.02466}

\bibitem[{{Rigby} {et~al.}(2015){Rigby}, {Bayliss}, {Gladders}, {Sharon}, {Wuyts}, {Dahle}, {Johnson}, \& {Pe{\~n}a-Guerrero}}]{rigby15}
{Rigby}, J.~R., {Bayliss}, M.~B., {Gladders}, M.~D., {et~al.} 2015, \apjl, 814, L6, \dodoi{10.1088/2041-8205/814/1/L6}

\bibitem[{{Roberts-Borsani} {et~al.}(2022){Roberts-Borsani}, {Morishita}, {Treu}, {Leethochawalit}, \& {Trenti}}]{rb22}
{Roberts-Borsani}, G., {Morishita}, T., {Treu}, T., {Leethochawalit}, N., \& {Trenti}, M. 2022, \apj, 927, 236, \dodoi{10.3847/1538-4357/ac4803}

\bibitem[{{Roberts-Borsani} {et~al.}(2023{\natexlab{a}}){Roberts-Borsani}, {Treu}, {Mason}, {Ellis}, {Laporte}, {Schmidt}, {Bradac}, {Fontana}, {Morishita}, \& {Santini}}]{rb23}
{Roberts-Borsani}, G., {Treu}, T., {Mason}, C., {et~al.} 2023{\natexlab{a}}, \apj, 948, 54, \dodoi{10.3847/1538-4357/acc798}

\bibitem[{{Roberts-Borsani} {et~al.}(2023{\natexlab{b}}){Roberts-Borsani}, {Treu}, {Chen}, {Morishita}, {Vanzella}, {Zitrin}, {Bergamini}, {Castellano}, {Fontana}, {Glazebrook}, {Grillo}, {Kelly}, {Merlin}, {Nanayakkara}, {Paris}, {Rosati}, {Yang}, {Acebron}, {Bonchi}, {Boyett}, {Brada{\v{c}}}, {Brammer}, {Broadhurst}, {Calabr{\'o}}, {Diego}, {Dressler}, {Furtak}, {Filippenko}, {Henry}, {Koekemoer}, {Leethochawalit}, {Malkan}, {Mason}, {Mercurio}, {Metha}, {Pentericci}, {Pierel}, {Rieck}, {Roy}, {Santini}, {Strait}, {Strausbaugh}, {Trenti}, {Vulcani}, {Wang}, {Wang}, \& {Windhorst}}]{rb23_nature}
{Roberts-Borsani}, G., {Treu}, T., {Chen}, W., {et~al.} 2023{\natexlab{b}}, \nat, 618, 480, \dodoi{10.1038/s41586-023-05994-w}

\bibitem[{{Roberts-Borsani} {et~al.}(2024){Roberts-Borsani}, {Bagley}, {Rojas-Ruiz}, {Treu}, {Morishita}, {Finkelstein}, {Trenti}, {Arrabal Haro}, {Ba{\~n}ados}, {Ch{\'a}vez Ortiz}, {Chworowsky}, {Hutchison}, {Larson}, {Leethochawalit}, {Leung}, {Mason}, {Somerville}, {Stiavelli}, {Yung}, {Kassin}, \& {Soto}}]{rb24_borg}
{Roberts-Borsani}, G., {Bagley}, M., {Rojas-Ruiz}, S., {et~al.} 2024, arXiv e-prints, arXiv:2407.17551, \dodoi{10.48550/arXiv.2407.17551}

\bibitem[{{Roberts-Borsani} {et~al.}(2016){Roberts-Borsani}, {Bouwens}, {Oesch}, {Labbe}, {Smit}, {Illingworth}, {van Dokkum}, {Holden}, {Gonzalez}, {Stefanon}, {Holwerda}, \& {Wilkins}}]{rb16}
{Roberts-Borsani}, G.~W., {Bouwens}, R.~J., {Oesch}, P.~A., {et~al.} 2016, \apj, 823, 143, \dodoi{10.3847/0004-637X/823/2/143}

\bibitem[{{Sanders} {et~al.}(2023{\natexlab{a}}){Sanders}, {Shapley}, {Topping}, {Reddy}, \& {Brammer}}]{sanders23}
{Sanders}, R.~L., {Shapley}, A.~E., {Topping}, M.~W., {Reddy}, N.~A., \& {Brammer}, G.~B. 2023{\natexlab{a}}, arXiv e-prints, arXiv:2301.06696, \dodoi{10.48550/arXiv.2301.06696}

\bibitem[{{Sanders} {et~al.}(2023{\natexlab{b}}){Sanders}, {Shapley}, {Topping}, {Reddy}, \& {Brammer}}]{sanders_metal}
---. 2023{\natexlab{b}}, arXiv e-prints, arXiv:2303.08149, \dodoi{10.48550/arXiv.2303.08149}

\bibitem[{{Santini} {et~al.}(2023){Santini}, {Fontana}, {Castellano}, {Leethochawalit}, {Trenti}, {Treu}, {Belfiori}, {Birrer}, {Bonchi}, {Merlin}, {Mason}, {Morishita}, {Nonino}, {Paris}, {Polenta}, {Rosati}, {Yang}, {Boyett}, {Bradac}, {Calabr{\`o}}, {Dressler}, {Glazebrook}, {Marchesini}, {Mascia}, {Nanayakkara}, {Pentericci}, {Roberts-Borsani}, {Scarlata}, {Vulcani}, \& {Wang}}]{santini23}
{Santini}, P., {Fontana}, A., {Castellano}, M., {et~al.} 2023, \apjl, 942, L27, \dodoi{10.3847/2041-8213/ac9586}

\bibitem[{{Saxena} {et~al.}(2023{\natexlab{a}}){Saxena}, {Bunker}, {Jones}, {Stark}, {Cameron}, {Witstok}, {Arribas}, {Baker}, {Baum}, {Bhatawdekar}, {Bowler}, {Boyett}, {Carniani}, {Charlot}, {Chevallard}, {Curti}, {Curtis-Lake}, {Eisenstein}, {Endsley}, {Hainline}, {Helton}, {Johnson}, {Kumari}, {Looser}, {Maiolino}, {Rieke}, {Rix}, {Robertson}, {Sandles}, {Simmonds}, {Smit}, {Tacchella}, {Williams}, {Willmer}, \& {Willott}}]{saxena23}
{Saxena}, A., {Bunker}, A.~J., {Jones}, G.~C., {et~al.} 2023{\natexlab{a}}, arXiv e-prints, arXiv:2306.04536, \dodoi{10.48550/arXiv.2306.04536}

\bibitem[{{Saxena} {et~al.}(2023{\natexlab{b}}){Saxena}, {Robertson}, {Bunker}, {Endsley}, {Cameron}, {Charlot}, {Simmonds}, {Tacchella}, {Witstok}, {Willott}, {Carniani}, {Curtis-Lake}, {Ferruit}, {Jakobsen}, {Arribas}, {Chevallard}, {Curti}, {D'Eugenio}, {De Graaff}, {Jones}, {Looser}, {Maseda}, {Rawle}, {Rix}, {Del Pino}, {Smit}, {{\"U}bler}, {Eisenstein}, {Hainline}, {Hausen}, {Johnson}, {Rieke}, {Williams}, {Willmer}, {Baker}, {Bhatawdekar}, {Bowler}, {Boyett}, {Chen}, {Egami}, {Ji}, {Kumari}, {Nelson}, {Perna}, {Sandles}, {Scholtz}, \& {Shivaei}}]{saxena_faint}
{Saxena}, A., {Robertson}, B.~E., {Bunker}, A.~J., {et~al.} 2023{\natexlab{b}}, \aap, 678, A68, \dodoi{10.1051/0004-6361/202346245}

\bibitem[{{Schaerer}(2003)}]{schaerer03}
{Schaerer}, D. 2003, \aap, 397, 527, \dodoi{10.1051/0004-6361:20021525}

\bibitem[{{Schaerer} {et~al.}(2011){Schaerer}, {de Barros}, \& {Stark}}]{schaerer11}
{Schaerer}, D., {de Barros}, S., \& {Stark}, D.~P. 2011, \aap, 536, A72, \dodoi{10.1051/0004-6361/201117685}

\bibitem[{{Schaerer} {et~al.}(2022){Schaerer}, {Izotov}, {Worseck}, {Berg}, {Chisholm}, {Jaskot}, {Nakajima}, {Ravindranath}, {Thuan}, \& {Verhamme}}]{schaerer22}
{Schaerer}, D., {Izotov}, Y.~I., {Worseck}, G., {et~al.} 2022, \aap, 658, L11, \dodoi{10.1051/0004-6361/202243149}

\bibitem[{{Schenker} {et~al.}(2014){Schenker}, {Ellis}, {Konidaris}, \& {Stark}}]{schenker14}
{Schenker}, M.~A., {Ellis}, R.~S., {Konidaris}, N.~P., \& {Stark}, D.~P. 2014, \apj, 795, 20, \dodoi{10.1088/0004-637X/795/1/20}

\bibitem[{{Schenker} {et~al.}(2012){Schenker}, {Stark}, {Ellis}, {Robertson}, {Dunlop}, {McLure}, {Kneib}, \& {Richard}}]{schenker12}
{Schenker}, M.~A., {Stark}, D.~P., {Ellis}, R.~S., {et~al.} 2012, \apj, 744, 179, \dodoi{10.1088/0004-637X/744/2/179}

\bibitem[{{Schmidt} {et~al.}(2017){Schmidt}, {Huang}, {Treu}, {Hoag}, {Brada{\v{c}}}, {Henry}, {Jones}, {Mason}, {Malkan}, {Morishita}, {Pentericci}, {Trenti}, {Vulcani}, \& {Wang}}]{schmidt17}
{Schmidt}, K.~B., {Huang}, K.~H., {Treu}, T., {et~al.} 2017, \apj, 839, 17, \dodoi{10.3847/1538-4357/aa68a3}

\bibitem[{{Senchyna} {et~al.}(2024){Senchyna}, {Plat}, {Stark}, {Rudie}, {Berg}, {Charlot}, {James}, \& {Mingozzi}}]{senchyna24}
{Senchyna}, P., {Plat}, A., {Stark}, D.~P., {et~al.} 2024, \apj, 966, 92, \dodoi{10.3847/1538-4357/ad235e}

\bibitem[{{Senchyna} {et~al.}(2019){Senchyna}, {Stark}, {Chevallard}, {Charlot}, {Jones}, \& {Vidal-Garc{\'\i}a}}]{senchyna19}
{Senchyna}, P., {Stark}, D.~P., {Chevallard}, J., {et~al.} 2019, \mnras, 488, 3492, \dodoi{10.1093/mnras/stz1907}

\bibitem[{{Shapley} {et~al.}(2023){Shapley}, {Sanders}, {Reddy}, {Topping}, \& {Brammer}}]{shapley23}
{Shapley}, A.~E., {Sanders}, R.~L., {Reddy}, N.~A., {Topping}, M.~W., \& {Brammer}, G.~B. 2023, \apj, 954, 157, \dodoi{10.3847/1538-4357/acea5a}

\bibitem[{{Shapley} {et~al.}(2003){Shapley}, {Steidel}, {Pettini}, \& {Adelberger}}]{shapley03}
{Shapley}, A.~E., {Steidel}, C.~C., {Pettini}, M., \& {Adelberger}, K.~L. 2003, \apj, 588, 65, \dodoi{10.1086/373922}

\bibitem[{{Shapley} {et~al.}(2022){Shapley}, {Sanders}, {Salim}, {Reddy}, {Kriek}, {Mobasher}, {Coil}, {Siana}, {Price}, {Shivaei}, {Dunlop}, {McLure}, \& {Cullen}}]{shapley22}
{Shapley}, A.~E., {Sanders}, R.~L., {Salim}, S., {et~al.} 2022, \apj, 926, 145, \dodoi{10.3847/1538-4357/ac4742}

\bibitem[{{Shivaei} {et~al.}(2018){Shivaei}, {Reddy}, {Siana}, {Shapley}, {Kriek}, {Mobasher}, {Freeman}, {Sanders}, {Coil}, {Price}, {Fetherolf}, {Azadi}, {Leung}, \& {Zick}}]{shivae18}
{Shivaei}, I., {Reddy}, N.~A., {Siana}, B., {et~al.} 2018, \apj, 855, 42, \dodoi{10.3847/1538-4357/aaad62}

\bibitem[{{Simmonds} {et~al.}(2024){Simmonds}, {Tacchella}, {Hainline}, {Johnson}, {McClymont}, {Robertson}, {Saxena}, {Sun}, {Witten}, {Baker}, {Bhatawdekar}, {Boyett}, {Bunker}, {Charlot}, {Curtis-Lake}, {Egami}, {Eisenstein}, {Hausen}, {Maiolino}, {Maseda}, {Scholtz}, {Williams}, {Willott}, \& {Witstok}}]{simmonds24}
{Simmonds}, C., {Tacchella}, S., {Hainline}, K., {et~al.} 2024, \mnras, 527, 6139, \dodoi{10.1093/mnras/stad3605}

\bibitem[{{Smit} {et~al.}(2014){Smit}, {Bouwens}, {Labb{\'e}}, {Zheng}, {Bradley}, {Donahue}, {Lemze}, {Moustakas}, {Umetsu}, {Zitrin}, {Coe}, {Postman}, {Gonzalez}, {Bartelmann}, {Ben{\'\i}tez}, {Broadhurst}, {Ford}, {Grillo}, {Infante}, {Jimenez-Teja}, {Jouvel}, {Kelson}, {Lahav}, {Maoz}, {Medezinski}, {Melchior}, {Meneghetti}, {Merten}, {Molino}, {Moustakas}, {Nonino}, {Rosati}, \& {Seitz}}]{smit14}
{Smit}, R., {Bouwens}, R.~J., {Labb{\'e}}, I., {et~al.} 2014, \apj, 784, 58, \dodoi{10.1088/0004-637X/784/1/58}

\bibitem[{{Speagle} {et~al.}(2014){Speagle}, {Steinhardt}, {Capak}, \& {Silverman}}]{speagle14}
{Speagle}, J.~S., {Steinhardt}, C.~L., {Capak}, P.~L., \& {Silverman}, J.~D. 2014, \apjs, 214, 15, \dodoi{10.1088/0067-0049/214/2/15}

\bibitem[{{Stark} {et~al.}(2014){Stark}, {Richard}, {Siana}, {Charlot}, {Freeman}, {Gutkin}, {Wofford}, {Robertson}, {Amanullah}, {Watson}, \& {Milvang-Jensen}}]{stark14}
{Stark}, D.~P., {Richard}, J., {Siana}, B., {et~al.} 2014, \mnras, 445, 3200, \dodoi{10.1093/mnras/stu1618}

\bibitem[{{Stark} {et~al.}(2015){Stark}, {Richard}, {Charlot}, {Cl{\'e}ment}, {Ellis}, {Siana}, {Robertson}, {Schenker}, {Gutkin}, \& {Wofford}}]{stark15}
{Stark}, D.~P., {Richard}, J., {Charlot}, S., {et~al.} 2015, \mnras, 450, 1846, \dodoi{10.1093/mnras/stv688}

\bibitem[{{Stark} {et~al.}(2017){Stark}, {Ellis}, {Charlot}, {Chevallard}, {Tang}, {Belli}, {Zitrin}, {Mainali}, {Gutkin}, {Vidal-Garc{\'\i}a}, {Bouwens}, \& {Oesch}}]{stark17}
{Stark}, D.~P., {Ellis}, R.~S., {Charlot}, S., {et~al.} 2017, \mnras, 464, 469, \dodoi{10.1093/mnras/stw2233}

\bibitem[{{Stefanon} {et~al.}(2023){Stefanon}, {Bouwens}, {Labb{\'e}}, {Illingworth}, {Gonzalez}, \& {Oesch}}]{stefanon23}
{Stefanon}, M., {Bouwens}, R.~J., {Labb{\'e}}, I., {et~al.} 2023, \apj, 943, 81, \dodoi{10.3847/1538-4357/aca470}

\bibitem[{{Steidel} {et~al.}(2016){Steidel}, {Strom}, {Pettini}, {Rudie}, {Reddy}, \& {Trainor}}]{steidel16}
{Steidel}, C.~C., {Strom}, A.~L., {Pettini}, M., {et~al.} 2016, \apj, 826, 159, \dodoi{10.3847/0004-637X/826/2/159}

\bibitem[{{Strait} {et~al.}(2020){Strait}, {Brada{\v{c}}}, {Coe}, {Bradley}, {Salmon}, {Lemaux}, {Huang}, {Zitrin}, {Sharon}, {Acebron}, {Andrade-Santos}, {Avila}, {Frye}, {Hoag}, {Mahler}, {Nonino}, {Ogaz}, {Oguri}, {Ouchi}, {Paterno-Mahler}, \& {Pelliccia}}]{strait20}
{Strait}, V., {Brada{\v{c}}}, M., {Coe}, D., {et~al.} 2020, \apj, 888, 124, \dodoi{10.3847/1538-4357/ab5daf}

\bibitem[{{Tamura} {et~al.}(2019){Tamura}, {Mawatari}, {Hashimoto}, {Inoue}, {Zackrisson}, {Christensen}, {Binggeli}, {Matsuda}, {Matsuo}, {Takeuchi}, {Asano}, {Sunaga}, {Shimizu}, {Okamoto}, {Yoshida}, {Lee}, {Shibuya}, {Taniguchi}, {Umehata}, {Hatsukade}, {Kohno}, \& {Ota}}]{tamura19}
{Tamura}, Y., {Mawatari}, K., {Hashimoto}, T., {et~al.} 2019, \apj, 874, 27, \dodoi{10.3847/1538-4357/ab0374}

\bibitem[{{Tang} {et~al.}(2021){Tang}, {Stark}, {Chevallard}, {Charlot}, {Endsley}, \& {Congiu}}]{tang21}
{Tang}, M., {Stark}, D.~P., {Chevallard}, J., {et~al.} 2021, \mnras, 503, 4105, \dodoi{10.1093/mnras/stab705}

\bibitem[{{Tang} {et~al.}(2023){Tang}, {Stark}, {Chen}, {Mason}, {Topping}, {Endsley}, {Senchyna}, {Plat}, {Lu}, {Whitler}, {Robertson}, \& {Charlot}}]{tang23}
{Tang}, M., {Stark}, D.~P., {Chen}, Z., {et~al.} 2023, arXiv e-prints, arXiv:2301.07072, \dodoi{10.48550/arXiv.2301.07072}

\bibitem[{{Tilvi} {et~al.}(2020){Tilvi}, {Malhotra}, {Rhoads}, {Coughlin}, {Zheng}, {Finkelstein}, {Veilleux}, {Mobasher}, {Wang}, {Probst}, {Swaters}, {Hibon}, {Joshi}, {Zabl}, {Jiang}, {Pharo}, \& {Yang}}]{tilvi20}
{Tilvi}, V., {Malhotra}, S., {Rhoads}, J.~E., {et~al.} 2020, \apjl, 891, L10, \dodoi{10.3847/2041-8213/ab75ec}

\bibitem[{{Topping} {et~al.}(2022){Topping}, {Stark}, {Endsley}, {Plat}, {Whitler}, {Chen}, \& {Charlot}}]{topping22}
{Topping}, M.~W., {Stark}, D.~P., {Endsley}, R., {et~al.} 2022, \apj, 941, 153, \dodoi{10.3847/1538-4357/aca522}

\bibitem[{{Topping} {et~al.}(2023){Topping}, {Stark}, {Endsley}, {Whitler}, {Hainline}, {Johnson}, {Robertson}, {Tacchella}, {Chen}, {Alberts}, {Baker}, {Bunker}, {Carniani}, {Charlot}, {Chevallard}, {Curtis-Lake}, {DeCoursey}, {Egami}, {Eisenstein}, {Ji}, {Maiolino}, {Williams}, {Willmer}, {Willott}, \& {Witstok}}]{topping23}
---. 2023, arXiv e-prints, arXiv:2307.08835, \dodoi{10.48550/arXiv.2307.08835}

\bibitem[{{Topping} {et~al.}(2024){Topping}, {Stark}, {Senchyna}, {Plat}, {Zitrin}, {Endsley}, {Charlot}, {Furtak}, {Maseda}, {Smit}, {Mainali}, {Chevallard}, {Molyneux}, \& {Rigby}}]{topping24}
{Topping}, M.~W., {Stark}, D.~P., {Senchyna}, P., {et~al.} 2024, arXiv e-prints, arXiv:2401.08764, \dodoi{10.48550/arXiv.2401.08764}

\bibitem[{{Treu} {et~al.}(2013){Treu}, {Schmidt}, {Trenti}, {Bradley}, \& {Stiavelli}}]{treu13}
{Treu}, T., {Schmidt}, K.~B., {Trenti}, M., {Bradley}, L.~D., \& {Stiavelli}, M. 2013, \apjl, 775, L29, \dodoi{10.1088/2041-8205/775/1/L29}

\bibitem[{{Treu} {et~al.}(2022){Treu}, {Roberts-Borsani}, {Bradac}, {Brammer}, {Fontana}, {Henry}, {Mason}, {Morishita}, {Pentericci}, {Wang}, {Acebron}, {Bagley}, {Bergamini}, {Belfiori}, {Bonchi}, {Boyett}, {Boutsia}, {Calabr{\'o}}, {Caminha}, {Castellano}, {Dressler}, {Glazebrook}, {Grillo}, {Jacobs}, {Jones}, {Kelly}, {Leethochawalit}, {Malkan}, {Marchesini}, {Mascia}, {Mercurio}, {Merlin}, {Nanayakkara}, {Nonino}, {Paris}, {Poggianti}, {Rosati}, {Santini}, {Scarlata}, {Shipley}, {Strait}, {Trenti}, {Tubthong}, {Vanzella}, {Vulcani}, \& {Yang}}]{treu22}
{Treu}, T., {Roberts-Borsani}, G., {Bradac}, M., {et~al.} 2022, \apj, 935, 110, \dodoi{10.3847/1538-4357/ac8158}

\bibitem[{Vijayan {et~al.}(2020)Vijayan, Lovell, Wilkins, Thomas, Barnes, Irodotou, Kuusisto, \& Roper}]{vijayan20}
Vijayan, A.~P., Lovell, C.~C., Wilkins, S.~M., {et~al.} 2020, Monthly Notices of the Royal Astronomical Society, 501, 3289, \dodoi{10.1093/mnras/staa3715}

\bibitem[{{Vikaeus} {et~al.}(2024){Vikaeus}, {Zackrisson}, {Wilkins}, {Nabizadeh}, {Kokorev}, {Abdurro'uf}, {Bradley}, {Coe}, {Dayal}, \& {Ricotti}}]{vikaeus24}
{Vikaeus}, A., {Zackrisson}, E., {Wilkins}, S., {et~al.} 2024, \mnras, \dodoi{10.1093/mnras/stae323}

\bibitem[{{Wang} {et~al.}(2022){Wang}, {Cheng}, {Ge}, {Meng}, {Daddi}, {Yan}, {Jones}, {Malkan}, {Arrabal Haro}, {Brammer}, \& {Oguri}}]{wang22}
{Wang}, X., {Cheng}, C., {Ge}, J., {et~al.} 2022, arXiv e-prints, arXiv:2212.04476, \dodoi{10.48550/arXiv.2212.04476}

\bibitem[{{Whitler} {et~al.}(2023{\natexlab{a}}){Whitler}, {Endsley}, {Stark}, {Topping}, {Chen}, \& {Charlot}}]{whitler23b}
{Whitler}, L., {Endsley}, R., {Stark}, D.~P., {et~al.} 2023{\natexlab{a}}, \mnras, 519, 157, \dodoi{10.1093/mnras/stac3535}

\bibitem[{{Whitler} {et~al.}(2023{\natexlab{b}}){Whitler}, {Stark}, {Endsley}, {Leja}, {Charlot}, \& {Chevallard}}]{whitler23}
{Whitler}, L., {Stark}, D.~P., {Endsley}, R., {et~al.} 2023{\natexlab{b}}, \mnras, 519, 5859, \dodoi{10.1093/mnras/stad004}

\bibitem[{{Wilkins} {et~al.}(2011){Wilkins}, {Bunker}, {Stanway}, {Lorenzoni}, \& {Caruana}}]{wilkins11}
{Wilkins}, S.~M., {Bunker}, A.~J., {Stanway}, E., {Lorenzoni}, S., \& {Caruana}, J. 2011, \mnras, 417, 717, \dodoi{10.1111/j.1365-2966.2011.19315.x}

\bibitem[{{Wilkins} {et~al.}(2024){Wilkins}, {Lovell}, {Irodotou}, {Vijayan}, {Vikaeus}, {Zackrisson}, {Caruana}, {Stanway}, {Conselice}, {Seeyave}, {Roper}, {Chworowsky}, \& {Finkelstein}}]{wilkins24}
{Wilkins}, S.~M., {Lovell}, C.~C., {Irodotou}, D., {et~al.} 2024, \mnras, 527, 7965, \dodoi{10.1093/mnras/stad3558}

\bibitem[{{Williams} {et~al.}(2022){Williams}, {Kelly}, {Chen}, {Brammer}, {Zitrin}, {Treu}, {Scarlata}, {Koekemoer}, {Oguri}, {Lin}, {Diego}, {Nonino}, {Hjorth}, {Langeroodi}, {Broadhurst}, {Rogers}, {Perez-Fournon}, {Foley}, {Jha}, {Filippenko}, {Strolger}, {Pierel}, {Poidevin}, \& {Yang}}]{williams22}
{Williams}, H., {Kelly}, P.~L., {Chen}, W., {et~al.} 2022, arXiv e-prints, arXiv:2210.15699, \dodoi{10.48550/arXiv.2210.15699}

\bibitem[{{Withers} {et~al.}(2023){Withers}, {Muzzin}, {Ravindranath}, {Sarrouh}, {Abraham}, {Asada}, {Bradac}, {Brammer}, {Desprez}, {Iyer}, {Martis}, {Mowla}, {Noirot}, {Sawicki}, {Strait}, \& {Willott}}]{withers23}
{Withers}, S., {Muzzin}, A., {Ravindranath}, S., {et~al.} 2023, arXiv e-prints, arXiv:2304.11181, \dodoi{10.48550/arXiv.2304.11181}

\bibitem[{{Witten} {et~al.}(2024){Witten}, {McClymont}, {Laporte}, {Roberts-Borsani}, {Sijacki}, {Tacchella}, {Simmonds}, {Katz}, {Ellis}, {Witstok}, {Maiolino}, {Ji}, {Hayes}, {Looser}, \& {D'Eugenio}}]{witten24}
{Witten}, C., {McClymont}, W., {Laporte}, N., {et~al.} 2024, arXiv e-prints, arXiv:2407.07937, \dodoi{10.48550/arXiv.2407.07937}

\bibitem[{{Yang} {et~al.}(2022){Yang}, {Leethochawalit}, {Treu}, {Roberts-Borsani}, {Brada{\v{c}}}, {Birrer}, {Castellano}, {Merlin}, {Fontana}, {Amorin}, \& {Trenti}}]{yang22}
{Yang}, L., {Leethochawalit}, N., {Treu}, T., {et~al.} 2022, \mnras, 514, 1148, \dodoi{10.1093/mnras/stac1236}

\bibitem[{{Zhu} {et~al.}(2023){Zhu}, {Becker}, {Christenson}, {D'Aloisio}, {Bosman}, {Bakx}, {D'Odorico}, {Bischetti}, {Cain}, {Davies}, {Davies}, {Eilers}, {Fan}, {Gaikwad}, {Haehnelt}, {Keating}, {Kulkarni}, {Lai}, {Ma}, {Mesinger}, {Qin}, {Satyavolu}, {Takeuchi}, {Umehata}, \& {Yang}}]{Zhu2023}
{Zhu}, Y., {Becker}, G.~D., {Christenson}, H.~M., {et~al.} 2023, \apj, 955, 115, \dodoi{10.3847/1538-4357/aceef4}

\end{thebibliography}
\bibliographystyle{aasjournal}

\appendix

\onecolumngrid

\section{Redshift Confirmations of $\lowercase{z}\geqslant5$ Sources}
Here we tabulate the spectroscopic redshifts and a number of spectrophotometric properties associated with our parent sample of confirmed $z\geqslant5$ sources. In Table~\ref{tab:redshifts1} we present all sources (star-forming sources and those flagged as potential AGN, see Section~\ref{sec:sample}) with secure redshifts and where sufficient photometric and spectroscopic coverage existed for a scaling of the spectrum in the rest-frame UV. In Table ~\ref{tab:redshifts2}, we adopt the same format as Table~\ref{tab:redshifts1} but list sources where either the redshift was uncertain (e.g., where the \texttt{msaexp}-derived redshift was $z\geqslant5$ but the spectrum lacked clearly identifiable features) and/or where UV coverage necessary for a scaling of the spectrum was lacking.

\startlongtable
\begin{deluxetable*}{ccccccccccc}
\tabletypesize{\footnotesize}
\tablewidth{0pt}
\tablecaption{The List of Spectroscopically Confirmed Sources Reduced and Compiled in This Study, along with Several Spectrophotometric Properties.}
\label{tab:redshifts1}
\tablehead{
\colhead{Prog. ID} &
\colhead{MPT ID} &
\colhead{RA} &
\colhead{Dec} &
\colhead{Field} &
\colhead{$z_{\rm spec}$} &
\colhead{$m_{\rm UV}$} &
\colhead{$M_{\rm UV}$} &
\colhead{$\beta$} &
\colhead{$\mu$} &
\colhead{flag$_{Ly\alpha}$} \\
\colhead{} &
\colhead{} &
\colhead{[J2000]} &
\colhead{[J2000]} &
\colhead{} &
\colhead{} &
\colhead{[AB]} &
\colhead{[AB]} &
\colhead{} &
\colhead{} &
\colhead{}}
\startdata
\\[-0.2cm]
\multicolumn{11}{c}{\textit{Star-forming galaxies}}\\[0.1cm]
\cline{4-6}\\[-0.1cm]
1210,3215 & 17400,20128771 & 53.14988 & -27.77650 & GOODS-S & 12.926 & 29.222 & -18.677 & -2.703 & 1.000 & 0 \\
2561 & 13077 & 3.57087 & -30.40159 & Abell 2744 & 12.908 & 28.363 & -18.259 & -2.385 & 2.422 & 0 \\
1210,3215 & 2773,20096216 & 53.16634 & -27.82156 & GOODS-S & 12.516 & 28.792 & -19.061 & -2.238 & 1.000 & 0 \\
2561 & 38766 & 3.51356 & -30.35680 & Abell 2744 & 12.393 & 28.408 & -18.824 & -2.634 & 1.790 & 0 \\
3073 & 22600 & 3.49898 & -30.32475 & Abell 2744 & 12.338 & 26.846 & -20.500 & -2.327 & 1.308 & 0 \\
1210,3215 & 10014220,20130158 & 53.16477 & -27.77463 & GOODS-S & 11.552 & 28.107 & -19.630 & -2.458 & 1.000 & 0 \\
2750 & 1 & 214.94315 & 52.94244 & CANDELS-EGS & 11.400 & 27.864 & -19.895 & -2.778 & 1.000 & 0 \\
2750 & 10 & 214.90663 & 52.94550 & CANDELS-EGS & 11.040 & 27.387 & -20.241 & -1.810 & 1.000 & 0 \\
1345 & 80073 & 214.93206 & 52.84187 & CANDELS-EGS & 10.810 & 27.770 & -19.911 & -3.096 & 1.000 & 1 \\
3073 & 16115 & 3.51374 & -30.35157 & Abell 2744 & 10.660 & 28.771 & -18.358 & -2.023 & 1.662 & 0 \\
\\[-0.2cm]
\multicolumn{11}{c}{\textit{AGN candidates}}\\[0.1cm]
\cline{4-6}\\[-0.1cm]
2561 & 26185 & 3.56707 & -30.37786 & Abell 2744 & 10.061 & 26.760 & -19.345 & -2.124 & 3.710 & 0 \\
1345 & 1019 & 215.03539 & 52.89066 & CANDELS-EGS & 8.679 & 25.512 & -21.772 & -2.116 & 1.000 & 1 \\
2561 & 20466 & 3.64041 & -30.38644 & Abell 2744 & 8.503 & 29.418 & -17.495 & -1.696 & 1.396 & 1 \\
2561 & 16594 & 3.59720 & -30.39433 & Abell 2744 & 7.040 & 29.019 & -16.311 & -1.121 & 3.172 & 1 \\
2561 & 13123 & 3.57983 & -30.40157 & Abell 2744 & 7.038 & 28.503 & -16.410 & -1.669 & 5.897 & 1 \\
2561 & 15383 & 3.58353 & -30.39668 & Abell 2744 & 7.038 & 26.767 & -17.721 & -1.757 & 10.222 & 1 \\
2561 & 28876 & 3.56960 & -30.37322 & Abell 2744 & 7.037 & 30.183 & -15.475 & -2.040 & 3.650 & 0 \\
1345 & 717 & 215.08141 & 52.97218 & CANDELS-EGS & 6.934 & 25.929 & -21.016 & -2.478 & 1.000 & 0 \\
2561 & 11254 & 3.58045 & -30.40502 & Abell 2744 & 6.871 & 27.028 & -18.356 & -2.323 & 4.498 & 0 \\
2561 & 41225 & 3.53400 & -30.35331 & Abell 2744 & 6.770 & 27.465 & -18.928 & -1.857 & 1.853 & 0
\enddata
\tablecomments{Star-forming sources are listed in the top half of the table, while AGN candidates (identified here and in other works) are listed in the bottom half. Only the first 10 sources from each classification are shown here, while the table is available in its entirety online. All apparent magnitudes remain \textit{uncorrected} for gravitational lensing, however absolute magnitudes are corrected for such effects. Objects for which coordinates are absent represent multiply-imaged sources where the apparent and absolute magnitudes are derived by de-magnifying the photometry of the individual images and taking the median value.}
\vspace{-0.5cm}
\end{deluxetable*}

\begin{deluxetable*}{ccccccccccc}
\tabletypesize{\footnotesize}
\tablewidth{0pt}
\tablecaption{The Same as Table~\ref{tab:redshifts1} But for Sources with Uncertain Redshifts and/or Where Rest-frame Ultraviolet Photometry or Prism fluxes Were Not Available for Normalization Purposes.}
\label{tab:redshifts2}
\tablehead{
\colhead{Prog. ID} &
\colhead{MPT ID} &
\colhead{RA} &
\colhead{Dec} &
\colhead{Field} &
\colhead{$z_{\rm spec}$} &
\colhead{$m_{\rm UV}$} &
\colhead{$M_{\rm UV}$} &
\colhead{$\beta$} &
\colhead{$\mu$} &
\colhead{flag$_{Ly\alpha}$} \\
\colhead{} &
\colhead{} &
\colhead{[J2000]} &
\colhead{[J2000]} &
\colhead{} &
\colhead{} &
\colhead{[AB]} &
\colhead{[AB]} &
\colhead{} &
\colhead{} &
\colhead{}}
\startdata
\\[-0.2cm]
\multicolumn{11}{c}{\textit{Star-forming galaxies}}\\[0.1cm]
\cline{4-6}\\[-0.1cm]
3073 & 14033 & 3.49371 & -30.36017 & Abell 2744 & 9.294 & 28.585 & -18.894 & -3.120 & 1.465 & 0 \\
1345 & 7 & 215.01171 & 52.98830 & CANDELS-EGS & 8.876 & -- & -- & -- & 1.000 \\
1345 & 2 & 214.99440 & 52.98938 & CANDELS-EGS & 8.809 & -- & -- & -- & 1.000 \\
3215 & 97586 & 53.15084 & -27.82014 & GOODS-S & 8.696 & 29.453 & -17.722 & -2.010 & 1.000 & 0 \\
1181 & 54165 & 189.27184 & 62.19518 & GOODS-N & 8.660 & -- & -- & -- & 1.000 \\
3073 & 14495 & 3.49417 & -30.35797 & Abell 2744 & 8.372 & 28.505 & -18.644 & -3.064 & 1.469 & 0 \\
3215 & 129731 & 53.13603 & -27.77523 & GOODS-S & 8.355 & 29.050 & -18.451 & -2.115 & 1.000 & 0 \\
3073 & 15840 & 3.51882 & -30.35288 & Abell 2744 & 8.273 & 29.981 & -16.610 & -3.910 & 1.744 & 0 \\
3073 & 21322 & 3.46106 & -30.31290 & Abell 2744 & 8.161 & 27.837 & -19.321 & -1.564 & 1.195 & 0 \\
3215 & 281814 & 53.16155 & -27.77017 & GOODS-S & 8.147 & 29.135 & -17.970 & -2.678 & 1.000 & 0 \\
\\[-0.2cm]
\multicolumn{11}{c}{\textit{AGN candidates}}\\[0.1cm]
\cline{4-6}\\[-0.1cm]
1210 & 10013704 & 53.12653 & -27.81809 & GOODS-S & 5.920 & 27.690 & -19.037 & -2.592 & 1.000 & 0 \\
1181 & 38147 & 189.27068 & 62.14842 & GOODS-N & 5.878 & -- & -- & -- & 1.000 \\
1433 & 1742 & 101.89250 & 70.21018 & MACS 0647 & 5.808 & -- & -- & -- & 1.216 \\
2565 & 13610 & 34.31482 & -5.23012 & MACS 0647 & 5.402 & -- & -- & -- & 1.000 \\
1181 & 68797 & 189.22914 & 62.14619 & GOODS-N & 5.037 & -- & -- & -- & 1.000
\enddata
\vspace{-0.5cm}
\end{deluxetable*}

\clearpage

\onecolumngrid

\section{Average Slit Sloss Correction Function}
The correction of wavelength-dependent slit losses (when present) in galaxy spectra represents a crucial step toward obtaining accurate measurements. As mentioned in Section~\ref{subsec:datacheck}, residual wavelength-dependent slit losses are found to be present in a number of our reduced spectra that are not fully accounted for in the path loss step of the data reduction pipeline. Here we expand on the determination of those slit losses and the derivation of an average correction function.

\begin{figure*}
\center
 \includegraphics[width=\textwidth]{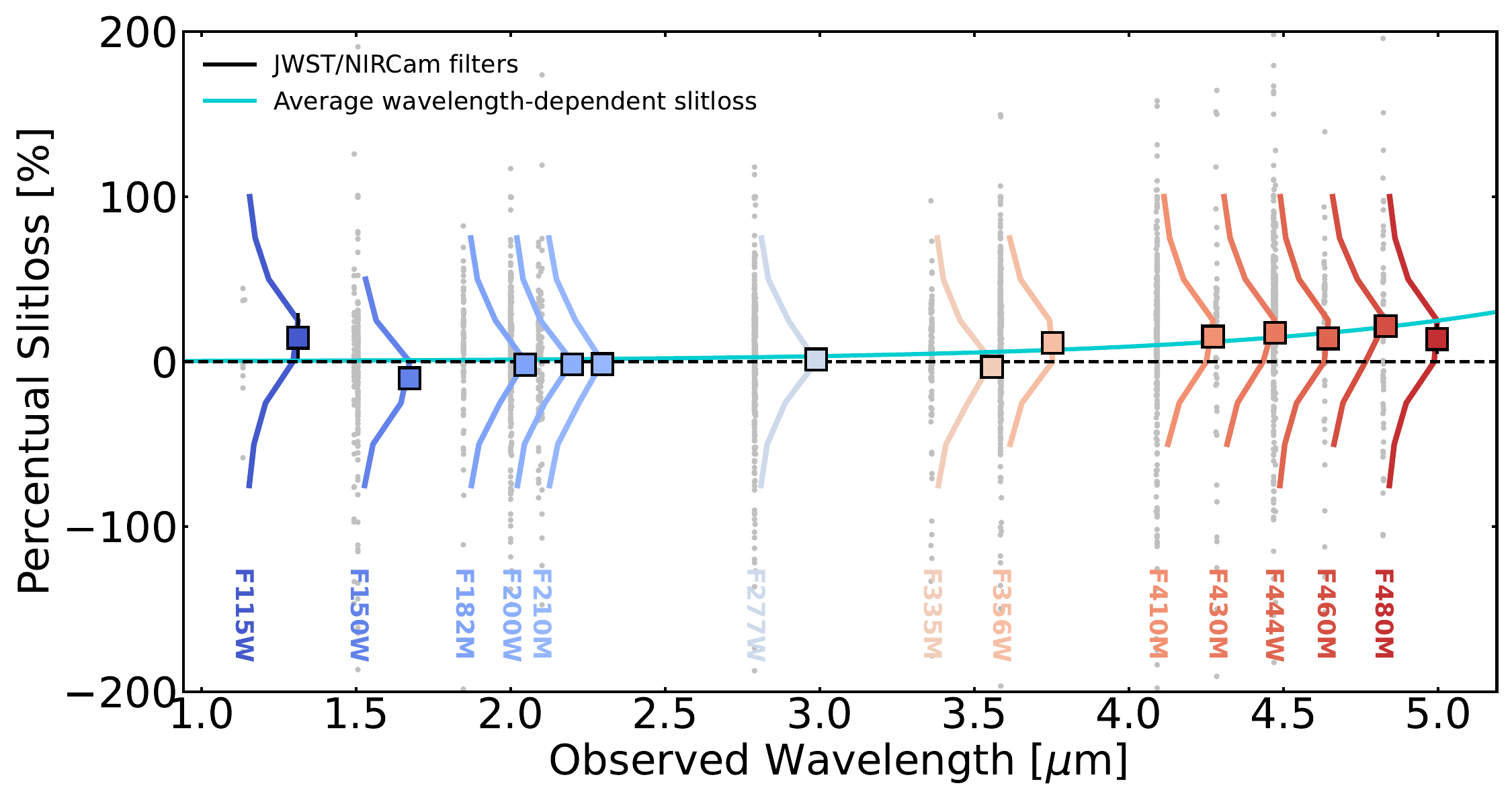}
 \caption{The percentage slit loss of sources observed with NIRSpec prism compared to NIRCam photometry. The slit loss is defined as the NIRCam photometry minus the NIRSpec pseudophotometry, divided by the NIRCam photometry, and multiplied by 100. Measurements for individual sources are plotted as grey points, while Gaussian fits to the density distribution in each filter are plotted as solid lines with normalized amplitudes. The centroid of each Gaussian is shown as a colored square, shifted by $+$0.2 $\mu$m for clarity. A best-fit exponential function to the squares is shown as a turquoise line. We note that the distributions of single-object points for virtually all bands are Gaussian in shape, except the F115W and F150W bands, which are less well defined and thus appear slightly up- or down-scattered compared to their redder counterparts.}
 \label{fig:slitloss}
\end{figure*}

For each photometric band redward of the adopted UV-scaling filter, we evaluate the percentage difference between the extracted photometry and prism pseudophotometry generated directly from the spectrum with the relevant filter response curves (i.e., NIRCam photometry minus prism photometry, divided by NIRCam photometry times 100). The results are plotted in Figure~\ref{fig:slitloss}, where for each source and photometric band the percentage of slit loss is plotted as a gray point. The distribution of points for each filter are approximately Gaussian, and thus to highlight their centroids and spread, we fit each density profile with a Gaussian function and plot the best fit as colored lines (out to their 5$\sigma$ width limits), along with a marker (shifted in wavelength for clarity) indicating the centroid and its $1\sigma$ uncertainty. We note a nonnegligible amount of wavelength-dependent slit loss, starting at approximately 1-2\% (absolute values) on average in the blue ($<2 \mu$m) bands (albeit with some scatter in the F115W and F150W bands) and increasing to 10-20\% on average in the red bands, undoubtedly reflecting loss of light from an increasing point-spread function width with wavelength. We note that the apparently larger deviations in the F115W and F150W centroids are due to poorly constrained density profiles, which begin to deviate from a well-defined Gaussian. The average wavelength-dependent slit loss is well fit by an exponential function (plotted as a turquoise line), which begins at 10-20\% slit loss in the reddest bands and progressively decreases to subpercent levels by $\sim3 \mu$m and shorter-wavelength bands. We use this parameterization to correct all composite spectra derived in this analysis (although we do not propagate uncertainties on the slit loss correction into our spectral measurements) and render the correction function publicly available for community use.

We also note there is nonnegligible scatter among individual sources, some of which is not uniquely attributable to loss of light due to the NIRSpes MSA shutter. These include, in a few cases, some possible oversubtraction of the sky background or overcorrection of the path loss by the data reduction pipeline, in which case the spectrum would lie below and above the photometric data points, respectively, and result in values beyond $\pm100$\%. Accurate corrections on an object-by-object basis are not trivial, requiring knowledge of the object size and position in the NIRSpec MSA shutter as well as a robust photometric baseline, which is not always the case in low SNR regimes where even HST and NIRCam photometry can display significant scatter. However, an average correction for stacked spectra is both possible and required: the high SNR of composite spectra ensures a more robust parameterization of the trend which is less sensitive to outliers (e.g., apparently faint galaxies or low SNR spectra) and encapsulating of various degrees of slit loss deriving from, e.g., size and morphology and/or position in the slit.

\clearpage


\section{Line Fluxes and Equivalent Widths}
We list here the H$\beta$-normalized integrated line fluxes and EW$_{0}$s, calculated as described in Section~\ref{sec:method} and discussed throughout the various sections of the paper.

\begin{deluxetable}{lllllllllll}[h!]
\tabletypesize{\footnotesize}
\tablecolumns{11}
\tablewidth{0pt}
\centering
\tablecaption{Emission line Ratios Relative to H$\beta$ and Rest-frame Equivalent Widths for the $z\sim5.5-9.5$ Redshift-binned and Ly$\alpha$-Emitter Composite Spectra listed in Table~\ref{tab:stacks}.}
\label{tab:line_ratios_redshift}
\tablehead{\colhead{} & \multicolumn{2}{c}{z5\_SF} & \multicolumn{2}{c}{z6\_SF} & \multicolumn{2}{c}{z7\_SF} & \multicolumn{2}{c}{z8\_SF} & \multicolumn{2}{c}{z9\_SF}\\
\colhead{Line} & \colhead{$\frac{F_{\rm Line}}{F_{\rm H\beta}}$} & \colhead{EW$_{0}$} & \colhead{$\frac{F_{\rm Line}}{F_{\rm H\beta}}$} & \colhead{EW$_{0}$} & \colhead{$\frac{F_{\rm Line}}{F_{\rm H\beta}}$} & \colhead{EW$_{0}$} & \colhead{$\frac{F_{\rm Line}}{F_{\rm H\beta}}$} & \colhead{EW$_{0}$} & \colhead{$\frac{F_{\rm Line}}{F_{\rm H\beta}}$} & \colhead{EW$_{0}$}}
\startdata
\ciii]$_{1907,1909}$ & 0.4$\pm$0.1 & 4.6$\pm$0.6 & 0.5$\pm$0.1 & 7.4$\pm$1.6 & 0.6$\pm$0.1 & 7.8$\pm$0.9 & 0.7$\pm$0.1 & 11.2$\pm$1.2 & 0.6$\pm$0.1 & 12.8$\pm$1.0 \\\relax
[\oii]$_{3727,3729}$ & 0.6$\pm$0.1 & 35.1$\pm$0.1 & 0.6$\pm$0.1 & 39.1$\pm$0.1 & 0.4$\pm$0.1 & 29.3$\pm$0.1 & 0.5$\pm$0.1 & 30.6$\pm$0.1 & 0.1$\pm$0.1 & 14.3$\pm$0.1 \\\relax
[\neiii]$_{3869}$ & 0.7$\pm$0.1 & 42.1$\pm$0.2 & 0.6$\pm$0.1 & 47.1$\pm$0.1 & 0.7$\pm$0.1 & 50.1$\pm$0.2 & 0.4$\pm$0.1 & 32.9$\pm$0.2 & 0.3$\pm$0.1 & 33.5$\pm$0.2 \\\relax
[\neiii]$_{3968}$ & 0.2$\pm$0.1 & 16.0$\pm$0.1 & 0.3$\pm$0.1 & 24.8$\pm$0.1 & 0.3$\pm$0.1 & 25.9$\pm$0.1 & 0.4$\pm$0.1 & 29.7$\pm$0.2 & 0.2$\pm$0.1 & 22.7$\pm$0.2 \\\relax
H$\delta_{4102}$ & 0.2$\pm$0.1 & 16.8$\pm$0.1 & 0.2$\pm$0.1 & 17.8$\pm$0.1 & 0.2$\pm$0.1 & 16.4$\pm$0.1 & 0.2$\pm$0.1 & 22.4$\pm$0.2 & 0.1$\pm$0.1 & 18.8$\pm$0.1 \\\relax
H$\gamma_{4341}$+[\oiii]$_{4364}$ & 0.6$\pm$0.1 & 48.0$\pm$0.1 & 0.6$\pm$0.1 & 61.7$\pm$0.2 & 0.5$\pm$0.1 & 57.1$\pm$0.3 & 0.7$\pm$0.1 & 76.3$\pm$0.6 & 0.3$\pm$0.1 & 55.4$\pm$0.7 \\\relax
H$\beta_{4863}$ & 1.00 & 116.2$\pm$4.2 & 1.00 & 136.9$\pm$4.5 & 1.00 & 153.1$\pm$6.9 & 1.00 & 139.5$\pm$14.4 & 1.00 & 157.1$\pm$16.6 \\\relax
[\oiii]$_{4960}$ & 1.6$\pm$0.1 & 181.1$\pm$5.3 & 1.6$\pm$0.1 & 226.1$\pm$8.4 & 1.9$\pm$0.1 & 299.5$\pm$11.1 & 2.1$\pm$0.3 & 295.3$\pm$31.0 & 1.4$\pm$0.3 & 215.3$\pm$36.7 \\\relax
[\oiii]$_{5008}$ & 4.1$\pm$0.1 & 464.8$\pm$9.3 & 4.8$\pm$0.2 & 685.5$\pm$19.5 & 5.3$\pm$0.3 & 837.6$\pm$24.5 & 5.9$\pm$0.8 & 858.2$\pm$87.7 & 4.0$\pm$0.8 & 603.9$\pm$103.1 \\\relax
H$\alpha_{6564}$+[\nii]$_{6585}$ & 2.7$\pm$0.1 & 546.2$\pm$8.4 & 3.4$\pm$0.1 & 912.3$\pm$39.4 & -- & --  & -- & --  & -- & --
\enddata
\tablecomments{The H$\beta$ fluxes are $3.3(\pm0.1)\times10^{-19}$, $3.1(\pm0.1)\times10^{-19}$, $2.9(\pm0.1)\times10^{-19}$, and $2.7(\pm0.3)\times10^{-19}$, and $2.8(\pm0.3)\times10^{-19}$ erg\,s$^{-1}$\,cm$^{-2}$ for the z5\_SF, z6\_SF, z7\_SF, z8\_SF, and z9\_SF stacks, respectively. All equivalent widths are quoted in the rest-frame and have units of \AA.}
\end{deluxetable}

\begin{deluxetable}{lllll}[h!]
\tabletypesize{\footnotesize}
\tablecolumns{5}
\tablewidth{0pt}
\centering
\tablecaption{The Same as Table~\ref{tab:line_ratios_redshift} But for the Ly$\alpha$ Emitter and non-Ly$\alpha$ Emitter Stacks from Table~\ref{tab:stacks}.}
\label{tab:lyaflux}
\tablehead{\colhead{} & \multicolumn{2}{c}{LAE} & \multicolumn{2}{c}{non-LAE}\\
\colhead{Line} & \colhead{$\frac{F_{\rm Line}}{F_{\rm H\beta}}$} & \colhead{EW$_{0}$} & \colhead{$\frac{F_{\rm Line}}{F_{\rm H\beta}}$} & \colhead{EW$_{0}$}}
\startdata
\civ$_{1548,1551}$ & 0.2$\pm$0.1 & 2.4$\pm$0.1 & -- & -- \\\relax
\heii$_{1640}$+[\oiii]$_{1660,1666}$ & 0.3$\pm$0.1 & 3.8$\pm$0.1 & -- & -- \\\relax
\ciii]$_{1907,1909}$ & 0.4$\pm$0.0 & 6.9$\pm$0.1 & 0.4$\pm$0.1 & 4.5$\pm$0.1 \\\relax
[\oii]$_{3727,3729}$ & 0.2$\pm$0.1 & 21.0$\pm$0.1 & 0.6$\pm$0.1 & 32.3$\pm$0.1 \\\relax
[\neiii]$_{3869}$ & 0.6$\pm$0.0 & 62.9$\pm$0.2 & 0.7$\pm$0.0 & 37.6$\pm$0.2 \\\relax
[\neiii]$_{3968}$ & 0.2$\pm$0.1 & 25.8$\pm$0.1 & 0.3$\pm$0.0 & 17.4$\pm$0.1 \\\relax
H$\delta_{4102}$ & 0.2$\pm$0.1 & 25.4$\pm$0.1 & 0.2$\pm$0.1 & 14.5$\pm$0.1 \\\relax
H$\gamma_{4341}$+[\oiii]$_{4364}$ & 0.6$\pm$0.1 & 86.7$\pm$0.4 & 0.6$\pm$0.1 & 46.1$\pm$0.2 \\\relax
H$\beta_{4863}$ & 1.00 & 215.8$\pm$7.3 & 1.00 & 118.7$\pm$3.0 \\\relax
[\oiii]$_{4960}$ & 1.5$\pm$0.1 & 342.0$\pm$13.5 & 1.7$\pm$0.1 & 168.0$\pm$4.5 \\\relax
[\oiii]$_{5008}$ & 4.4$\pm$0.2 & 1014.9$\pm$30.3 & 4.3$\pm$0.1 & 446.9$\pm$9.4 \\\relax
H$\alpha_{6564}$+[\nii]$_{6585}$ & 2.8$\pm$0.1 & 1080.9$\pm$52.7 & 2.7$\pm$0.1 & 469.0$\pm$8.4 
\enddata
\tablecomments{The H$\beta$ fluxes are $4.0(\pm0.1)\times10^{-19}$ erg\,s$^{-1}$\,cm$^{-2}$ for the LAE stack and $3.0(\pm6.0)\times10^{-19}$ erg\,s$^{-1}$\,cm$^{-2}$ for the non-LAE stack. All equivalent widths are quoted in the rest-frame and have units of \AA.}
\end{deluxetable}

\end{document}